%% file: main.tex
\documentclass[namedreferences]{solarphysics}
\pdfoutput=1
\usepackage[hyperref,optionalrh,solaromanenum,linksfromyear]{spr-sola-addons} 
\usepackage{graphicx}                    
\usepackage{color}                       

\begin{document}

\begin{article}
\include{defs}

\begin{opening}

\title{A Global Survey of EUV Coronal Power Spectra}

\author[addressref={aff1},corref,email={karl.battams@nrl.navy.mil}]{\inits{K.}\fnm{Karl}~\lnm{Battams}\orcid{0000-0002-8692-6925}}
\author[addressref={aff1,aff2},email={brendan.gallagher@nrl.navy.mil}]{\inits{B. M.}\fnm{Brendan~M.}~\lnm{Gallagher}\orcid{0000-0002-8353-5865}}
\author[addressref={aff2},email={rweigel@gmu.edu}]{\inits{R. S.}\fnm{Robert~S.}~\lnm{Weigel}\orcid{0000-0002-9521-5228}}

\address[id=aff1]{US Naval Research Laboratory, Washington D.C., USA}
\address[id=aff2]{George Mason University, Fairfax VA, USA}
%
\runningauthor{Battams et al.}
\runningtitle{EUV Coronal Power Spectra}



\begin{abstract}
We present results of an investigation of single-pixel intensity power spectra from a 12-hour time period on 26 June 2013 in a 1600$\times$1600-pixel region from four wavelength channels of NASA's \textit{Solar Dynamics Observatory Atmospheric Imaging Assembly}. We extract single-pixel time series from derotated image sequences, fit two models as a function of frequency $[\nu]$ to their computed power spectra, and study the spatial dependence of the model parameters: i) a three-parameter power-law + tail, $A\nu^{-n}+C$, and ii) a power-law + tail + three-parameter localized Lorentzian, $A\nu^{-n} + C + \alpha/\left(1 + \left(\ln\nu-\beta)^2\right/\delta^2\right)$, the latter to model periodicity. Spectra are well-described by at least one of these models for all pixel locations, with the spatial distribution of best-fit model parameters shown to provide new and unique insights into turbulent, quiescent and periodic features in the EUV corona and upper photosphere. Findings include: individual model parameters correspond clearly and directly to visible solar features; detection of numerous quasi-periodic three- and five-minute oscillations; observational identification of concentrated magnetic flux as regions of largest power-law indices $[n]$; identification of sporadically located five-minute oscillations throughout the corona; detection of the known global $\approx4.0$-minute chromospheric oscillation; 2D spatial mapping of ``coronal bullseyes'' appearing as radially decaying periodicities over sunspots and sporadic foot-point regions, and of ``penumbral periodic voids'' appearing as broad rings around sunspots in 1600 and 1700\,{\AA} in which spectra contain no statistically significant periodic component.
\end{abstract}

%

\end{opening}

%

\section{Introduction}

Studies of power spectra are common in solar physics and are used to provide insight into physical processes from the photosphere out to the solar-wind. Many of these studies rely on magnetohydrodynamic (MHD) or reduced MHD (RMHD) simulations of small regions of the corona. Central to these investigations is the study of turbulence or, more specifically in the solar case, MHD turbulence, which is frequently proposed as a mechanism for transferring energy from the denser and cooler photospheric regions to the diffuse and super-heated outer corona. While MHD turbulence differs from fluid turbulence, a Kolmogorov-like turbulent cascade is frequently cited as a plausible means for coronal heating \citep{Ballegooijen86,Heyvaerts92,Einaudi96}. Furthermore, turbulence -- and specifically turbulence driven by wave reflection within the corona -- has been proposed to play a key role in fast solar-wind acceleration and can be related to the turbulent power spectra observed in the solar-wind further out in the heliosphere \citep{Verdini09}. On small spatial scales, several authors have studied spectra from coronal-loop simulations \citep[e.g.][]{Dmitruk97,Dmitruk03,Rappazzo10,Taroyen10}, generally finding power spectra in these regions to be well described by large power-law indices (corresponding to slopes of -2 to -3). A high-level overview of the characterization of coronal turbulence is given by \cite{Zhou04}, who review the energy spectra of MHD turbulence in several special cases and provide references to theoretical derivations of spectra properties in different turbulence regimes under many complex scenarios.

Observation-driven spectral analysis includes the works of \cite{McIntosh04}, \cite{McIntosh08}, \cite{Reznikova12a} \cite{Reznikova12b} and \cite{Jess2012a}, who leveraged high-resolution space-based observations to focus on spatially small regions (\textit{e.g.} sunspots and loops), typically examining only limited frequency ranges of the full power spectrum (\textit{e.g.} three- to five-minute oscillations). Ground-based observations have been used to study localized chromospheric regions (\textit{e.g.} \cite{Reardon08, Tziotziou07} and many more), but the emphasis is primarily on isolation of specific frequencies, rather than a broad analysis of the entire available power spectrum. Observations from the Transition Region and Coronal Explorer (TRACE: \cite{Handy99}) were put to similar use by \cite{Muglach03}, who performed a detailed study of the oscillatory nature of sunspot regions, and by \cite{McIntosh04} who applied wavelet-based techniques to study oscillations in limited regions of interest in the photosphere. Furthermore, the energy of MHD waves propagating upward from the chromosphere to the corona has long been believed to be a contributor to coronal heating (\cite{Alfven47}, recently summarized by \cite{Arregui15} and references therein), and thus studies of coronal oscillations and waves, and their damping properties, is of key importance to the understanding of the very broad topic of coronal heating \citep{Parnell12}. 

With the advent of high-temporal resolution observations from the NASA Solar Dynamics Observatory (SDO) \citep{Pesnell12}, a number of new studies relating to power spectra (or Fourier transform and wavelets) have arisen. For example, \cite{Reznikova12a} and \cite{Reznikova12b} considered the periodic spectral properties of a sunspot across multiple wavelengths, and they have exploited the high temporal and spatial resolution of the SDO Atmospheric Imaging Assembly (AIA: \cite{Lemen11}). \cite{Threlfall17} combined SDO/AIA disk observations with white light observations from the Solar Terrestrial Relations Observatory (STEREO) to trace periodic signatures from the solar disk out into the heliosphere. Meanwhile, the power-law behavior of solar power spectra has been increasingly investigated in a number of studies such as \cite{Inglis15} and \cite{Auchere16a}. A more general approach by \cite{Ireland15} used a power-law + tail + localized Gaussian model to describe solar power spectra from SDO/AIA images and noted that select solar features appear to have different characteristic power spectra. Despite an increased focus on power spectral behavior in the solar corona, there have been no published global studies beyond \cite{Ireland15}, who considered average spectral properties in large regions across broad frequencies and for two wavelength channels, to investigate the broad spectral properties of all solar coronal features across spatially large regions and multiple wavelength channels. 

In summary, theoretical or simulation-based models of turbulence (MHD or Reduced MHD) in the corona, and the power spectra that result from these models, apply directly to many key unanswered questions in solar physics involving energy transfer, wave propagation, coronal heating, and solar-wind turbulence.  Such studies require both high temporal and spatial cadence measurements sufficient to enable high-precision spectral analysis of coronal observations. In recent years, most notably with the launch of SDO, data of sufficient temporal and spatial resolution have become available, but techniques needed to derive the properties of power spectra over both large regions and at high spatial resolution have not been developed.

Our work begins by extending \cite{Ireland15} by averaging spectra over much smaller regions (3$\times$3 vs. $\approx$50$\times$50 pixels). In addition, instead of considering select sub-regions containing a given feature in the AIA images and then computing a best-fit model for the average power spectra of all pixels in the sub-regions, we compute a best-fit model for each pixel at the center of a 3$\times$3 pixel sub-region in a 1600$\times$1600 pixel ($\approx$ 1000$\times$1000 arc-second) AIA image and then study the relationship between the spatial distribution of the best-fit model parameters and features in the AIA image. We also provide an  improvement to the model by replacing the Gaussian term with a more physically meaningful Lorentzian to better describe the damped oscillatory features observed in many coronal power spectra. The presented methodology can be used to i) support simulation-driven studies with observation-based spectra at high spatial and temporal resolution, ii) provide observation-based studies with a framework that enables the exploration of spatially large regions at full spatial and temporal resolution and across multiple wavelengths, and iii) enable the parameterization of large regions of the solar disk based on the properties of a two- or three-component power spectral model. 

In this work, we provide details on the methodology, apply it to a 12-hour time interval of images from four wavelength channels of the SDO/AIA instrument, and then highlight a number of key discoveries and observations found in this application.

\section{Data and Methodology}

\subsection{Data Selection}

Our analysis uses data recorded by SDO/AIA on 26 June 2013, from 00:00:00 to 11:59:59\,{UT}. The AIA instrument observes at wavelengths of 94, 131, 171, 193, 211, 304, 335, 1600, and 1700\,{\AA} at 4096$\times$4096 pixel resolution (corresponding to a spatial resolution of approximately 1.67 arc-seconds per pixel) and with a nominal cadence of 12 seconds in each wavelength channel (24 seconds for 1600 and 1700\,{\AA}).

\begin{figure}[t]
\centering

{\includegraphics[width = 3.3in]{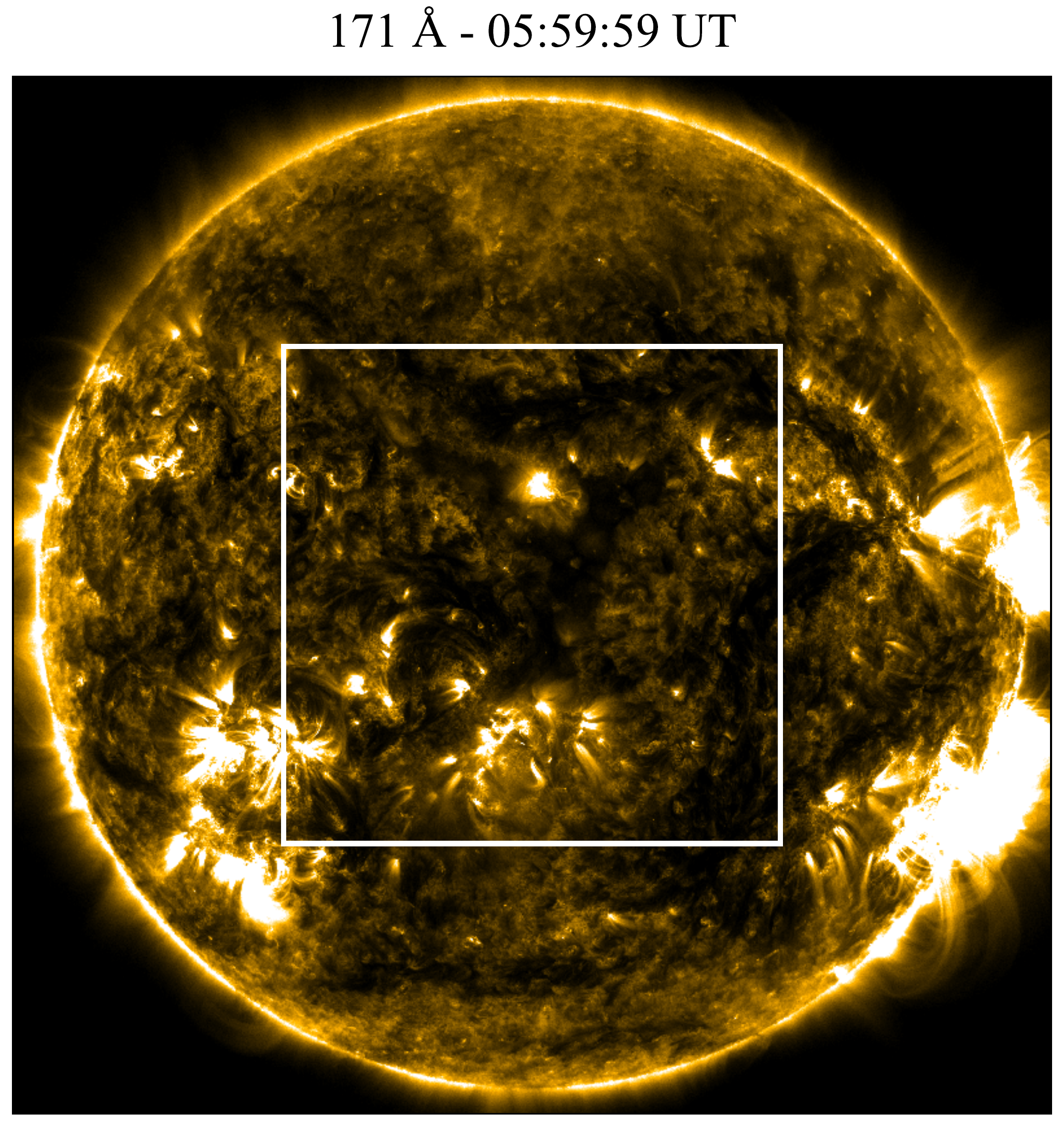}} 
\caption{AIA 171\,{\AA} full-disk image indicating the sun-centered 1600$\times$1600-pixel region of analysis considered for AIA wavelength channels 171, 193, 304, and 1700\,{\AA}.  This image was recorded on 26 June 2013 at 05:59:59\,{UT}.}
\label{f:Full-Disk-Box}
\end{figure}

The selection of this time interval was motivated by a desire to study a morphologically diverse corona, on a spatially global scale, in a single analysis; the selected region contains three diverse solar features: an active region (AR-1777), a coronal hole, and a filament. Also a factor in the choice of the time interval was that no major flares or eruptions occurred, as we are still investigating the impact of such events on our analyses. A small B9.2 flare did occur around 04\,{UT} during our identified time period, but as discussed later this did not have any impact on our results.  Figure~\ref{f:Full-Disk-Box} shows a full-disk 171\,{\AA} image during the selected time interval, with a white box indicating our specific sub-region of interest.

We present results obtained from AIA images in the 171, 193, 304, and 1700\,{\AA} wavelength channels. We omitted analysis of 131, 335, and 94\,{\AA} because their higher noise levels created problems with spectral curve fitting and we omit analysis of 211\,{\AA} because of the similarity of results with 193\,{\AA}. We also omit a full presentation of the 1600\,{\AA} results primarily because of the similarity with that of 1700\,{\AA}, but also because 1700\,{\AA} has less transition region contamination than 1600\,{\AA}, making it a better representation of the chromospheric continuum \citep{Lemen11}. For each of the wavelengths used.., we selected a 1600$\times$1600 pixel sub-region centered on the solar disk, as indicated in Figure~\ref{f:Full-Disk-Box}. To aid with interpretation of our results, we also obtained a single SDO Helioseismic Magnetic Imager (HMI) magnetogram and HMI continuum image from the middle of our sequence (26 June 2013 06:00\,UT), but note that these were used for qualitative analyses only, and thus no calibrations were performed on these data.

\subsection{Data Preparation}

Data were obtained from the Virtual Solar Observatory (VSO: \cite{Hill09}) using routines in the \textsf{SunPy} package \citep{Sunpy15}. All analyses were performed using Python along with the \textsf{NumPy} and \textsf{SunPy} packages, with visualizations created with the \textsf{Matplotlib} package. 

For each of the considered AIA wavelengths, the observed cadence was 12 or 24 seconds, although sporadic data gaps of up to $\approx$90 seconds were encountered. For each wavelength, we extracted the 1600$\times$1600-pixel Sun-centered region for each time step, compiled them into a single data cube, and then applied \textsf{SunPy}'s $\sf{differential\_derotation}$ function\footnote{Available in the latest release of \textsf{Sunpy}, v0.8.5} so that each image is presented as if from the perspective of an observer rotating at the same rate as the solar Equator, with corrections applied to account for differential rotation of the solar atmosphere. All images were divided by their exposure time to create what we refer to as normalized intensity images. 

\begin{figure}
  \centering
  \setlength{\tabcolsep}{1.0mm}
  \begin{tabular}{cc}
  \includegraphics[scale=0.203]{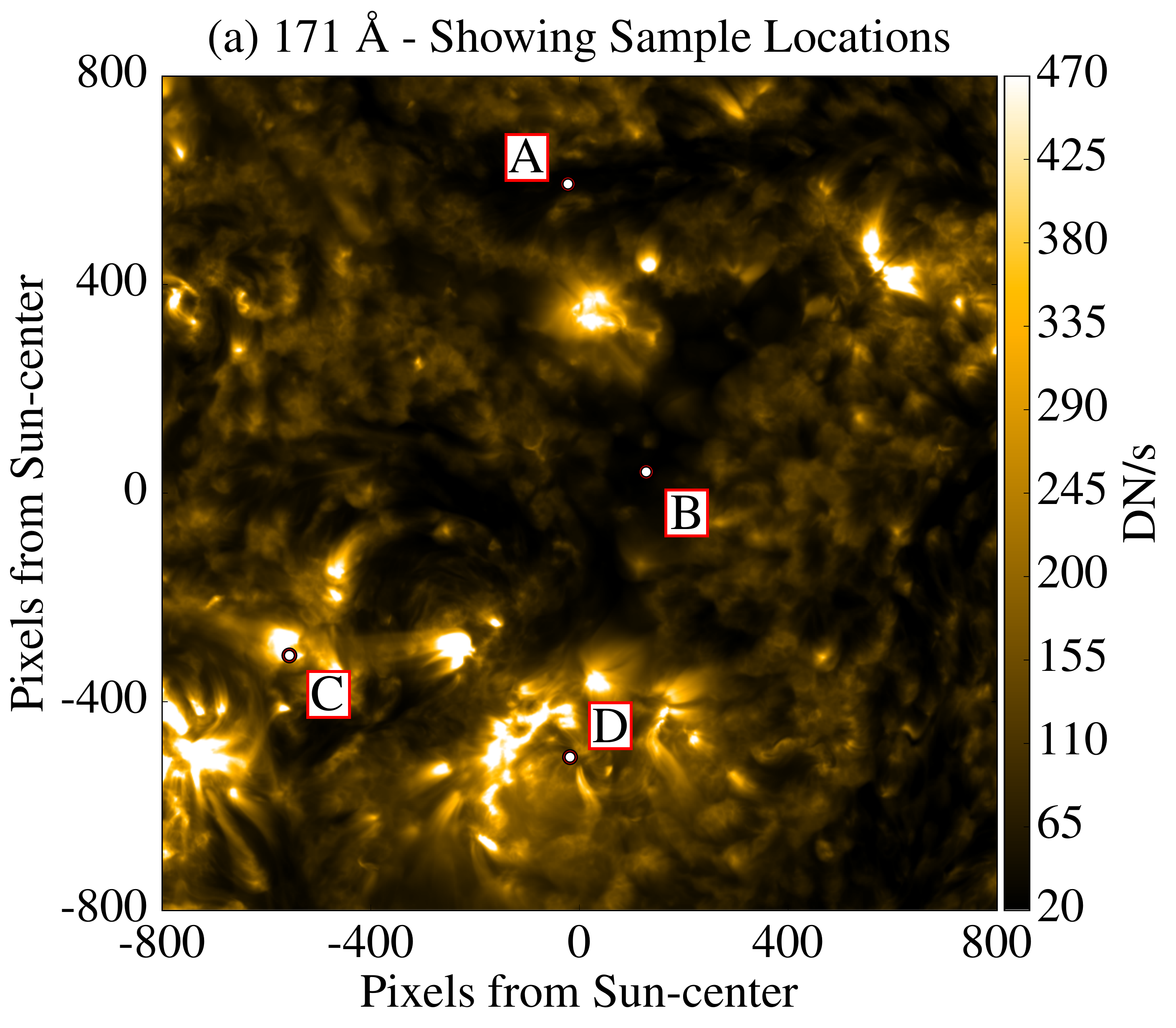}&\includegraphics[scale=0.203]{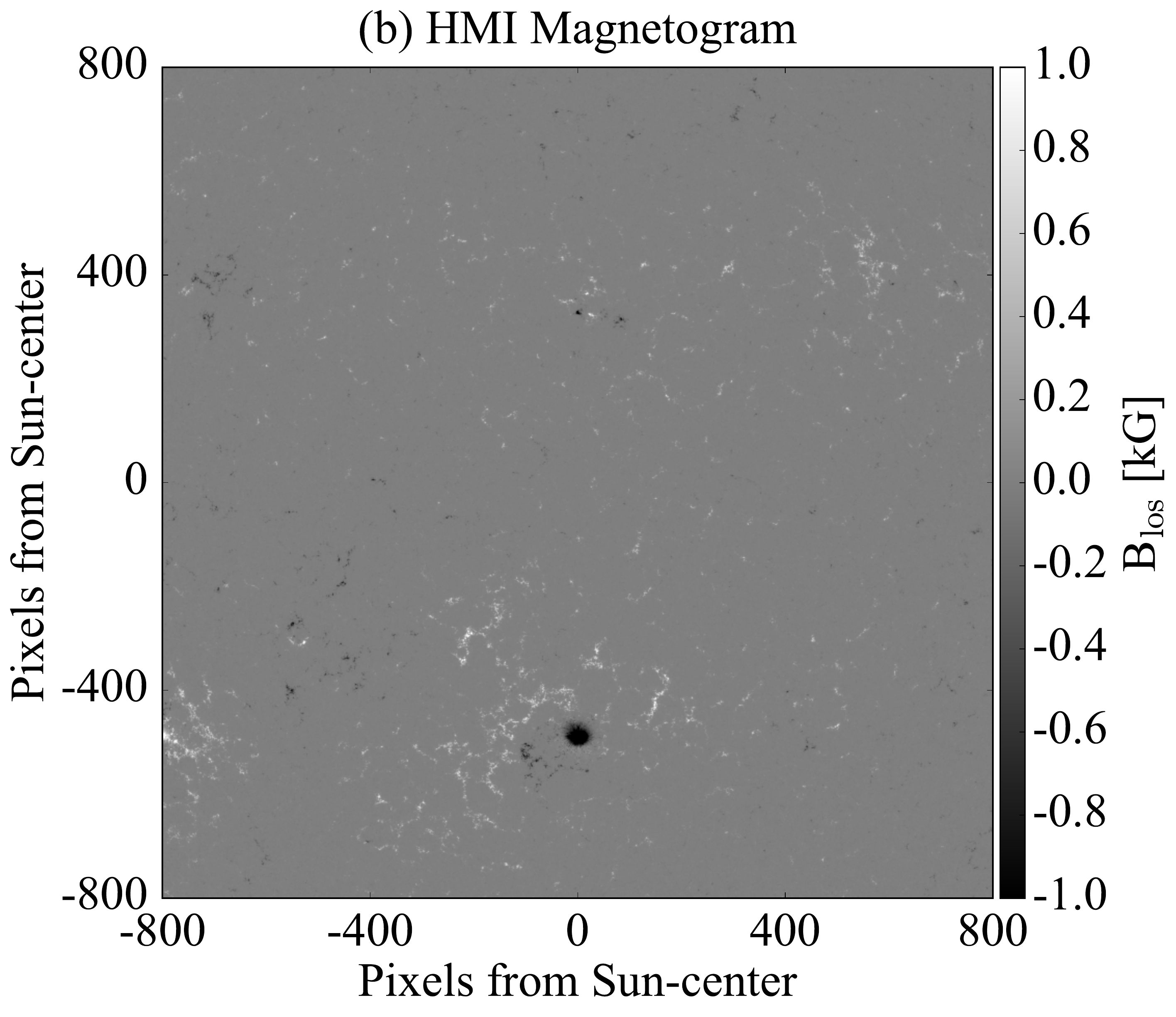}\\
  \includegraphics[scale=0.203]{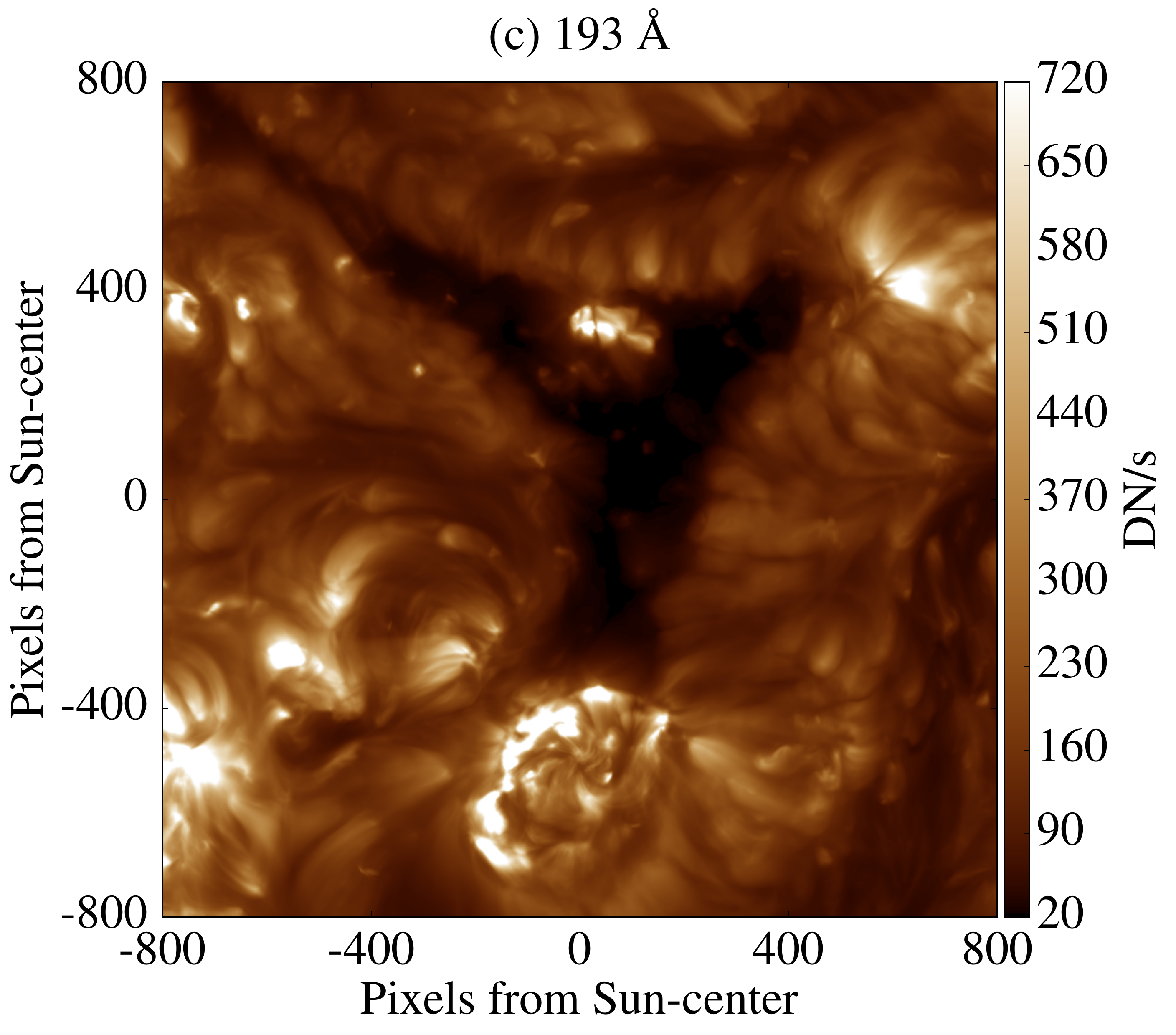}&\includegraphics[scale=0.203]{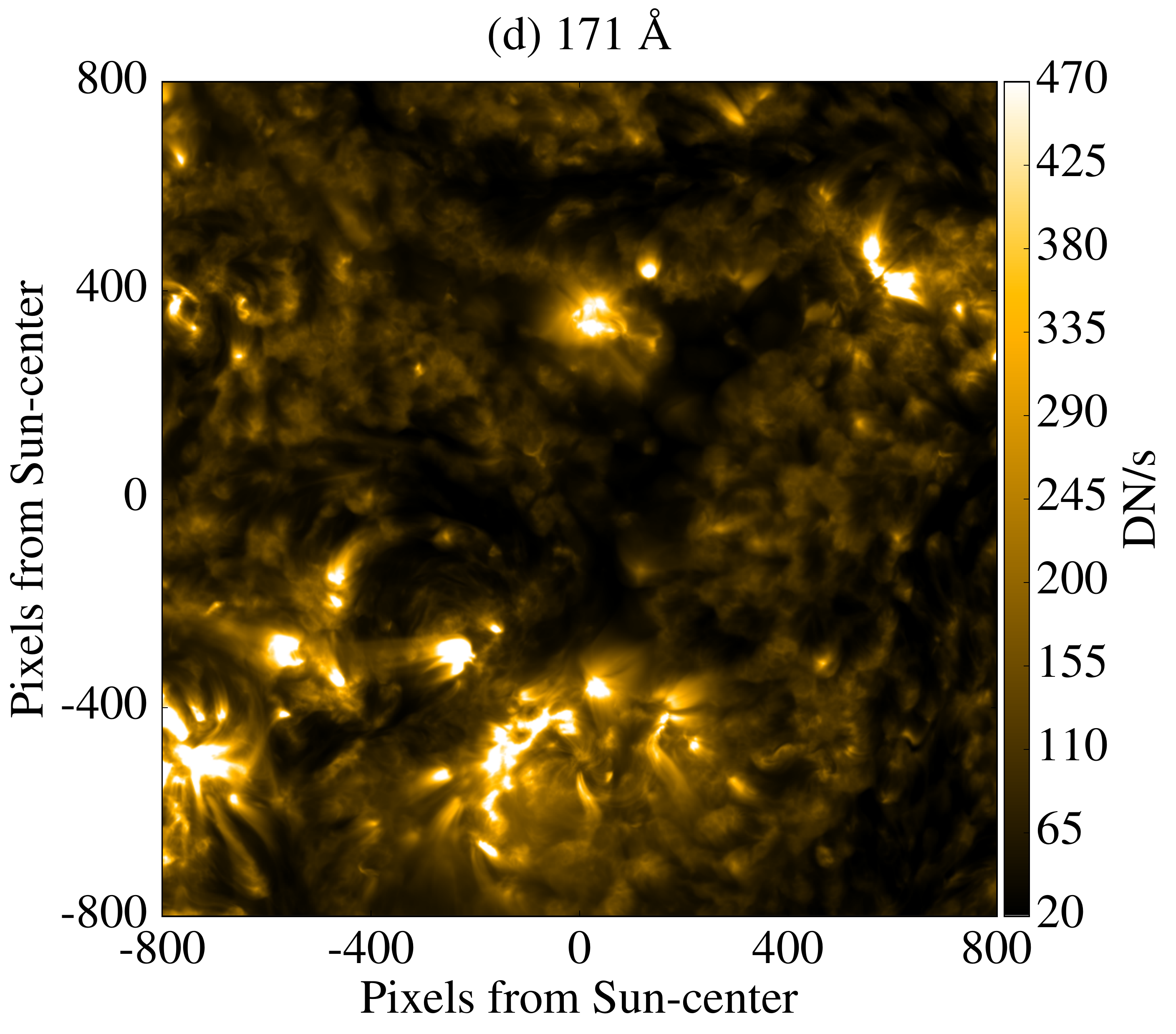} \\
  \includegraphics[scale=0.203]{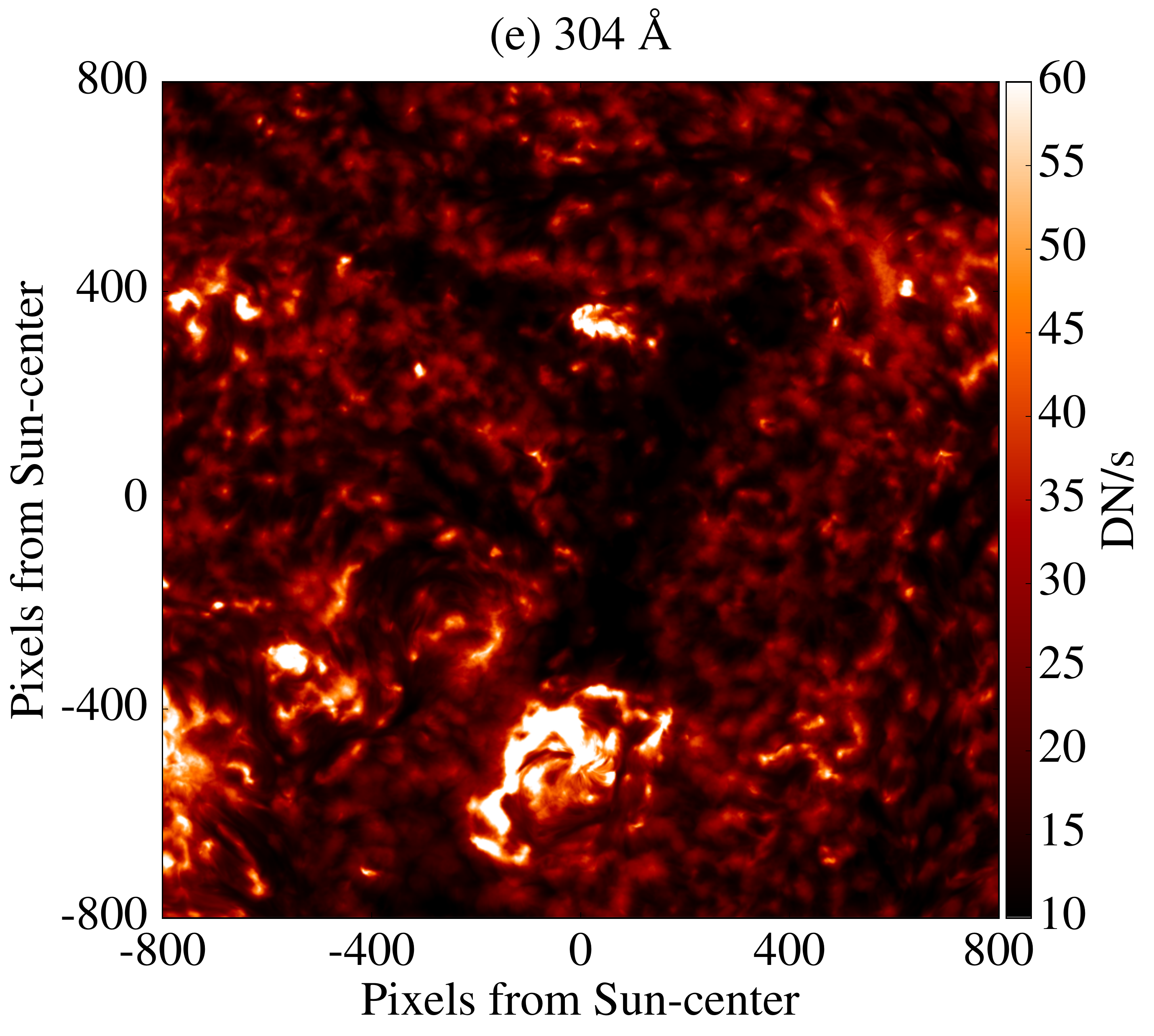}&\includegraphics[scale=0.203]{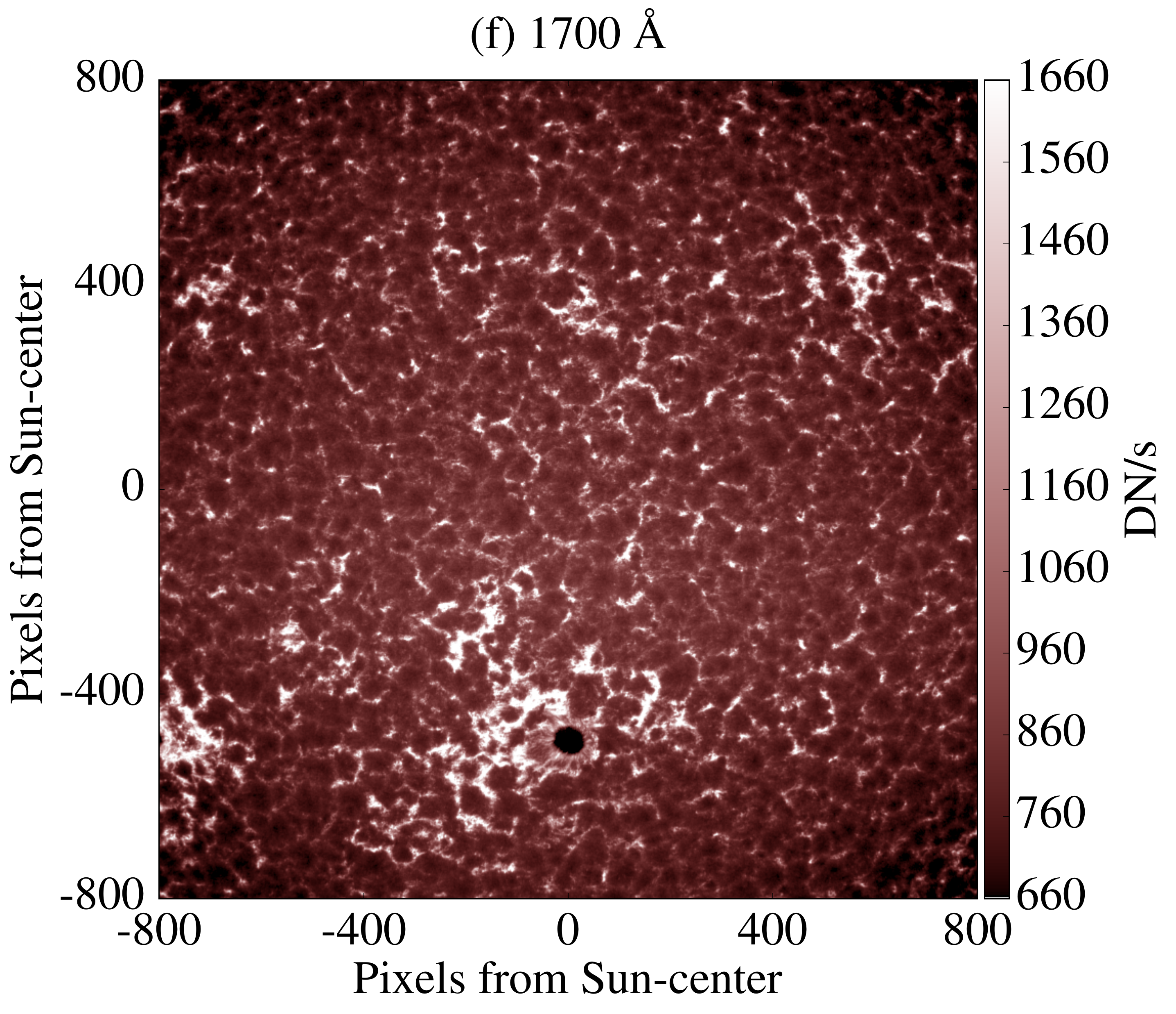}
  \end{tabular}
  \caption{Average of normalized intensity images after deroration for each of the studied AIA channels from 26 June 2013 from 00:00:00 to 11:59:59 on a 1600$\times$1600 pixel region centered on Sun-center for the 06\,{UT} observation. Panel a indicates the locations associated with the sample time series and spectra presented in Figure~\ref{f:All_TS} and~\ref{f:All_Fits}, Panel b shows the corresponding HMI Magnetogram (06:00\,{UT} observations) for this region of interest, and panels \,c--\,f show the arithmetic average for each wavelength channel. The $x$- and $y$-axes have labels for pixel value relative to Sun-center. The units on the colorbar [DN\,s$^{-1}$] are described in the text.}
\label{f:All_viz}
\end{figure}

\begin{figure}
  \centering
  \setlength{\tabcolsep}{1.0mm}
  \begin{tabular}{cc}
  \includegraphics[scale=0.205]{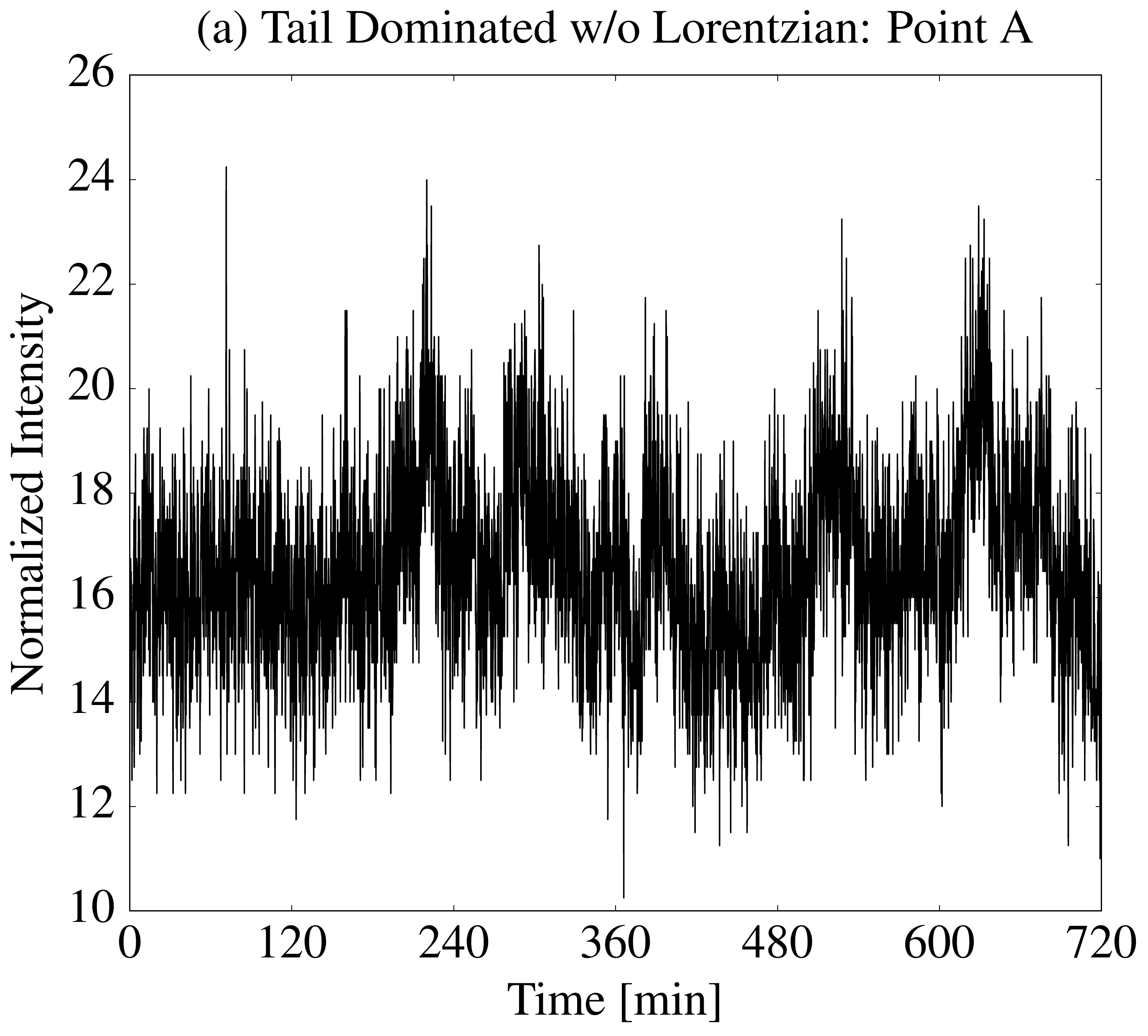}&\includegraphics[scale=0.205]{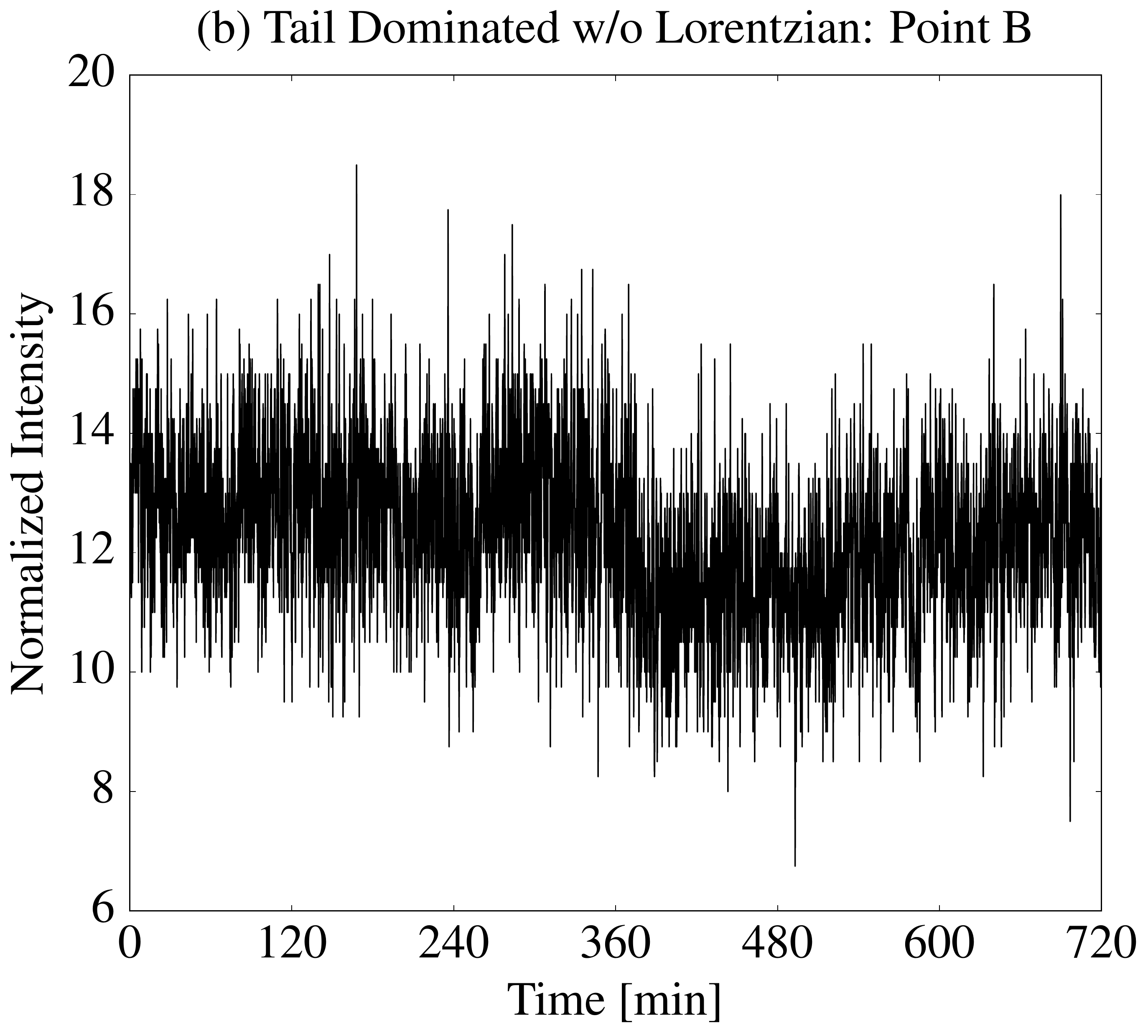}\\
  \includegraphics[scale=0.205]{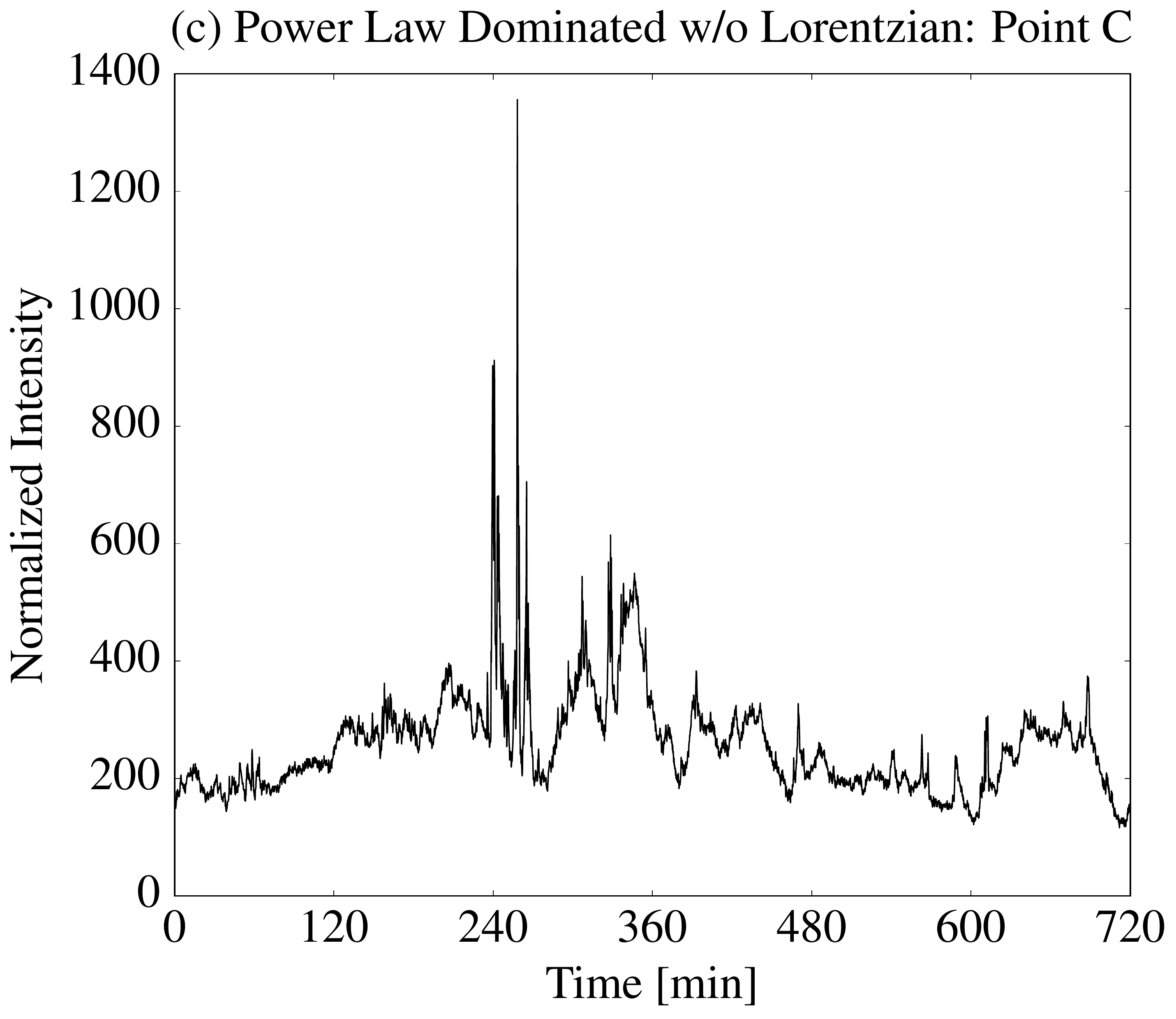}&\includegraphics[scale=0.205]{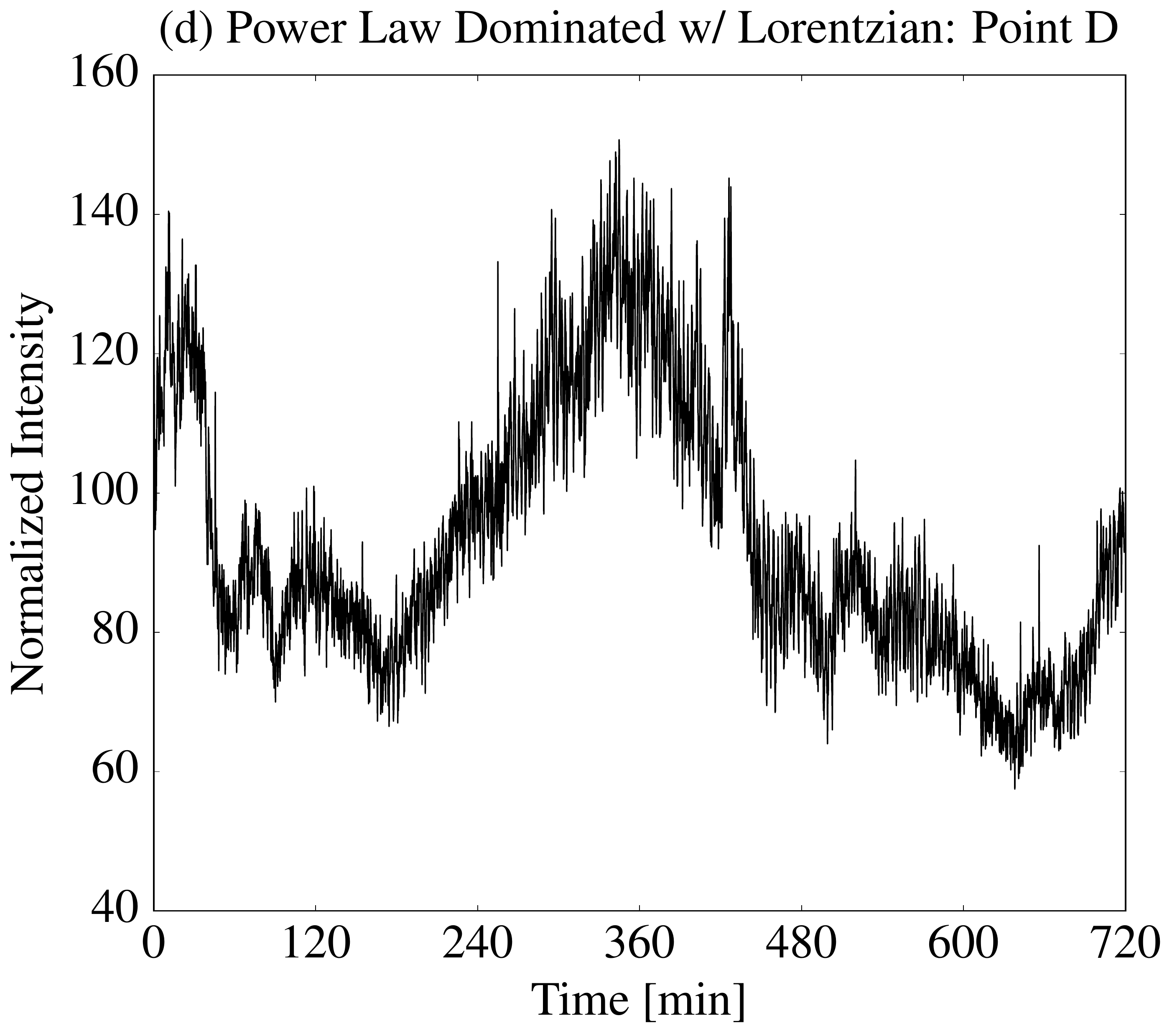} 
  \end{tabular}
  \caption{Time series for the sample points \,A--\,D indicated in Figure~\ref{f:All_viz}a. The titles in each panel include our characterizations of each point that are defined in Section~\ref{s:Categories}.}
\label{f:All_TS}
\end{figure}

Figure~\ref{f:All_viz} shows the arithmetic mean of the normalized intensity images of all four wavelength channels under investigation, as well as the corresponding region in HMI magnetogram observations (Panel b). (All observations in a given sequence were corrected for exposure time, then summed into a single image and divided by the number of images in the sequence to produce the images shown in Figure~\ref{f:All_viz}a and \,c--\,f.) Figure~\ref{f:All_viz}a shows sample locations referenced in later Figures~\ref{f:All_TS} and~\ref{f:All_Fits}; these locations were selected as representative of four broad categorizations of power spectra and are located on specific coronal features: a filament (Point A), a coronal hole (Point B), a small bright loop foot-point (Point C), and a sunspot umbra (Point D). Figure~\ref{f:All_viz}b shows the HMI Magnetogram observation corresponding to this region of interest. Figures~\ref{f:All_viz}\,c--\,f are the four wavelengths discussed in Section~\ref{s:results} shown in the order that they are discussed.

Following the differential rotation correction, we extracted the 12-hour time series of pixel intensities for each pixel in the 1600$\times$1600 region for each wavelength, and each time series was converted to its spectral equivalent by way of a Fast Fourier Transform (FFT). The spectra of these extracted time series (or \textit{light-curves}) are the focus of this investigation. Data were \textit{not} calibrated via the IDL-based $\sf{aia\_prep.pro}$ routine that is used for partial calibration of AIA observations. The omission of this preprocessing step means that, at the highest spatial resolution, we cannot precisely co-align images obtained from different AIA wavelength channels. However, the analyses presented here do not require such co-alignment.

Before applying the FFT, we linearly interpolated across missing time values onto a uniform time grid corresponding to each wavelength's nominal cadence, \textit{i.e.} a time series obtained at the nominal 12-second cadence, but with sporadic 24- or 36-second gaps, was linearly interpolated onto a 12-second time grid. Figure~\ref{f:All_TS} shows four time series extracted at the sample locations indicated in Figure~\ref{f:All_viz}a following interpolation. No preprocessing of the time series beyond linear interpolation was performed. We experimented with the use of various forms of apodization and found that our interpretation and results were unchanged. Similarly, we found that the use of the Lomb--Scargle algorithm for computing the power spectrum (instead of using linear interpolation to fill data gaps and using the standard FFT algorithm) gave spectra that were effectively indistinguishable.

There exist a number of different methods for investigating time series in the frequency domain, with wavelet-based methods particularly common. To assess the compatibility between the often-used Morlet wavelet transformation and the FFT used here, we computed spectra from both of these methods, looking at their output over a broad range of spectra covering each spectral type discussed in our analysis.  In general, the spectra returned by each method had largely the same shape and essentially identical periodicity locations (\textit{described in Section~\ref{s:fitting}}) for spectra in which a statistically significant Lorentzian fit was identified. Therefore,  our full-scale analysis is expected to yield comparable overall results if replicated using wavelet-based methods of spectra computation. However, given that the transient behavior of interest is often due to sinusoidal waves over many periods (as opposed to periodic spikes with broadband spectral content), the Fourier approach seems to be a natural choice.  In addition, the Fourier approach has advantages for computation speed.

\subsection{Spectra Calculation}
\label{s:fitting}
For reliable model-fitting, we found it necessary to first perform a set of averaging steps to reduce the noise in the spectra. In the first averaging step, each 12-hour time series for a pixel is split into six non-overlapping two-hour segments and the arithmetic average of the six power spectra associated with these segments is computed. This averaging (``segmenting'') is a necessary trade-off that allows us to reduce the noise in power spectra while still retaining the gross spectral behavior over the 12-hour time period. We discuss the limitations of this approach in Section~\ref{s:limitations}. In the second averaging step, the arithmetic average of the previously computed spectra in 3$\times$3 pixel-boxes around each pixel is computed, with the result saved as the final spectra for the center pixel in the 3$\times$3 box. To avoid edge effects from the 3$\times$3 averaging procedure, a one-pixel border was eliminated from the final data product, resulting in 1598$\times$1598 spectra.

Without these averaging steps for the spectra, many model fits ``failed" in the sense that they visually did not capture the gross features of the spectra, two visually similar spectra had very different fits, with one fit having a significantly larger data/model error, or the optimization routine would not converge at all. Details on these issues are given in Appendix~\ref{Appendix:SpectralFitting}, and figures showing the effect of the segment averaging are shown in Appendix~\ref{Appendix:seg_avg}. It can be seen in Appendix~\ref{Appendix:seg_avg} that different levels of segmenting can produce different values of both Lorentzian location and power-law slopes for any given feature. In Section~\ref{s:segment_discuss} we discuss our exploration of these variations as a function of segmenting, finding that power laws are indeed systematically biased by averaging, although not excessively, and that Lorentzian locations for all good M2 fits are largely unchanged. We also find that the 3$\times$3-pixel averaging has a negligible impact on either parameter value, and it only makes it easier for the curve-fitting algorithm to converge to a good fit. Thus our chosen approach employing these two averaging steps is an approach that resulted in few failed model fits for nearly all pixel locations. 

The average spectra for the time series shown in Figure~\ref{f:All_TS} are shown in Figure~\ref{f:All_Fits} as solid black lines. The other curves in Figure~\ref{f:All_Fits} are discussed in the following section.

\subsection{Spectral Fitting}

The spectral models used in this work are based on those used by \cite{Ireland15}, and we will follow the same parameter and model naming convention. The ``M1'' model for the power spectra [$P$] consists of a power law with a ``tail' and contains three parameters: the amplitude coefficient [$A$], the power-law index [$n$], and the `tail' coefficient [$C$]:
\begin{equation}
\mbox{M1:}\qquad P_{1}(\nu) = A\nu^{-n} + C.
\label{eq:M1_revised}
\end{equation}

This model is not a true power law, but has a flattening above a frequency that is approximately determined by what we define as the rollover frequency,

\begin{equation}
\nu_r = (C/A)^{-1/n},
\label{eq:rollover_freq}
\end{equation}

\noindent or equivalently, the rollover period,

\begin{equation}
T_r = (C/A)^{1/n}.
\label{eq:rollover_period}
\end{equation}

\noindent the value of which is related to the photon noise amplitude in the observations. For values of $\nu$ below $\nu_r$, model M1 approaches a true power-law.

The M1 model is `nested' within model M2. \cite{Ireland15} presents the M2 as the sum of M1 and an additional Gaussian component. Our early work followed this same convention, but it was recently modified to replace the Gaussian with a Lorentzian function. The motivation here is that the Lorentzian is more physically meaningful in the context of damped oscillations, as the amplitude of a damped harmonic oscillator are represented by a Lorentzian curve \citep[e.g.][]{Nakariakov16}, with several authors having employed such harmonic oscillator models in studies of damped oscillations in coronal loops \citep{Nistico13,Anfinogentov15}. Thus our M2 model takes the form:

\begin{equation}
\mbox{M2:}\qquad P_{2}(\nu) = {A\nu^{-n}}+\frac{\alpha}{1+(\ln\,\nu -\beta)^2/\delta^2}+C
\label{eq:M2}
\end{equation}

The additional parameters correspond to the Lorentzian component amplitude [$\alpha$], location [$\beta$], and width [$\delta$].

\begin{figure}
  \centering
  \setlength{\tabcolsep}{1.0mm}
  \begin{tabular}{cc}
  \includegraphics[scale=0.21]{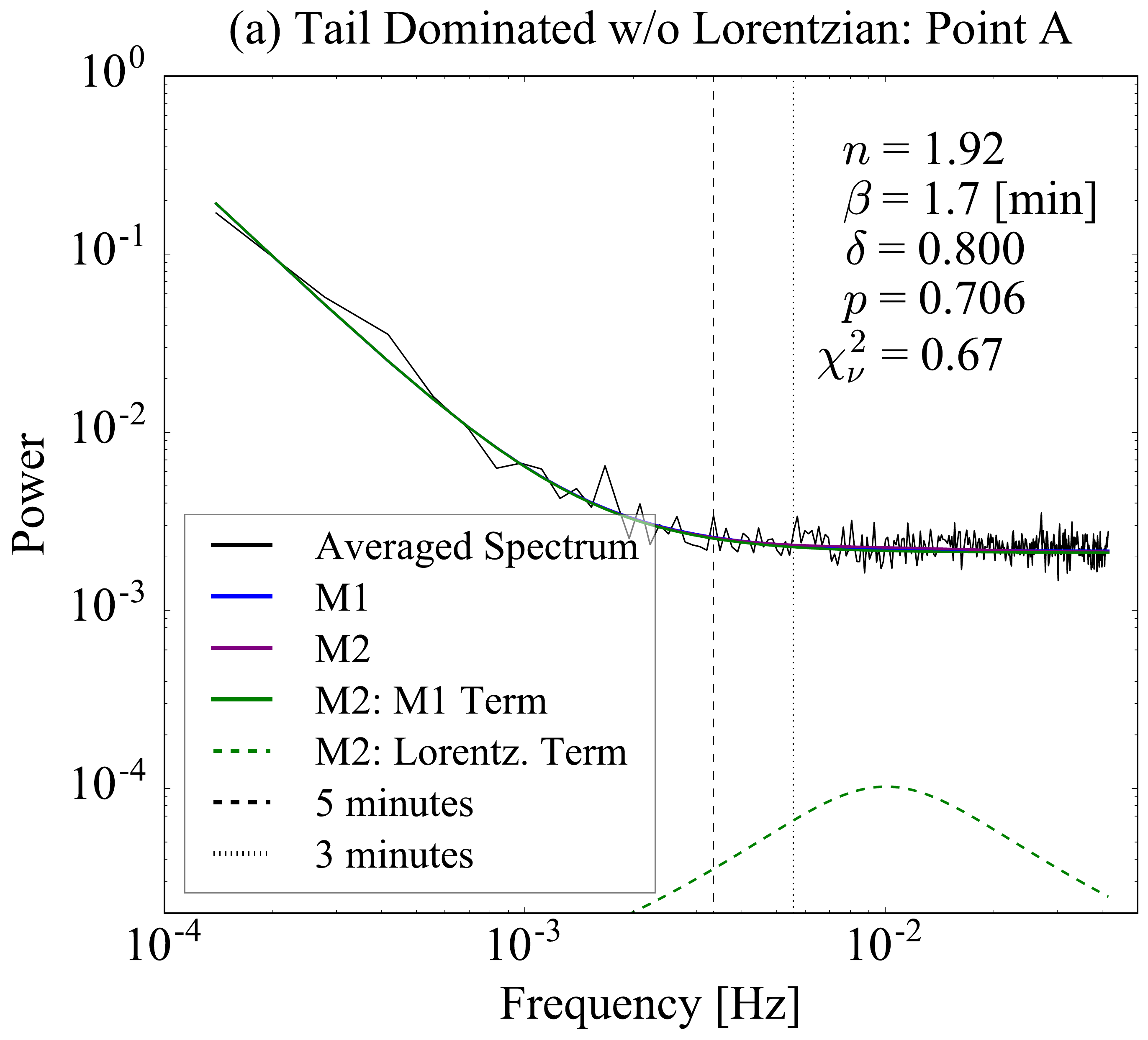}&\includegraphics[scale=0.21]{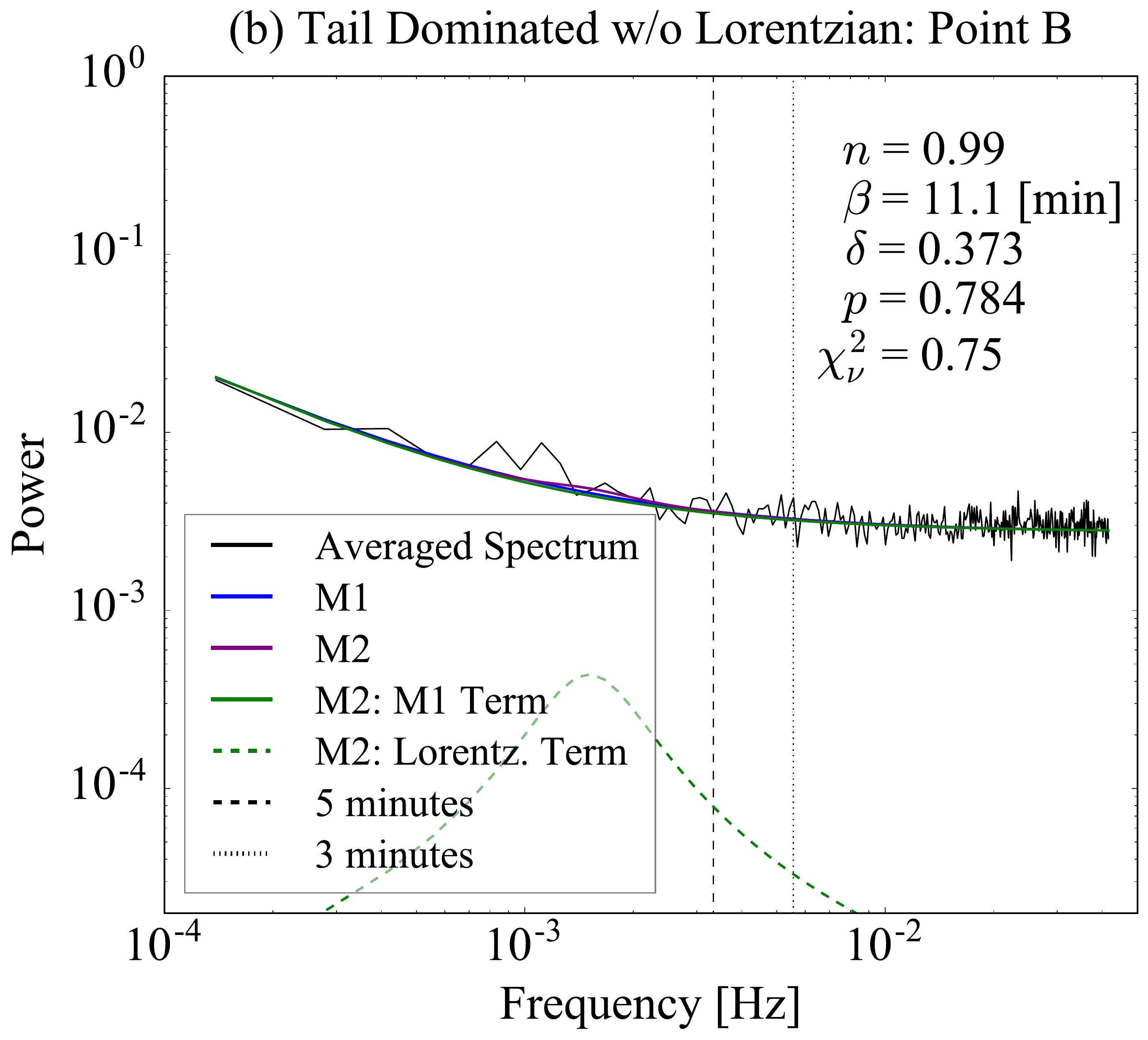}\\
  \includegraphics[scale=0.21]{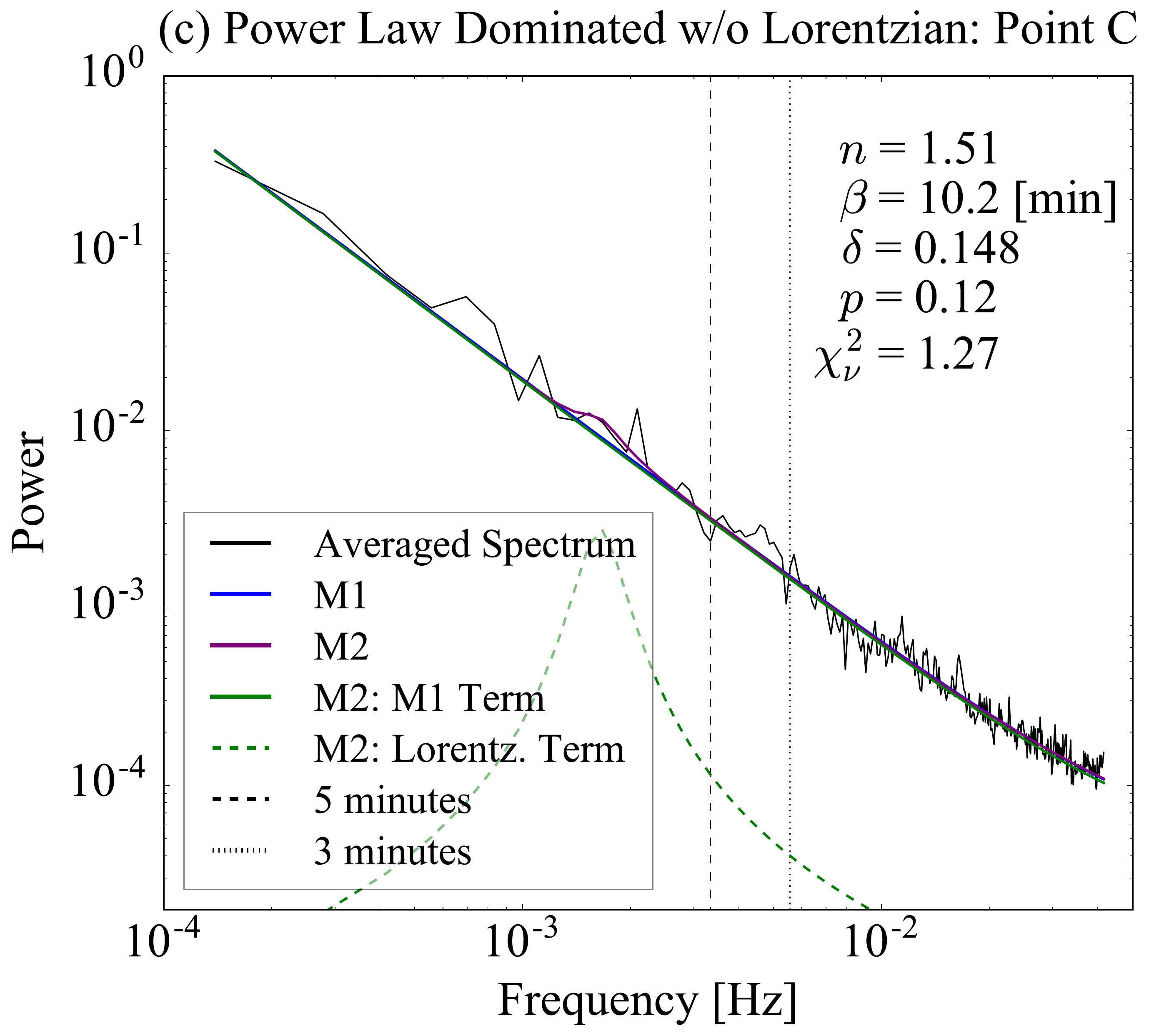}&\includegraphics[scale=0.21]{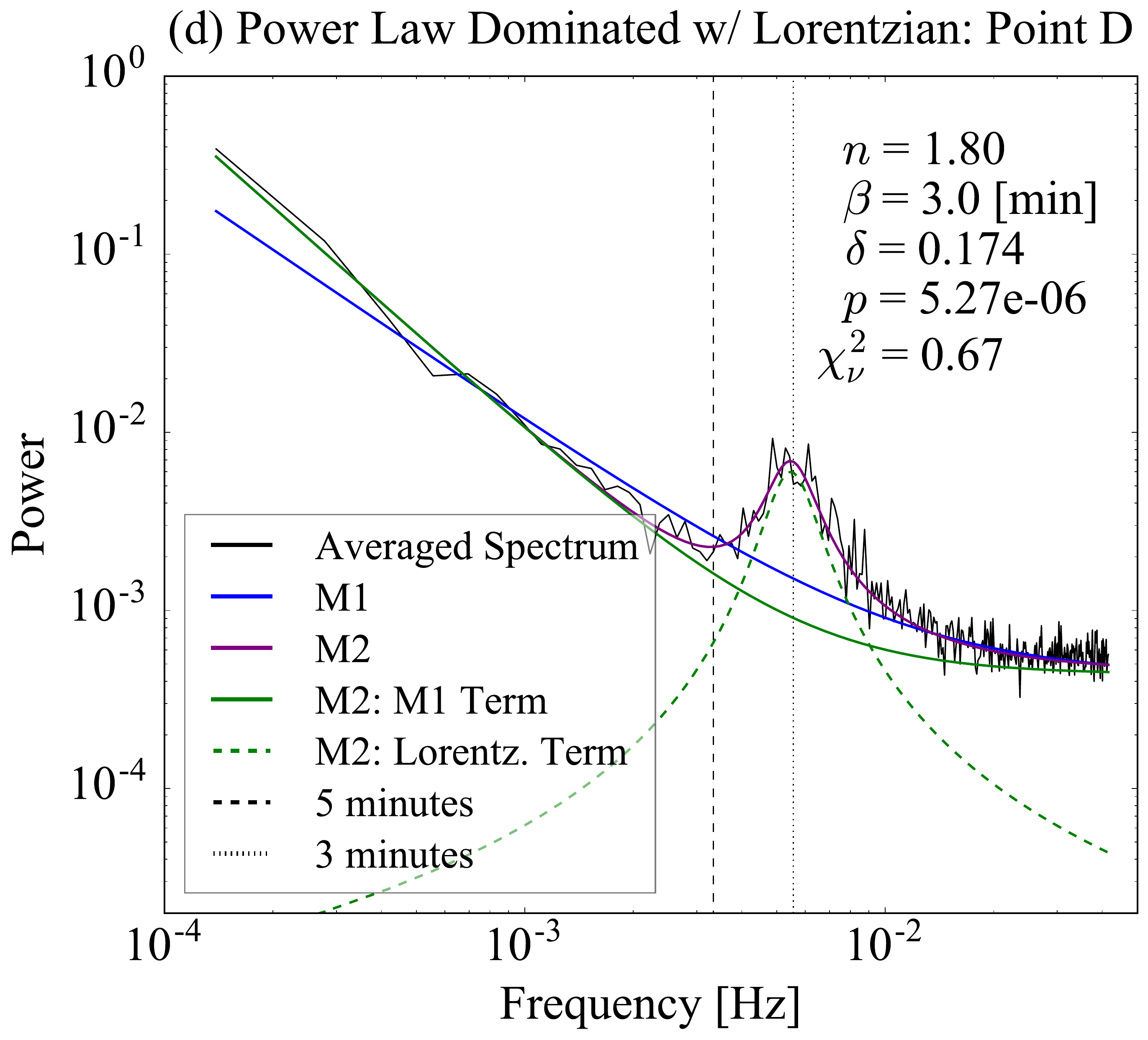} 
  \end{tabular}
  \caption{Model fits and averaged spectra computed from the sample time series shown in Figure~\ref{f:All_TS}\,a--\,d. The titles in each panel include characterizations of each point that are defined in Section~\ref{s:Categories}. }
\label{f:All_Fits}
\end{figure}

We found that in general the use of a Gaussian \textit{versus} a Lorentzian does not produce significant differences in fit quality, with both functions able to describe the broad noise-dominated curves such as that observed in Figure~\ref{f:All_Fits}d. The Lorentzian, however, generally provides a significantly better fit over that of the Gaussian in regions immediately surrounding sunspots, and it provides sporadic improvements elsewhere throughout the corona. Other models may also work, with (e.g.) \cite{Auchere16b} employing a Kappa function to model spectral humps in coronal time series, and in our early efforts we explored variations of Gaussian and Lorentzian models. In practice, we found that the primary limitation in model-fitting is that of noise in the data rather than the specifics of the spectral-bump model, with our exploration finding that it is difficult to clearly distinguish superiority of one model against another in all but a few cases (\textit{e.g.} in the case of sunspots, which produce clearly superior Lorentzian \textit{versus} Gaussian fits).

Given that certain periodic spectral features can be interpreted as the result of damped oscillations, with the full width at half maximum (FWHM) of the curve serving as a proxy for the amount of wave damping, we elected to use the Lorentzian for all analyses presented here for consistency. A key point that we emphasize here is that although damped harmonic oscillations can result in Lorentzian-like spectra, a spectrum with a significant `M2' component does not necessarily imply that it is explained by damped oscillations. Furthermore, as detailed by \cite{Auchere16a}, for example, spectral humps may be the result of background power rather than a true oscillatory signal, and we can not state with certainty that the features we observe are not the result of such processes except in cases where we can visually observe such oscillations (\textit{e.g.} sunspots). Thus, while we use the terms `periodicities' and `oscillations' throughout this article, we emphasize here that the true nature of some of these features is indeterminate.

\begin{equation}
\label{eq:FWHM}
FWHM = 1/(\textrm{e}^{\beta + \delta} - \textrm{e}^{\beta - \delta})
\end{equation}
\noindent and a large $FWHM$ value corresponds to a small spread in the Lorentzian component of the $P_2(\nu)$ spectra (small damping).

Many difficulties were encountered in attempting to produce a reliable algorithm to fit the spectra. We explored several fitting options, and the results presented here are based on the options that gave few failed fits and for which the successful spectral fits matched the data in a visually reasonable way and with a high data/model correlation.\footnote{To visualize the fits, a tool was developed that allowed us to select a pixel on a visual image and observe the corresponding spectral fit. This tool, and related code developed for this survey, will be released to the community via GitHub and a corresponding publication.}

We note here and detail in Appendix~\ref{Appendix:SpectralFitting} that the values of the parameters computed in this work depend slightly on how many spectra were averaged for a given pixel, how many spectra from neighboring pixels were averaged, assumptions about the noise, and the optimization method used. Thus, different approaches to computing best-fit models may produce different best-fit parameters for the models, although we found that the Lorentzian component parameters of the M2 model were insensitive to the choice of averaging methods. A primary objective of this work was to identify spatial and relative-value differences in the best-fit model parameters with few or no `failed' fits for a given averaging scheme, ensuring that the averaging used still retained the true shape of each unique power spectrum and reduced noise such that the computational curve-fitting codes reliably returned model fits that adhered strongly to the shape of the data. Identification of a more accurate value of the M1 model parameters was less of a concern, and arguably it was not possible given the features of the time series described above and the assumptions that would be required for such an analysis. In Appendix~\ref{Appendix:seg_avg} we provide examples of a power spectrum obtained with different averaging schemes and demonstrate that segmented averaging does not substantially alter the shape of the spectra and results in fewer `failed' model fits.

Our model-fitting procedure performs fits to both the M1 and M2 models and first compares the chi-squared values. If the $\chi^2$ value of M1 is less than that for M2, the M2 fit is labeled as `failed'. The M2 model is expected to produce $\chi^2$ values that are less than or equal to that of M1 because M2 is identical to M1 if the $\alpha$ parameter in M2 is set equal to zero. However, due to termination of the fitting procedure when changes in the error functions drop below our specified threshold, some spectra produce better fits to M1 than M2, and thus for those spectra we record only the three M1 model parameters. A successful fit of the M2 model does not necessarily imply that the included Lorentzian feature is meaningful in a statistical sense, and thus a second stage to our process uses a hypothesis test to determine if M2, which has more parameters than M1, is better than M1. In Appendix~\ref{Appendix:SpectralFitting} we detail the model-fitting steps, and in Appendix~\ref{Appendix:Masking} we detail the Lorentzian significance calculation. In Appendix~\ref{Appendix:example_spectra} we provide some examples of poor and good model fits in the 171 and 1700\,{\AA} observations.

Figure~\ref{f:All_Fits} shows the computed power spectra and resulting model fits of the sample time series shown in Figure~\ref{f:All_TS}. The four panels show the spectrum (black), the M1 model (blue) and the M2 model (purple), as well as the best-fit parameter values. The dashed vertical lines indicate the frequency corresponding to periods of three- and five-minutes, and \textit{r} is the correlation coefficient between the best-fit model (M1 or M2) and the averaged spectra. The parameter $p$ is the probability of rejecting the null hypothesis that M2 produces an equivalent (least-squares) fit to M1 when it is true.  In this work $p<0.005$ is the threshold at which we conclude that the M2 fit is statistically better than that of M1; the calculation of $p$ is described in Appendix~\ref{Appendix:Masking}. Thus, in Figure~\ref{f:All_Fits}, a Lorentzian component (green dashed curve) is shown for all panels, but only that shown in panel d is statistically significant and thus indicative of a periodic feature in the spectra.

In figures shown later in this article, values for the Lorentzian location parameter in the M2 model, and the derived FWHM, are only shown at locations where $p<0.005$. The titles of these figures include the percentage of the region in which the Lorentzian location values have been omitted (masked).

The spectral fitting procedure applied to each of the 1598$\times$1598 pixels is computationally expensive, initially accounting for the majority of the approximately 30 hours required for processing each wavelength. Given the complexity of this process we used the $\sf{MPI4Py}$ package to implement a simple multi-threaded scheme in which the fitting calculations were distributed evenly among 16 virtual processing cores (8 hyper-threaded Intel i7 4.1~Ghz (4.3~GHz single-core turbo) cores on a Linux-based desktop workstation computer. After parallelization, the total processing time was $\approx 4$ hours for each wavelength, 80\,\% of which was required for the spectral fits, roughly 10\,\% to load data from the FITS-format image files and derotate, and 10\,\% to compute and average the spectra.

\section{Results}
\label{s:results}

The fitted spectral models provided us with the six parameters shown in Equation~\ref{eq:M2} for each investigated wavelength. In this article, we focus on the results for three key parameters:
\begin{enumerate}
\item the power law index [$n$] representing the slope of the power law;
\item the Lorentzian location [$\beta$] representing the frequency of the peak of the fitted Lorentzian component of model M2; and 
\item the full-width at half-maximum (FWHM) of the fitted Lorentzian curve. 
\end{enumerate}

As noted previously, the FWHM is a value derived from model fits, representing the width of the Lorentzian (analogous to a damping coefficient) in the temporal domain, with units of minutes. Other parameters, such as the Lorentzian amplitude and the M1 amplitude coefficient, as well as various quality-of-fit metrics, are retained but not presented. The results from these parameters, which do have certain unique features, will be the focus of subsequent investigations.

In the following subsections, we present the results for the four different AIA wavelengths in Figures~\ref{f:results_193}, \ref{f:results_171}, \ref{f:results_304}, and \ref{f:results_1700}. Panel a of each figure shows the arithmetically averaged visual image at that wavelength followed by the best-fit power law indices in panel b, the Lorentzian locations in panel c, and Lorentzian FWHM in panel d.

\subsection{193\,{\AA}}

\begin{figure}
  \centering
  \setlength{\tabcolsep}{1.0mm}
  \begin{tabular}{cc}
  \includegraphics[scale=0.21]{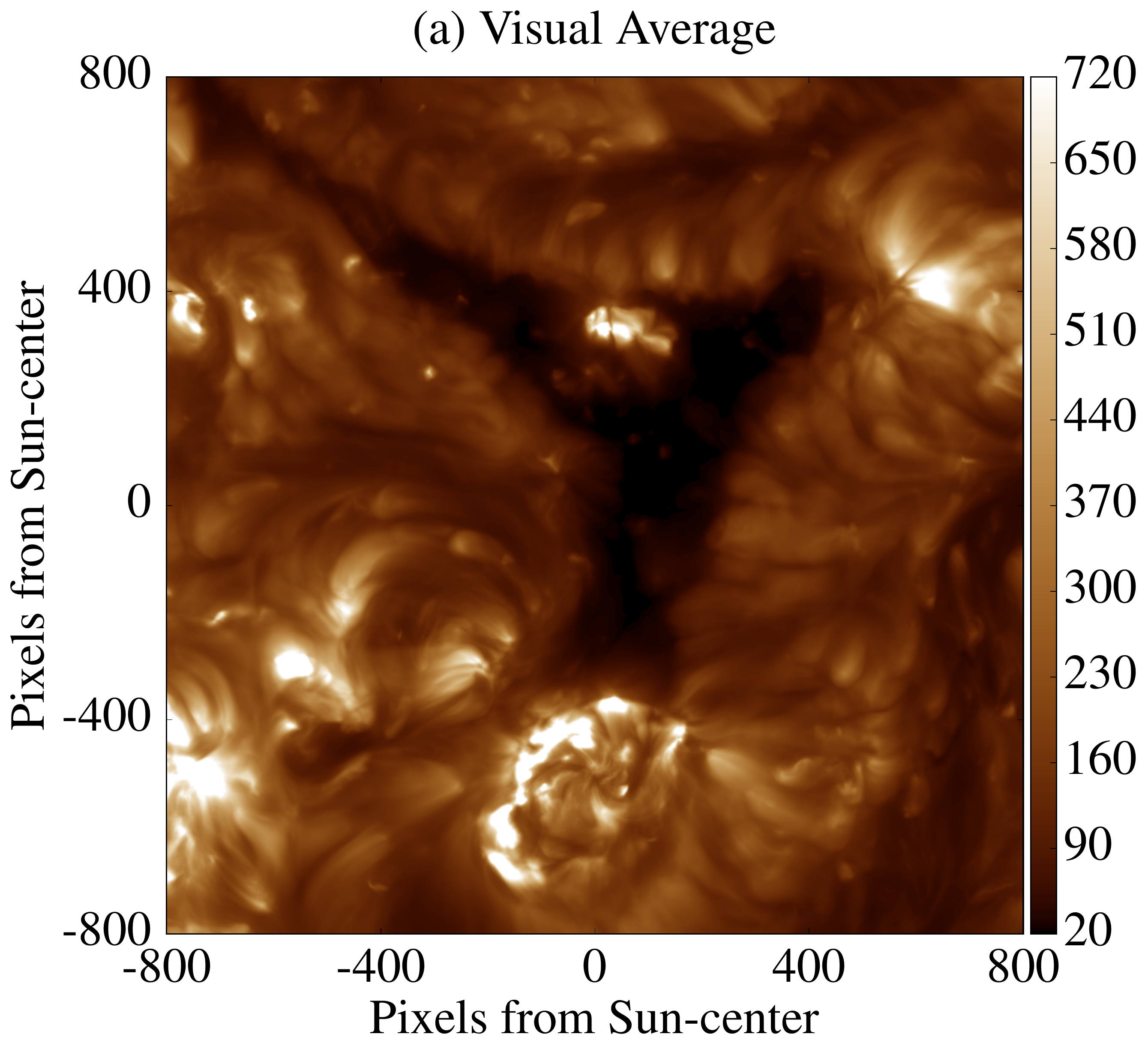}&\includegraphics[scale=0.21]{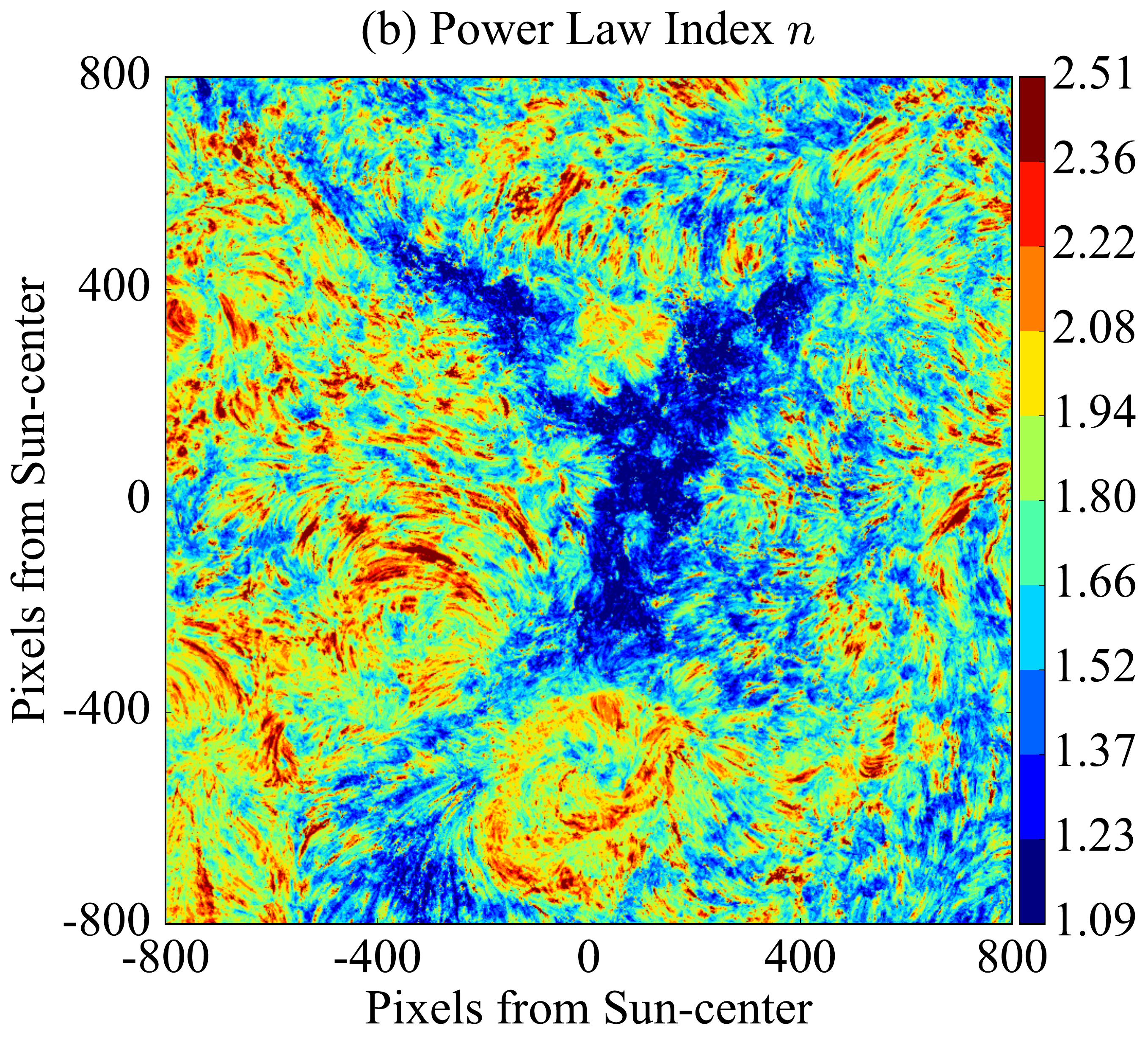}\\
  \includegraphics[scale=0.21]{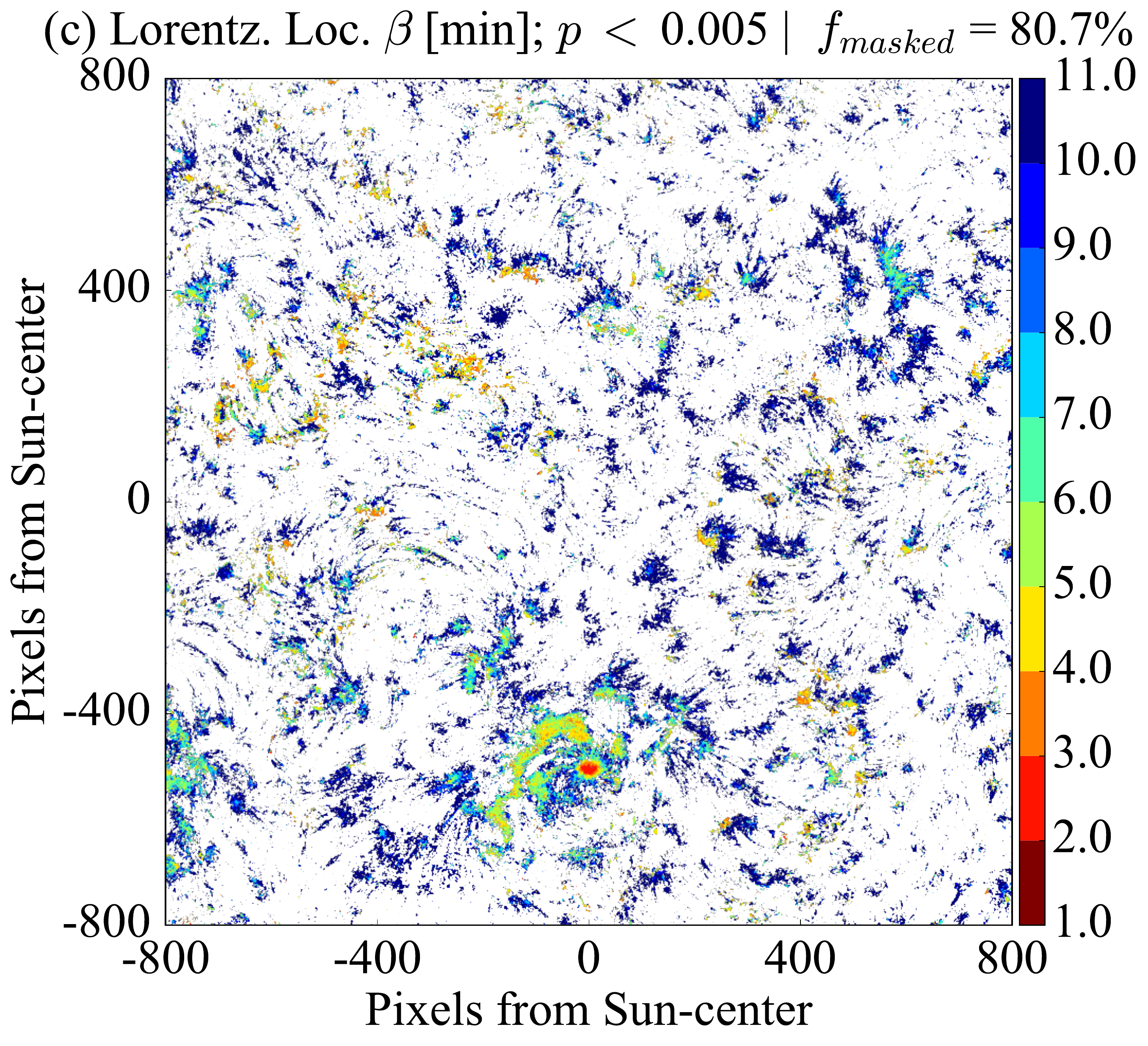}&\includegraphics[scale=0.21]{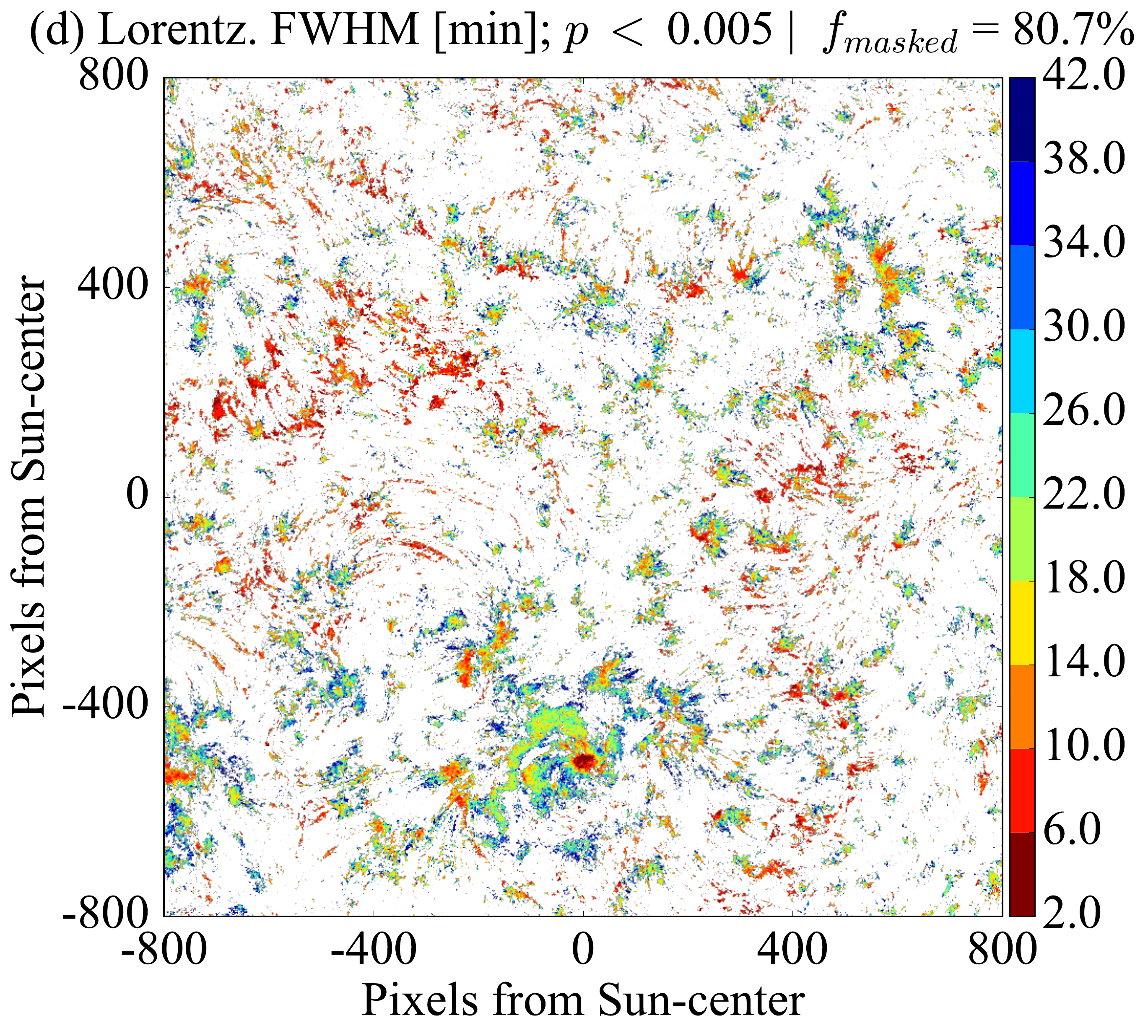} 
  \end{tabular}
  \caption{Results for the AIA-193\,{\AA} wavelength channel using 6$\times$2-hour time segment and 3$\times$3-pixel averaging, showing a the average of the normalized intensity images in the time interval \,00:00:00--\,11:59:59\,{UT} on 26 June 2013, plotted on a linear scale, (b) the power law index, (c) the masked Lorentzian location, and (d) the masked full-width at half maximum (FWHM) of the Lorentzian. The ``f$_{masked}$'' value for Lorentzian location is the percentage values masked due to having low statistical significance.  To increase the contrast in the heatmap displayed in panel b, the colorscale shown was generated using a range that excludes the top/bottom 1\,\% of pixel values.  Those pixels in the top/bottom 1\,\% are colored the same as the highest/lowest bins.}
\label{f:results_193}
\end{figure}

Figure~\ref{f:results_193} shows the results for the 193\,{\AA} channel, in which we see that the visualization of all derived spectral parameters relates clearly, directly, and uniquely to features visible in the observational data. This result applies equally to every wavelength under consideration.

In Figure~\ref{f:results_193}b, we observe a clear difference between the coronal hole and surrounding active areas. The former has spectra that are far flatter than surrounding loops regions as a consequence of increased instrument noise (low signal) in the coronal hole. Magnetic-loop structures are observed throughout the region with high power-law indices approaching 2.5, with high indices also in the center of coronal cells. Likewise, immediately below the coronal hole, the active region AR-1777 is well defined with high power-law indices. In short, high power-law indices relate directly to a a concentrated magnetic field in this and all other channels investigated here, with the inference that flow along these loops is enabling fast energy cascades. A histogram of the power-law indices in this region is bell-shaped, and the peak value (mode) is 1.77, implying that classical Kolmogorov-like turbulence is pervasive throughout the hot corona.

The Lorentzian location and width panels in Figure~\ref{f:results_193}c and d show the frequency (time) locations and widths (FWHM) of statistically significant ($p<0.005$) Lorentzian component in our spectral model. Note that at this significance level, we expect $0.5$\,\% false alarms. We found 19.3\,\% of this region to contain a statistically significant Lorentzian signal. This percentage increases in the characteristically cooler AIA channels, reaching a peak of essentially 100\,\% in the 1700\,{\AA}.

Inspection of the sunspot in Figure~\ref{f:results_193}c reveals a central $\approx$ three-minute periodicity with a surrounding ring of  $\approx$ five-minute periodicity. We define this structure as a ``\textit{coronal bullseye}'' -- an approximately circular region of concentric rings of specific unique oscillatory periods, decaying radially from the center. (The discrete nature of the concentric rings is a function of the binning used for the colorbar; in actuality the periodicity falloff is continuous.) Coronal bullseyes are features that we have observed in many other datasets, not only surrounding sunspots but also at the foot-points of sporadic loop structures. We discuss these features further in Section~\ref{s:Categories}. 

Interpretation of panels c and d is not trivial. The sunspot core and certain points surrounding the active region (AR 1777) exhibit clear short-period ($\approx$ three- to four-minute) oscillations, and these oscillations correspond to narrow Lorentzians in the spectra (in the temporal domain), with FWHM values approximately on the same order as the temporal frequency. Surrounding the AR we observe a fan-like structure of periodicities near 11-minutes, that themselves contain irregularly shaped structures of approximately five-minute periodicities. These oscillations correspond to much wider Lorentzians, with widths on the order of 18--30 minutes. Scattered around the upper-left of Panel c we see a number of sporadic $\approx$ four-minute locations that do not appear to correspond to any obvious magnetic structures in the visual observation (\textit{Panel a}). These sporadic oscillatory features, also observed in the same locations in 171 and 304\,{\AA}, uniformly have FWHM values around six- to ten-minutes, \textit{i.e.} approximately twice the oscillation period. 

More broadly scattered throughout Panel c are a number of $\approx $ 11-minute oscillations, with those that relate to bright points in the visual observations having central regions of $\approx$ five-minutes periodicity. The FWHM values in the center of these regions are on the order of 10--14 minutes, and closer to 20--30 minutes around their edges, but the high degree of scatter in these features makes this observation somewhat less certain.

Faintly observable in panel c are narrow oscillatory structures ($\approx$ 11-minute periods) that seemingly trace magnetic field lines. These structures, an example of which can be seen around pixel location [-300,-100], have Lorentzians with widths around six- to ten-minutes \textit{i.e.} less than one cycle, naturally raising the question of how such an oscillation can be established in less than one cycle. The apparent oscillations (spectral humps) in these (and likely other) features may quite possibly arise from a combination of smaller pulses as described by \cite{Auchere16b} (Kappa function) rather than a single discrete oscillation. A detailed studying incorporating and comparing a Kappa and Lorentzian function would be necessary to address this question but is beyond the scope of this initial survey article.

We note finally that if we relax our significance constraint to, say, $p<0.01$, then many of the features seen here gain increased definition. However, the relaxation of our $p$-value leads to excess noise in later figures, and thus for this article, we have chosen to retain a consistent (but perhaps overly conservative) value of $p<0.005$.

\subsection{171\,{\AA}}

\begin{figure}
  \centering
  \setlength{\tabcolsep}{1.0mm}
  \begin{tabular}{cc}
  \includegraphics[scale=0.21]{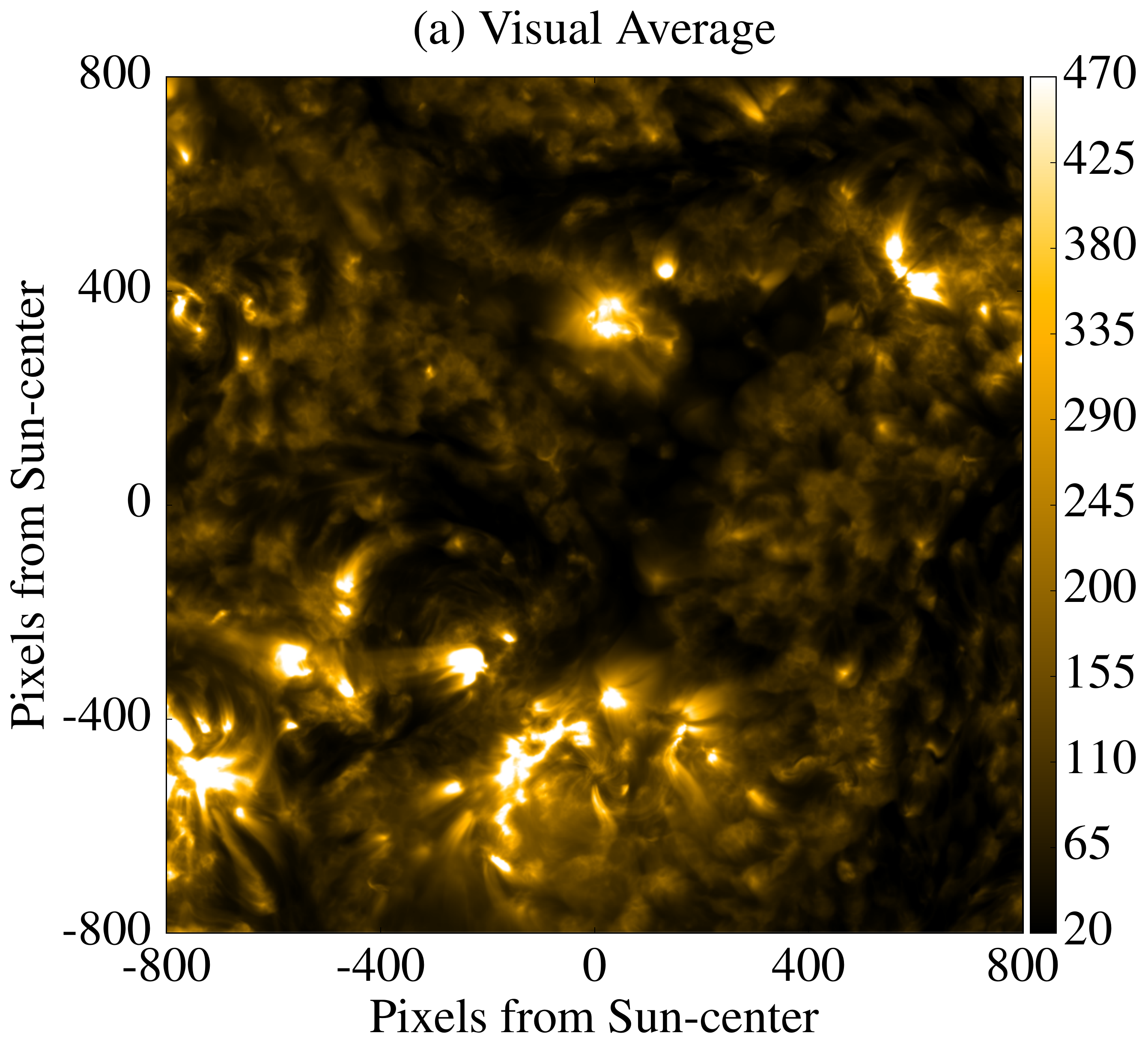}&\includegraphics[scale=0.21]{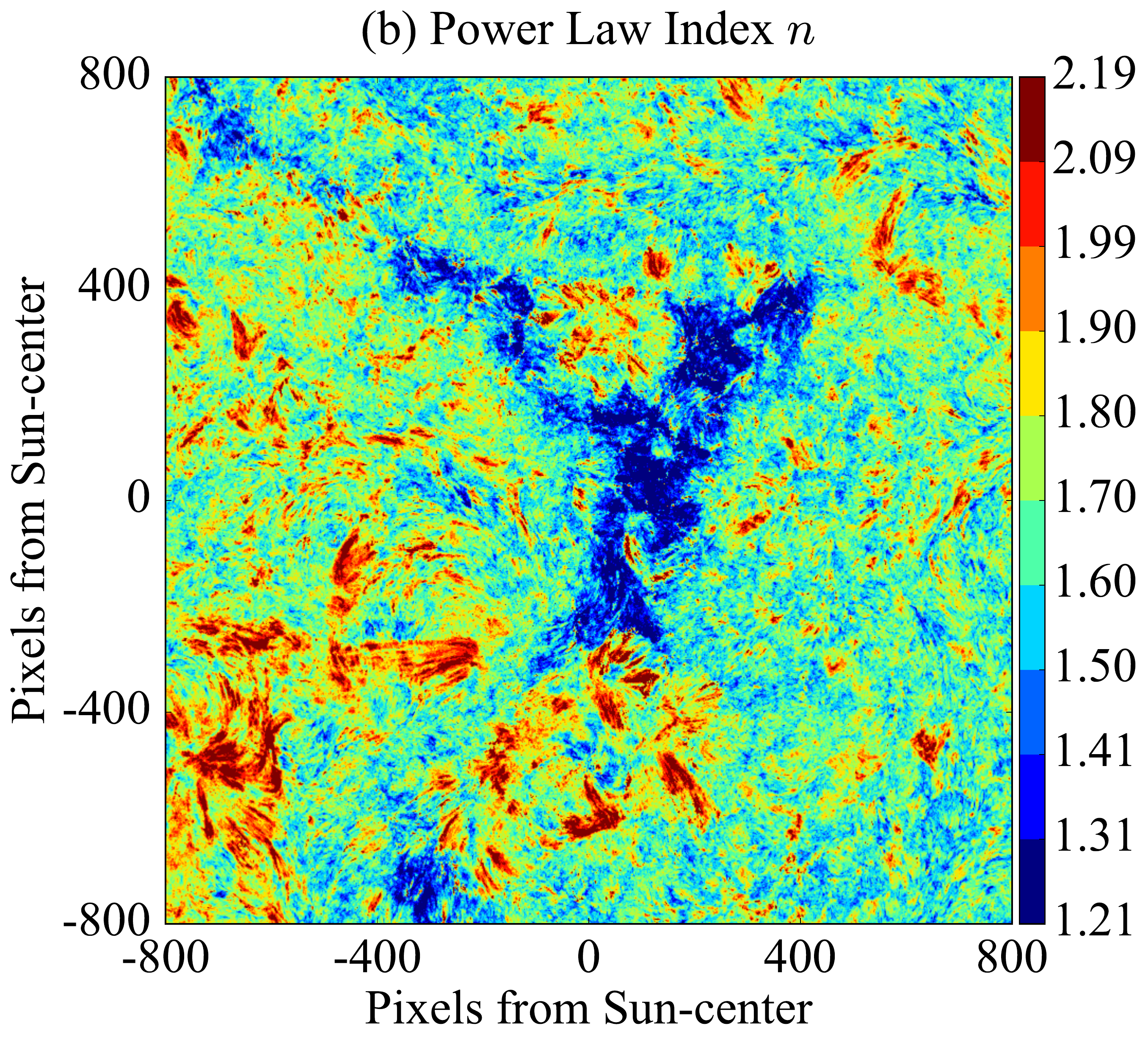}\\
  \includegraphics[scale=0.21]{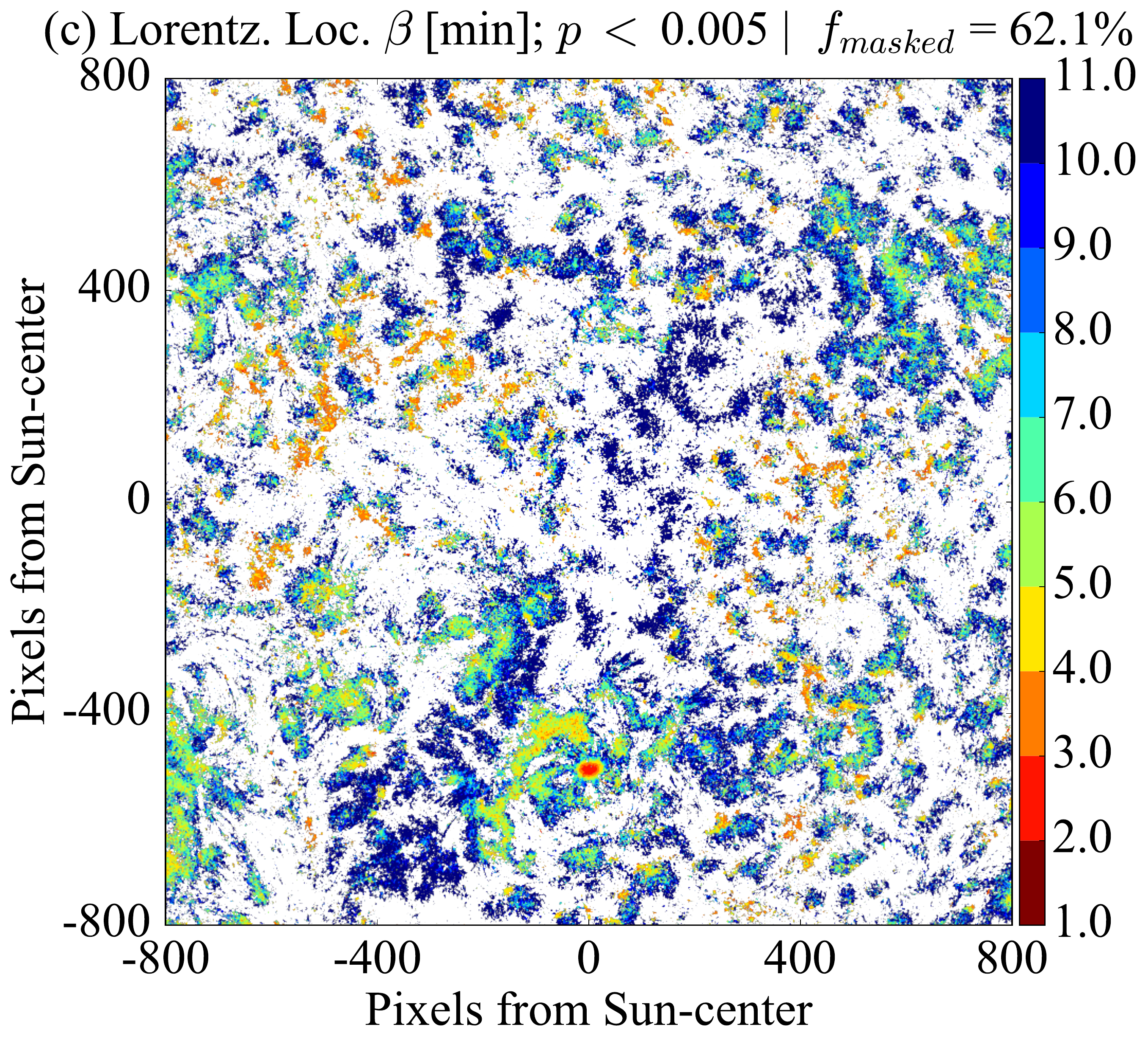}&\includegraphics[scale=0.21]{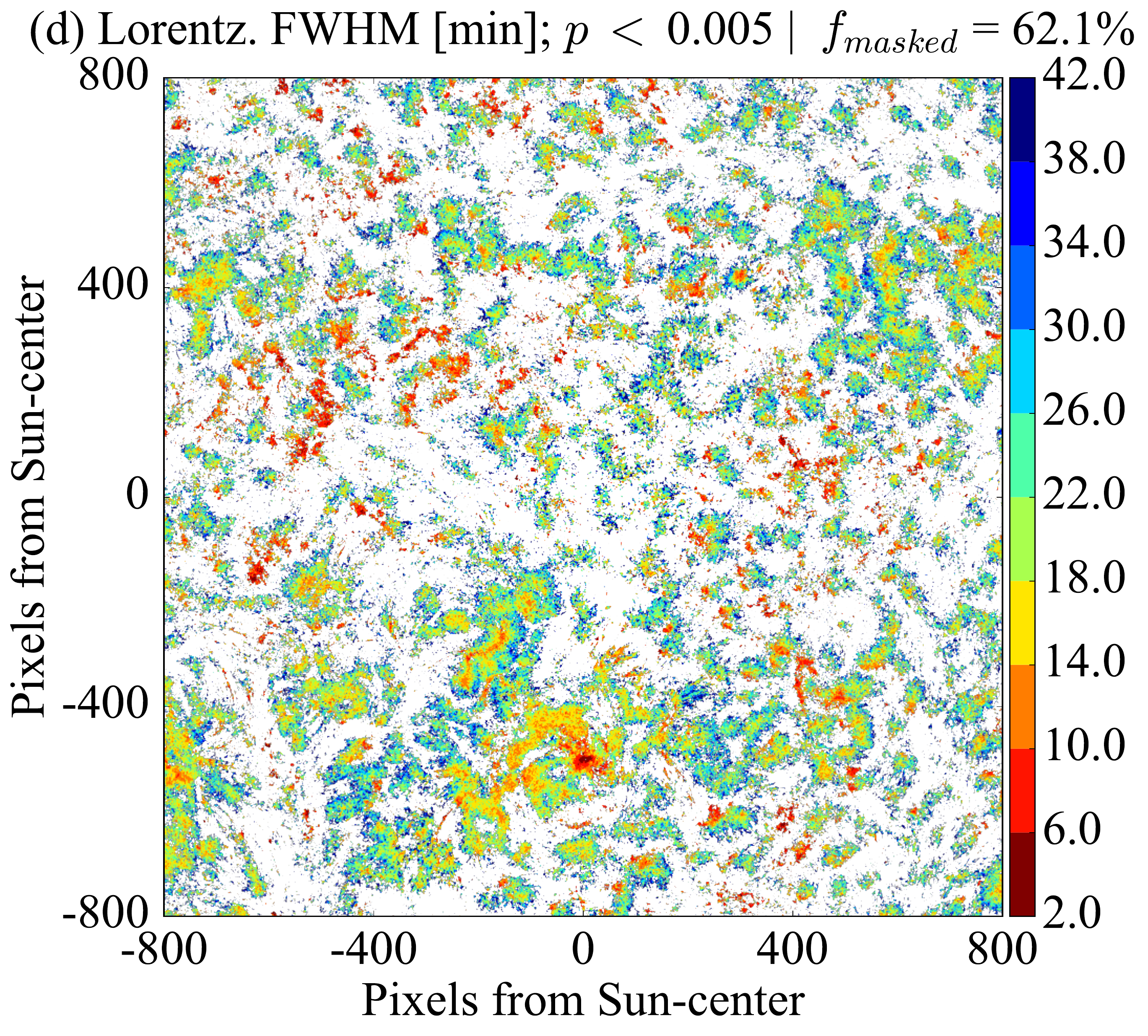} 
  \end{tabular}
  \caption{Results for the AIA-171\,{\AA} channel using 6$\times$2-hour time segment and 3$\times$3-pixel averaging, showing a the average of the normalized intensity images in the time interval \,00:00:00--\,11:59:59\,{UT} on 26 June 2013, plotted on a linear scale, (b) the power law index, (c) the masked Lorentzian location, and (d) the masked full-width at half maximum (FWHM) of the Lorentzian. The ``f$_{masked}$'' value for Lorentzian location is the percentage values masked due to having low statistical significance.  To increase the contrast in the heatmap displayed in panel b, the colorscale shown was generated using a range that excludes the top/bottom 1\,\% of pixel values.  Those pixels in the top/bottom 1\,\% are colored the same as the highest/lowest bins.}
\label{f:results_171}
\end{figure}

Figure~\ref{f:results_171} shows the results for the 171\,{\AA} observations. These observations correspond to the upper transition region to the solar corona, and accordingly, there are notable changes in the observed spectral properties concurrent with the changed visible features seen at 171\,{\AA}, including contributions from hot coronal lines as well as the underlying chromosphere. 

The map of power-law indices shown in Figure~\ref{f:results_171}b is similar in appearance and interpretation to that of the 193\,{\AA} observations. The peak power-law indices are almost exclusively in areas of a concentrated magnetic field (\textit{i.e.} loops and foot-points), reaching values of around 2.2, with the coronal hole and sporadic low-signal areas continuing to show low power-law indices. In general, the distribution of power-law indices is shifted towards lower values (distribution peak at 1.67) and is of a slightly narrower range than in 193\,{\AA}, but with a longer tail for high index values than observed at 193\,{\AA}.

Figure~\ref{f:results_171}c and d, however, show a marked difference to 193\,{\AA}, with the overall structure appearing to mirror that of the underlying chromosphere, and noting an increase to approximately 37.9\,\% coverage of statistically significant Lorentzian components. This growth comes partially from an apparent expansion of the $\approx$ 11-minute features observed in 193\,{\AA}, although we can no longer resolve the $\approx$ 11-minute ``loop-like'' structures. Many of the sporadic, isolated shorter-period features seen in 193\,{\AA} are now expanded and more clearly visible, but again it remains unclear if their period of oscillation is ``true'', or a consequence of many smaller combined pulses (with the exception of the sunspot, where our observed oscillations are supported by published literature and visual identification of oscillations). 

The Lorentzian widths (FWHM) seen in this channel have essentially identical properties to those discussed for 193\,{\AA}, with narrow widths in the sunspot core and in the sporadic $\approx$ four-minute points, more of which are now apparent in this channel (likely the result of chromospheric ``leakage'' in this channel), and the broader $\approx$ 18- to 30-minute widths elsewhere.   

\subsection{304\,{\AA}}

\begin{figure}
  \centering
  \setlength{\tabcolsep}{1.0mm}
  \begin{tabular}{cc}
  \includegraphics[scale=0.21]{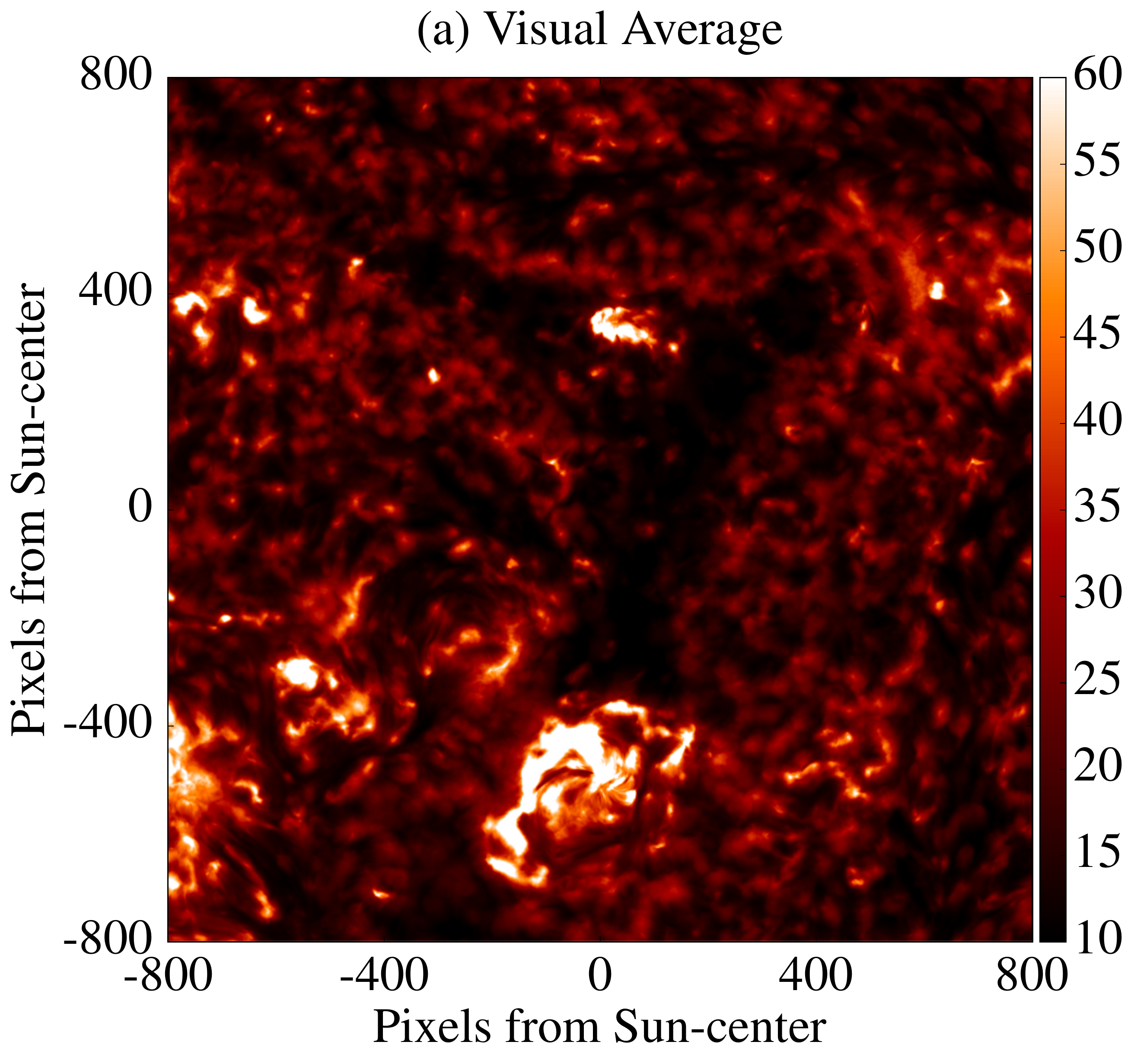}&\includegraphics[scale=0.21]{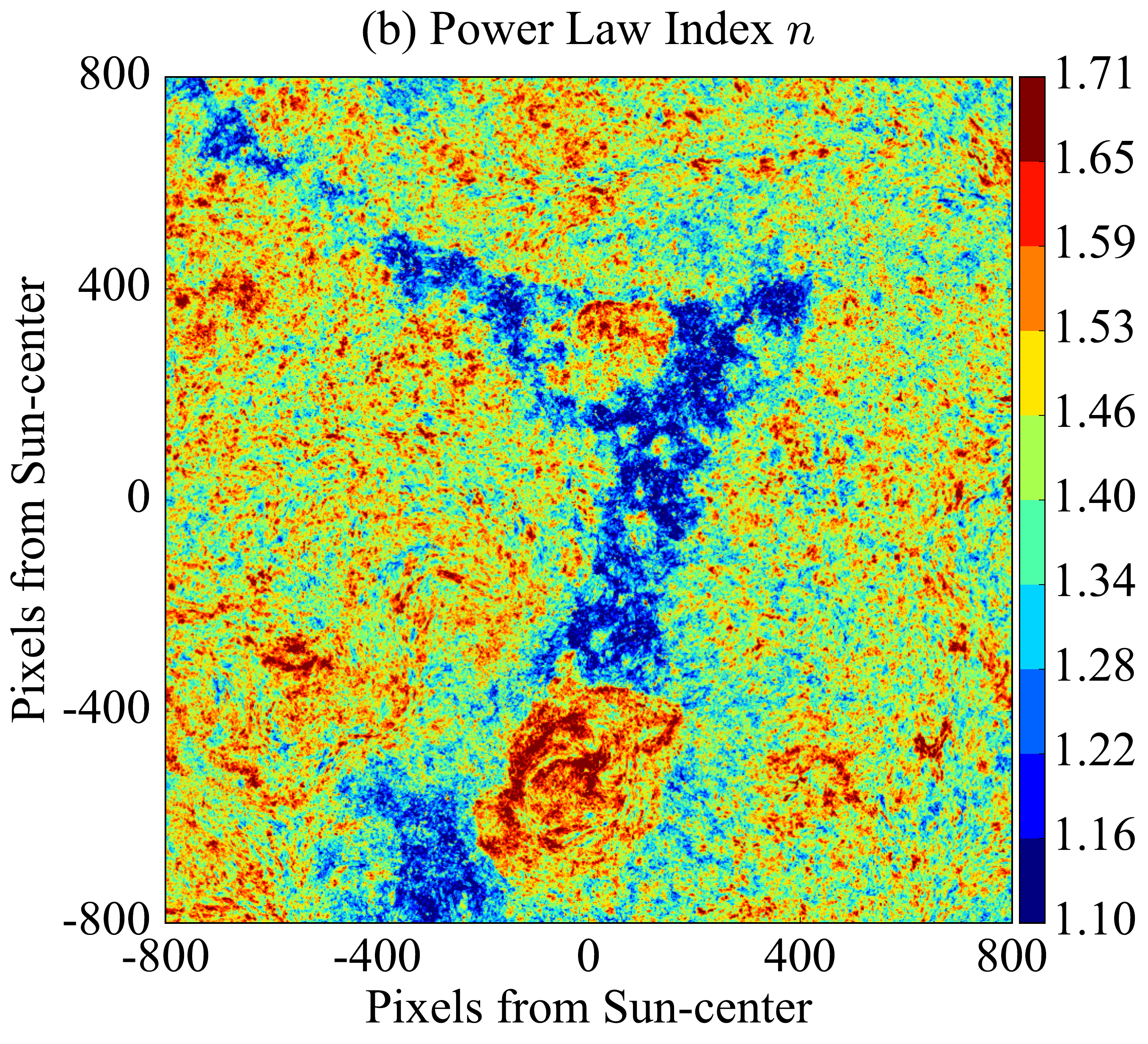}\\
  \includegraphics[scale=0.21]{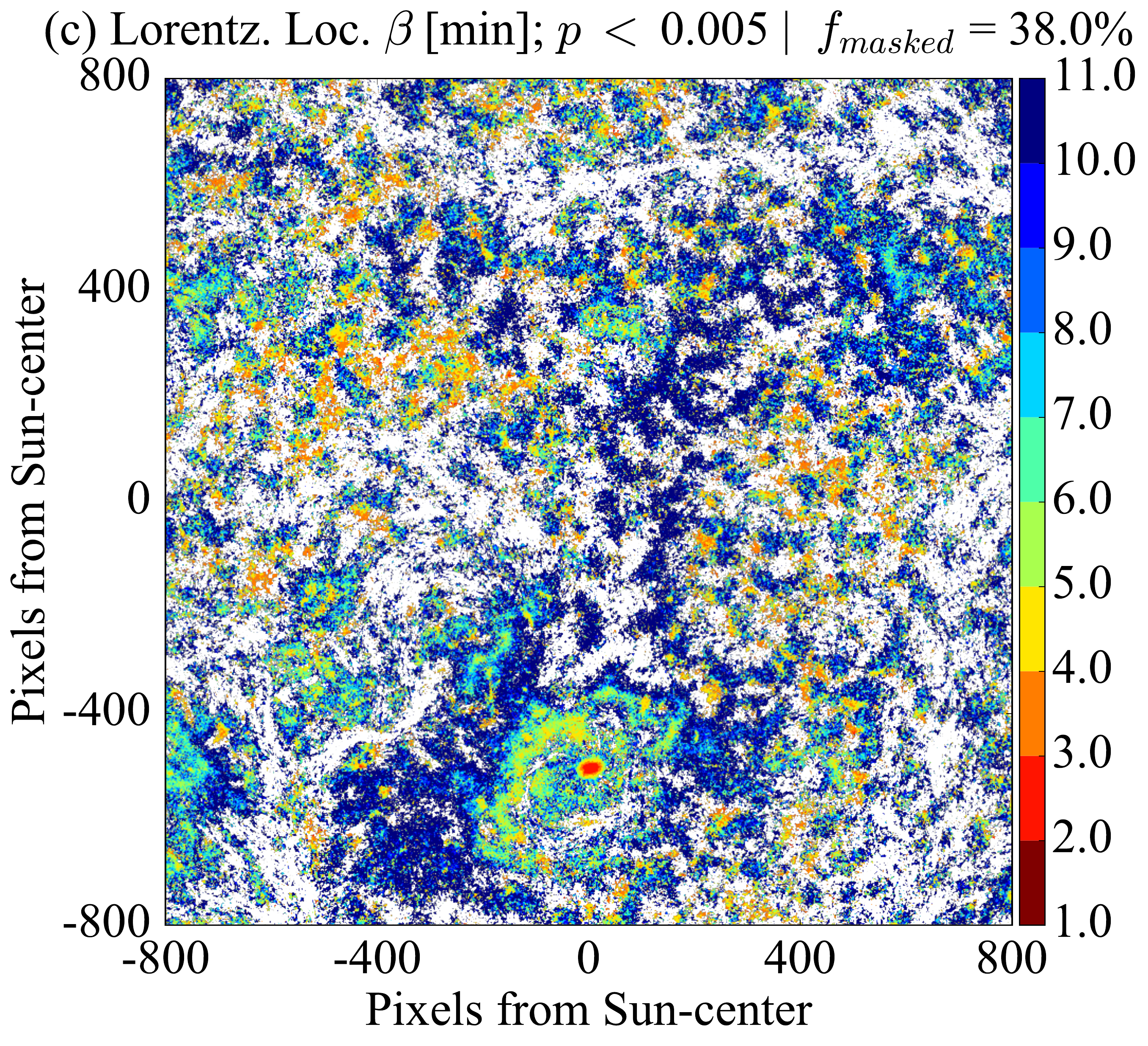}&\includegraphics[scale=0.21]{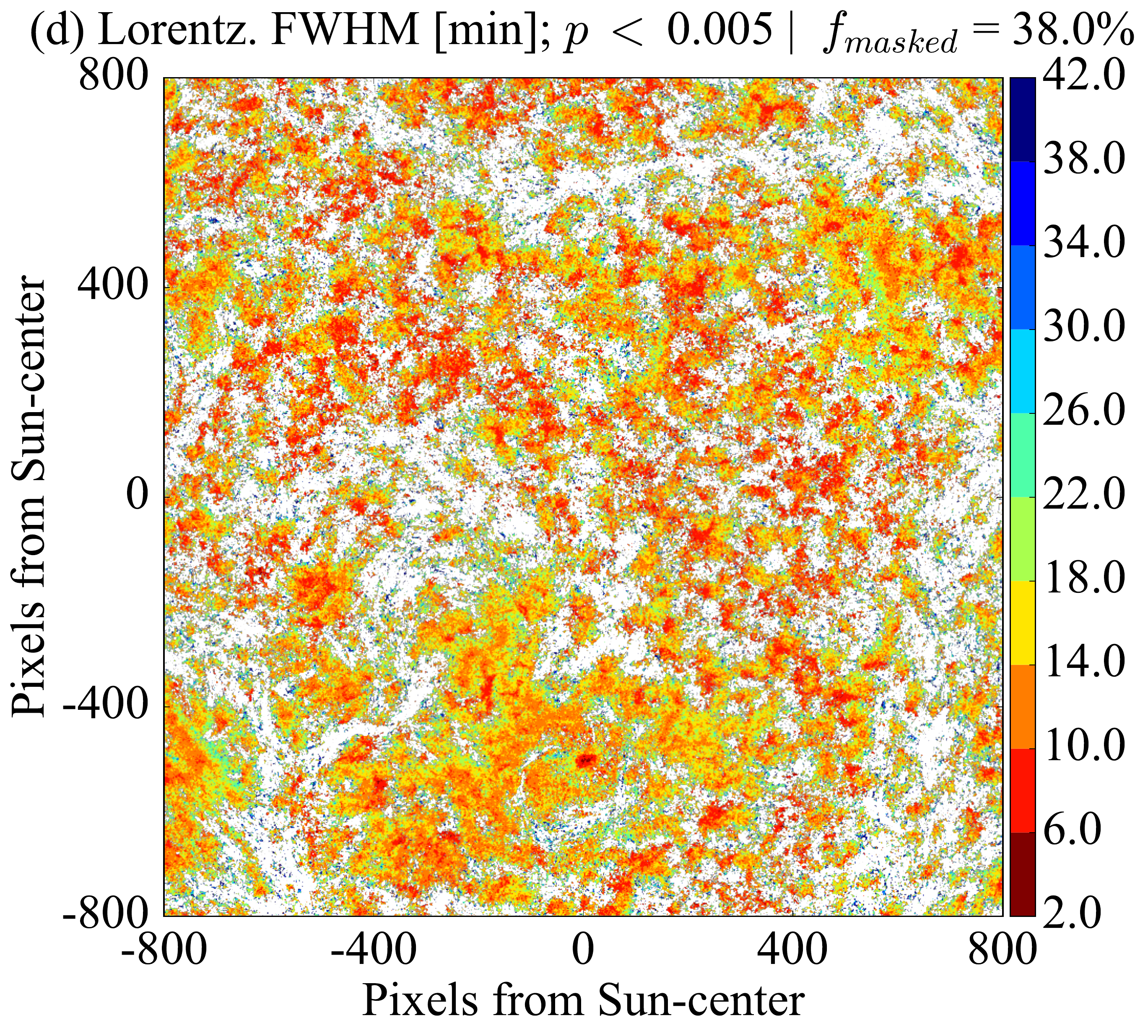} 
  \end{tabular}
  \caption{Results for the AIA-304\,{\AA} channel using 6$\times$2-hour time segment and 3$\times$3-pixel averaging, showing a the average of the normalized intensity images in the time interval \,00:00:00--\,11:59:59\,{UT} on 26 June 2013, plotted on a linear scale, (b) the power law index, (c) the masked Lorentzian location, and (d) the masked full-width at half maximum (FWHM) of the Lorentzian. The ``f$_{masked}$'' value for Lorentzian location is the percentage values masked due to having low statistical significance.  To increase the contrast in the heatmap displayed in panel b, the colorscale shown was generated using a range that excludes the top/bottom 1\,\% of pixel values.  Those pixels in the top/bottom 1\,\% are colored the same as the highest/lowest bins.}
\label{f:results_304}
\end{figure}

Figure~\ref{f:results_304} shows the results for the 304\,{\AA} observations. These observations represent much lower temperatures (approximately 50,000\,K) and processes more like photospheric/chromospheric than hot corona. 

Accordingly, in Figure~\ref{f:results_304}b we see globally lower power-law indices throughout this channel, with a range of $\approx$ 1.1\--1.7, and a distribution peak at 1.44. The structure in this image closely mirrors the magnetic network structures observed in chromospheric and magnetogram observations. Again, the peak index values correspond to the brightest regions in the visual observations and relate primarily to active region structures and loops (\textit{i.e.} concentrated magnetic fields). These power-law index results mirror those found in simulations \citep[e.g.][]{Kitiashvili15} at these characteristic temperatures. We note also a broken chain of high power-law indices that follow the filament across the upper part of this region. The coronal hole remains clearly defined, demonstrating that regions with higher levels of instrument noise in low signal areas typically present low power-law values (a notable exception being solar filaments, which have high power-law slopes but also low signal/high noise). As noted earlier, the rollover frequency ([$\nu_r$], not presented) is an excellent metric for determining the point at which photon noise dominates a given spectrum, and it highlights extremely well the low-signal features such as coronal holes; this result may have useful feature detection applications but is beyond the scope of this article.

In Figures~\ref{f:results_304}c and d, the Lorentzian parameters now occupy 62\,\% of the region, again appearing as a natural ``growth'' of similar features seen in other channels. The scattered $\approx$ five-minute periodicities are widespread and larger in area, though upon close inspection they appear quite disordered in their apparent periodicity at the pixel level. This may be a resolution limitation of our approach, but it could also be indicative of regions that have only quasi-stable periodicities or periodicities that exist over relatively short timescales. It is no longer possible to discern any clear pattern regarding the FWHM of the respective Lorentzians, but broadly we see that the distribution of widths is now dominated by much narrower $\approx$ six- to ten-minute FWHM, some of which again may not relate to ``true'' oscillations.

The coronal bullseye in panel c has a remarkably well-defined structure, with a central three-minute and outer ring of five-minute periodicities, and corresponds to the very narrowest end of the Lorentzian width scale. The $\approx$ five- six-minute periodicities near the active region are better defined, partly due to a growth in the surrounding $\approx$ 11-minute region.

\subsection{1700\,{\AA}}

\begin{figure}
  \centering
  \setlength{\tabcolsep}{1.0mm}
  \begin{tabular}{cc}
  \includegraphics[scale=0.21]{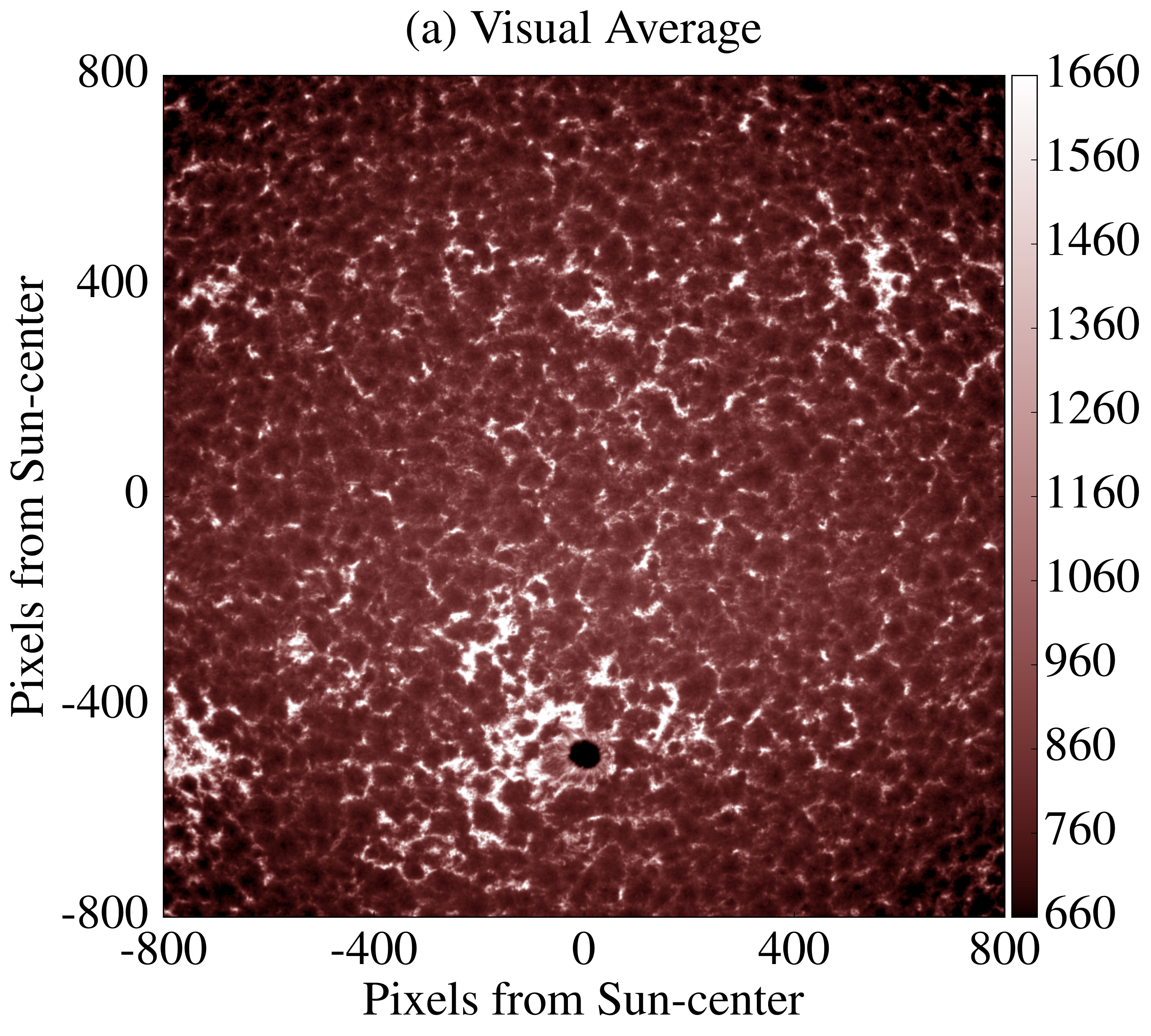}&\includegraphics[scale=0.21]{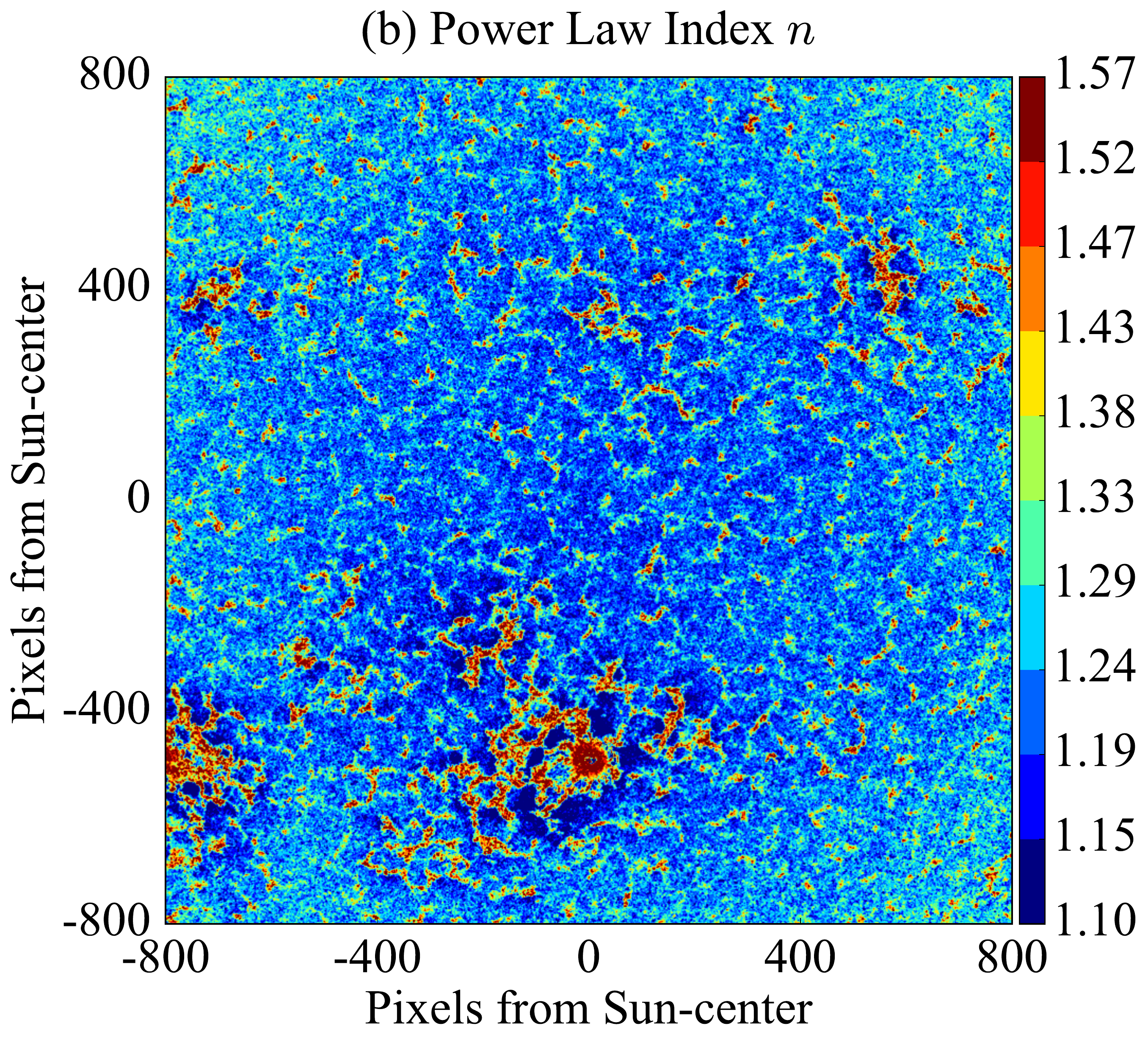}\\
  \includegraphics[scale=0.21]{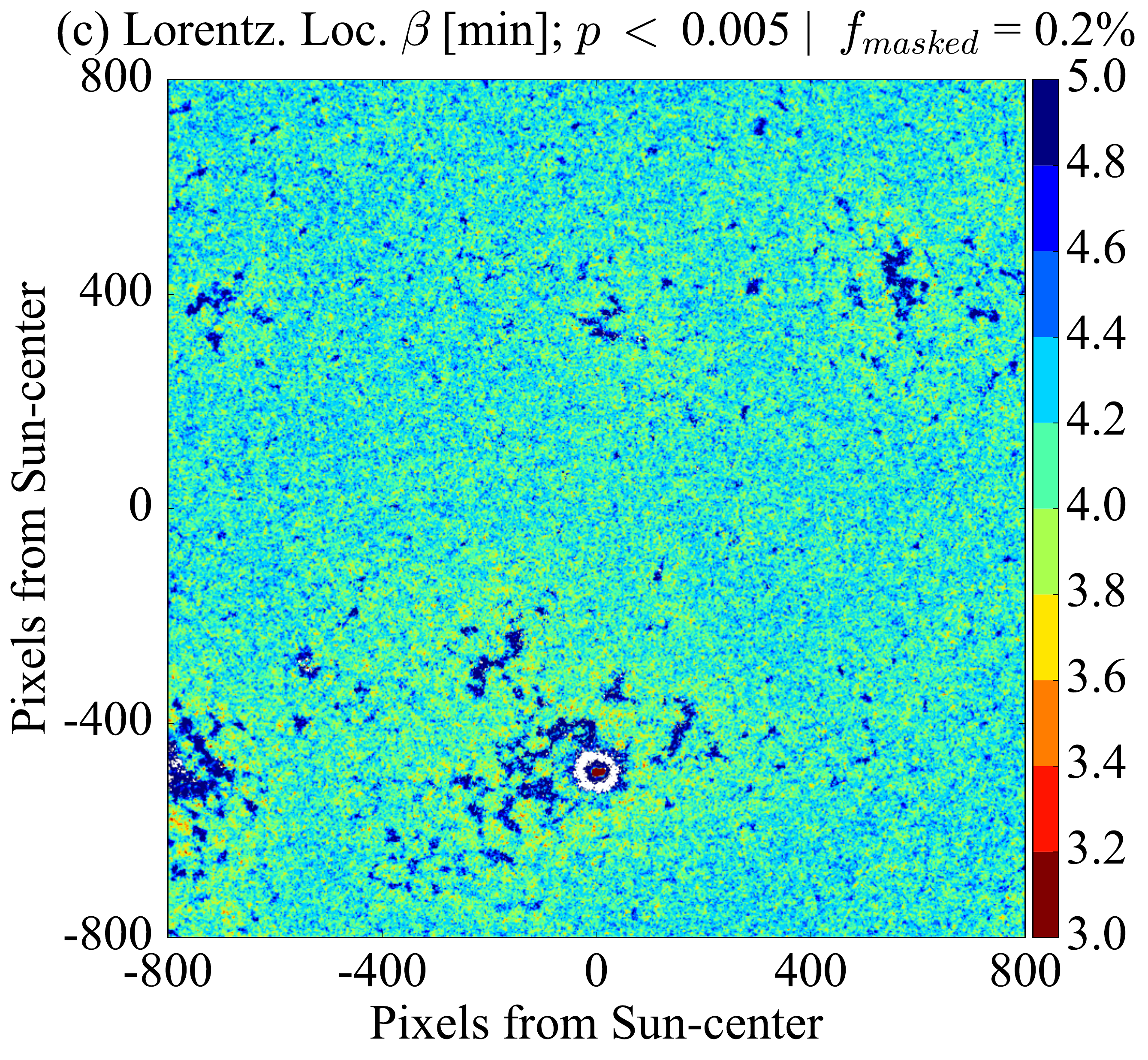}&\includegraphics[scale=0.21]{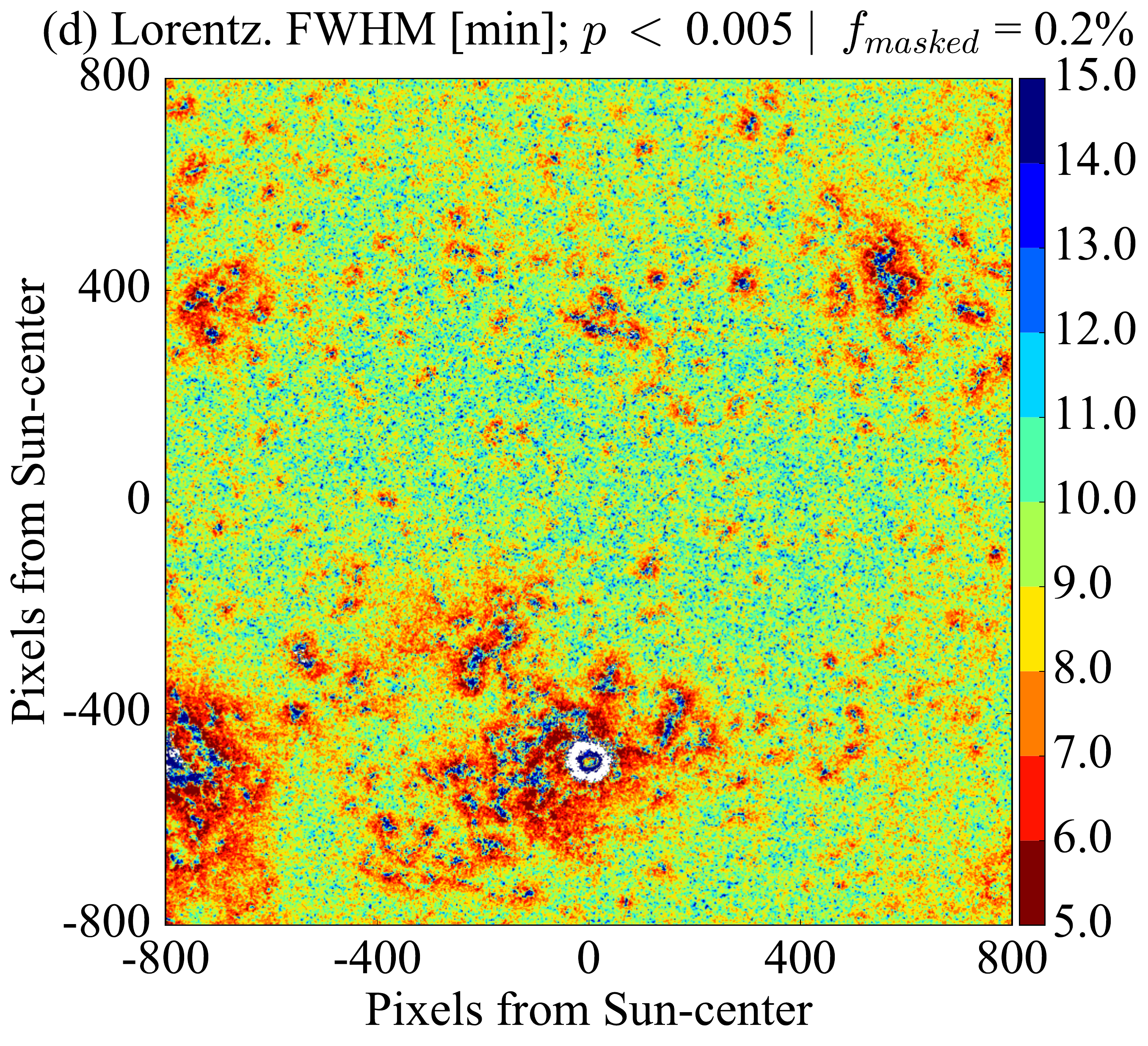} 
  \end{tabular}
  \caption{Results for the AIA-1700\,{\AA} channel using 6$\times$2-hour time segment and 3$\times$3-pixel averaging, showing a the average of the normalized intensity images in the time interval \,00:00:00--\,11:59:59\,{UT} on 26 June 2013, plotted on a linear scale, (b) the power law index, (c) the masked Lorentzian location, and (d) the masked full-width at half maximum (FWHM) of the Lorentzian. The ``f$_{masked}$'' value for Lorentzian location is the percentage values masked due to having low statistical significance.  To increase the contrast in the heatmap displayed in panel b, the colorscale shown was generated using a range that excludes the top/bottom 1\,\% of pixel values.  Those pixels in the top/bottom 1\,\% are colored the same as the highest/lowest bins.}
\label{f:results_1700}
\end{figure}

Figure~\ref{f:results_1700} shows the results for the 1700\,{\AA} (upper photosphere/chromosphere) observations, which are markedly different from the previous channels. While not shown here for reasons of brevity, the results for 1600\,{\AA} and 1700\,{\AA} observations are essentially identical, except where noted in the text.

In Figure~\ref{f:results_1700}b, we see the power-law indices to be lower across the entire region with a range of $\approx$1.1--1.5, and a histogram distribution peak at $1.22$.  Reinforcing our previous observations, the largest index values are found along magnetic network structures, with this parameterization producing a map remarkably resemblant to the magnetic network.

The Lorentzian results shown in Figures~\ref{f:results_1700}c and d, with only 0.2\,\% masking, are consistent with \cite{Leighton62} -- namely that the photosphere is dominated by a near-uniform, global oscillation. Our model fits consistently show a global oscillation near four-minutes, which, upon inspection of the histogram of Lorentzian location values in this image shows a very clear distribution peak centered at 4.19-minutes ($\sigma=0.33$~minutes). An equivalent analysis of the 1600\,{\AA} observations for this region have a similar distribution, with a peak at 4.07~minutes ($\sigma=0.30$~minutes). These results are consistent with \cite{McIntosh04}, who reported a global wavelet power peak of 4\,{mHz} (4.13 minutes) in 1700\,{\AA} observations from TRACE, using a spectral (wavelet-based) technique. 

As with the power-law indices, both Lorentzian components in panels c and d clearly define the magnetic network. However, we note that the Lorentzians in 1700\,{\AA} are (relatively) broader in regions of strong magnetic field (\textit{i.e.} magnetic network). These broader Lorentzians also correspond to regions of five-minute oscillations, with a possible interpretation being that the five-minute oscillations originate from the photosphere \citep[per][]{Leighton62}, using the magnetic network as the conduit, \textit{i.e.} these structures facilitate the passage of slow magnetoacoustic waves from the underlying photosphere \citep{Bogdan00, Roberts06, Vecchio06}. The power-law slopes here may therefore result from fundamentally different processes than those in the ``hot'' corona. 

Global oscillations distant from the magnetic network are more controversial in origin, although studies suggest they may be the result of waves reflected from the temperature gradient of the overlying transition region, and becoming trapped within a so-called ``chromospheric cavity'' \citep{Carlsson99, Taroyan08, Zhugzhda08, Botha11}. These oscillations, including the determination of a constant offset of peak dominant oscillations between 1600\,{\AA} and 1700\,{\AA}, are the focus of an article in preparation \citep{Battams18a}. These regions also tend to show the lowest power-law index value over this channel, consistent with the narrative that high power-law indices relate to the magnetic structure.  

The coronal bullseye is no longer evident in the 1700\,{\AA} observations. Instead, a $\approx 3$-minute umbra periodicity is sharply bordered by a narrow umbral/penumbral transition ring that transitions to an outer (penumbral) region of $\approx$ five-minute periodicity. The three- and five-minute features are expected and well documented and discussed in the literature \citep[e.g.][]{Yuan14,Bogdan06,DeMoortel09} and serve as validation that our technique is accurately describing oscillatory phenomena. Despite low signal inside the sunspot umbra, we are confident that the values presented here are both real and correct.

At the outer edge of the coronal bullseye, there is a circular region (white ring) surrounding the sunspot, corresponding approximately to the sunspot penumbra, in which no Lorentzian component of significance is found. This feature constitutes another key observation of this investigation -- a feature we label ``penumbral periodic voids'' (PPVs). We have observed such PPVs surrounding every sunspot that we have investigated in 1700\,{\AA} and 1600\,{\AA} observations. (Currently more than two dozen other sunspots have been investigated, the results of which will be presented in a follow-up article.) The PPVs we observe are not due to low signal, as the signal amplitude in the PPV is comparable to other regions (and much higher than that within the sunspot), or due to poor fits, as these regions are particularly well-described by our models. We discuss the PPV features further in Section~\ref{s:Categories}.

The spectra in the umbral/penumbral transition ring between the three-minute circle and five-minute ring appear fundamentally different from those in the PPV in that they seem to be a result of a broad mix of several periodicities in the three- to five-minute range, whereas the PPVs are simply an absence of any statistically significant periodicity. The PPV and this transition ring may be related to the ``regions of lower power'' reported by \cite{Muglach03}, although \cite{Howe12} report a contrasting result obtained from SDO observations in which spectral power is observed to be enhanced in a region surrounding a sunspot. Our results hint that both could be correct, as we do indeed see very closely bordering weak and strong spectral signatures at certain frequencies. The PPV features may also be a manifestation of ``acoustic moats'' \citep{Lindsey98} or ``acoustic halos'' \citep[e.g.][]{Hanson15}, believed to be a consequence of convective flows around sunspots. A detailed study of this, and comparison with these previous related studies, is beyond the scope of this initial results article.

\section{Discussion of Results}
\label{s:discussion}

\subsection{Model Fitting}
\label{s:model-fitting}
In Table~\ref{t:r_values}, we provide for the power spectra models for each wavelength the \textit{modal} reduced $\chi^2$-values: $\chi_{\nu}^2$. Although this metric comes with many caveats for nonlinear fits \citep{Andrae10}, it is consistent with previous studies. However, it is important to note two things. First, we report the mode of $\chi_{\nu}^2$ instead of its mean because the distribution of $\chi_{\nu}^2$-values for all regions is a long-tailed distribution, with a small number of high values (relatively poor fits) that bias the means, whereas the mode value provides a better representation of the majority of fits. The second point we note is that our estimates of the variance used in the reduced $\chi^2$-values are perhaps somewhat poor, as we derive the variance from the statistics of each 3$\times$3-pixel averaging region. The diversity of spectra are such that using a variance value obtained from the entire region is not appropriate. 

From this table, we see that we can obtain excellent model fits throughout the EUV corona (193, 171, and 304\,{\AA}), with $\chi^2$-values consistent with those reported by \cite{Ireland15}, supporting our visual observation that the M2 model performs well at describing essentially all EUV coronal power spectra. The $\chi_\nu^2$ of 79.50 reported in Table~\ref{t:r_values} for 1700\,{\AA} is somewhat misleading regarding the quality of fits in that wavelength channel, which are highly feature-dependent although admittedly far poorer than those of the EUV channels. In the region surrounding the sunspot, including the coronal bullseye region, and in essentially all regions corresponding to the underlying magnetic network, the model fits are excellent, with $\chi^2$ similar to that of EUV. The values are pulled down mainly by the relatively poor fits found in the centers of convective cell structures, the power spectra of which are not well described by the model M2 (either with a Gaussian or a Lorentzian, see Appendix~\ref{Appendix:example_spectra}), and also in the center of sunspot umbrae where the Lorentzian component dominates the spectra. Furthermore, we note that in 1700\,{\AA} (and 1600\,{\AA}) model fits frequently fail to flatten out at the very high-frequency end of the spectrum (period $\lesssim$ two-minutes) and thus often ``miss'' the entire remainder of the high frequency end of the spectrum, despite providing an excellent fit to the three- to five-minute region of the same spectrum (see panels a,b, and d of Figure~\ref{f:1700_examples} in Appendix~\ref{Appendix:example_spectra}.) The relative abundance of missed high-frequency points therefore biases the $\chi_\nu^2$-values. We plan to investigate adding a third model to future iterations of our technique designed to better fit the power-law component of the 1700\,{\AA} observations, which may be more accurately represented by a broken power-law, and to better capture the very high frequency end of the spectrum (although the 24-second cadence of these observations is somewhat limiting). Related to this, and again regarding the chromospheric fits, in particular, we note that preliminary tests of a Kappa function appear to better capture the very highest frequencies in the chromosphere, although with trade-offs in other parts of the spectrum. Further investigation of this facet of the model fitting is beyond the scope of this article. 

\begin{table}
\caption{Reduced $\chi^2$-values [$\chi_{\nu}^2$] and their standard deviation [$\sigma$] for each wavelength considered, and the mean values across EUV wavelengths and all wavelengths.}
\label{t:r_values}
\begin{tabular}{c||cc}     
  \hline                   
\multicolumn{3}{c}{$\chi_{\nu}^2$ for all channels} \\
\hline 
Wavelength\,[\AA] & $\chi_{\nu}^2$ & 1$\sigma$  \\
  \hline
193 & 1.14 & 0.53  \\
171 & 1.42 & 0.80  \\
304 & 1.88 & 0.77  \\
1700 & 79.5 & 44.5  \\
\hline
EUV channels & 1.48 & 0.70\\
All channels & 20.99 & 11.65 \\
  \hline
\end{tabular}
\end{table}

As noted, a certain number of fitting attempts failed to produce a meaningful fit to the M2 model, and instead we reverted to the parameters from the simpler M1 model. In theory, M2 should be adequate for all spectra as it is equal to M1 when Lorentzian components are set equal to zero. However, in cases where the spectrum in question is essentially a pure power-law, the curve fitting algorithm still attempts to incorporate the Lorentzian model components, and thus results in a poorer fit. The percentage of M1 fits in each channel were as follows: 193: 21\,\%; 171: 11\,\%; 304: 3\,\% and 1700: 0\,\%. These values are supportive of a narrative of power-law processes being relatively more dominant in the upper hot corona \textit{versus} the power-law + oscillations dominating the lower/cooler/denser corona (adopting an admittedly naive stratified coronal model). Finally, we note that in none of these ``rejected M2'' spectra is a legitimate Lorentzian feature being erroneously omitted, and thus we currently have no plans to address the sporadic bad fits to M2.

\subsection{Spectra Averaging}
\label{s:segment_discuss}
Spectra obtained from full, unmodified 12-hour time series contain a significant amount of noise, resulting in very slow and low-quality model fits. Therefore our methodology employs ``segment averaged' spectra in which a 12-hour time series is separated into six two-hour time series and their spectra are averaged. In Appendix~\ref{Appendix:seg_avg} we provide an example of the impact of this averaging upon the resulting spectra and model fits. In the specific example given in that Appendix, we see the Lorentzian location is almost entirely unaffected by segmenting but the power-law slope is reduced from 2.58 to 1.71, giving the impression that the slope value is highly subjective based on data preparation. However, that example is just one of more than six million spectra encountered in this investigation, and as discussed below it is not entirely representative of the behavior of all power spectra.

To derive metrics on the variations of power-law indices as a function of segmenting, we applied our described methodology to each AIA channel using segmenting of 12$\times$1-hour, 3$\times$4-hour, and 2$\times$6-hour (in addition to the 6$\times$2-hour segmented presented in Section~\ref{s:results}). From each resulting power-law index map, in each of the four AIA channels, we derived a `range` map representing the range (max - min) of values for each parameterization and examined the statistics (summarized in Table~\ref{t:segment_stats}), histograms, and spatial properties of these maps. From this analysis, we make the following observations.

First, the 193\,{\AA} channel is most affected by segmenting, with power-law indices tending to decrease on average as segmenting is increased (a total change of $\approx 10$\,\%). In the other three channels, this trend is reversed: the power-law indices tend to be slightly increased on average with increased segmenting (a total change of up to $\approx 15$\,\%). This effect is seen in the median values reported in Table~\ref{t:segment_stats}. 

Second, we observe that as we progress to characteristically cooler AIA channels, the impact of segmenting lessens in the sense that the standard deviations of power-law indices for all channels get smaller, going from 0.43 in 193\,{\AA} to just 0.08 in 1700\,{\AA}, although this will be partially related to the decreased range of power law indices we see in these channels. 

Third, histograms for these ``range'' maps all tend to appear as narrow, slightly skewed Gaussians with heavy but thin tails, centered on the median values stated in Table~\ref{t:segment_stats}. This means that in some extreme circumstances, a minority of pixel locations may have power-law ranges in excess of $\pm$3.0 as a consequence of segmenting, most commonly in 193\,{\AA}. In that channel in particular, such high-variation features are almost exclusively observed at the borders of loop structures, and are likely a consequence of loops ``swaying'' and occupying different pixels during the 12-hour sequence. However, it is important to note that most loop structures themselves are largely very stable in terms of power-law indices, particularly in 193\,{\AA}. In the other three channels, we again observe that the largest power-law variation occurs in regions bordering the most dynamic features (\textit{e.g.} near loops, active regions, magnetic networks) but again note that in general these features themselves have largely stable power-law indices. 

In summary, we find that for the great majority of pixel locations in our regions of interest, the variations in the power-law indices due to the amount of segmented averaging used is on the order of $10$\,\%. The median (and mode) values of the resulting histograms are all close to zero, and the spread is relatively narrow for the great majority of pixels. The largest variations we observe appear primarily as a result of the fundamental dynamic nature of the observations, with our segmenting process effectively just changing the sampling of these non-stationary features. As noted in the following Section, this is largely unavoidable in any study such as this. We conclude that the standard deviations that we observed are not unreasonable given the nature of the observations, but we reiterate remarks made elsewhere in this article that we do not claim to be able to determine the ``true'' power-law slope of any given location in the solar corona, should such a true value even exist, but that our slope values are largely consistent with those found in other published literature. 

\begin{table}
\caption{Uncertainties in derived power-law indices as a function of segmentation (averaging), with results here reflecting the difference between the most (12$\times$1-hour) and least (2$\times$6-hour) amounts of averaging.}
\label{t:segment_stats}
\begin{tabular}{c||cc}     
\multicolumn{3}{c}{Power-Law Indices [\textit{n}]} \\
\hline 
AIA Channel & Median & Std. Dev. \\
\hline 
$193$ & -0.1 & 0.43  \\
$171$ & 0.04 & 0.3  \\
$304$ & 0.09 & 0.22  \\
$1700$ & 0.11  & 0.08  \\
  \hline
\end{tabular}
\end{table}

Exploration of the averaging-induced variation in $\beta$ reveals that for most spectra fits, the value of $\beta$ remains essentially unchanged, per Appendix C. However, we do note that certain spectra that are always poorly represented by both M1 and M2 will spuriously report changes in $\beta$ of up to $\pm$~nine-minutes (i.e. the full parameter range) as a function of averaging, due simply to the fitting algorithm converging to different local minima within a very broad range of possible values. Such locations appear to be predominantly those that we suspect will be fit better by the Kappa model. Thus we are confident in our reassertion that the segmenting makes a negligible impact on the Lorentzian parameter. 

Finally, for completeness, we also investigated the impact of the 3$\times$3 averaging on our parameterizations finding a negligible change in both power-law indices (\textit{e.g.} median=-0.01, $\sigma$=0.08 in 193\,{\AA}) and Lorentzian location (\textit{e.g.} median=0.01 minutes, $\sigma$=0.05 minutes in 1700\,{\AA}) when compared to no spatial averaging. This 3$\times$3 averaging procedure aids in smoothing out pixel-level variations and produce more reliable spectra fits. An interesting investigation beyond the scope of this study would be an exploration of different smoothing kernels (\textit{e.g.} 3$\times$3, 5$\times$5) with different weighting parameters, although again the impact of this process is certainly secondary to that of the temporal (segment) averaging procedure. 

\subsection{Limitations of Approach}
\label{s:limitations}

While the approach presented here holds promise for a variety of studies of solar-atmospheric dynamics, turbulence, and wave propagation, there are limitations of note. Most of these limitations may be mitigated through improved analysis procedures, but some are inherent to the raw observations.

Raw, unsmoothed coronal power spectra are inherently noisy, leading to high uncertainties in model fit parameters. We reduced the noise by averaging spectra temporally in two-hour time series sequences and spatially over 3$\times$3-pixel regions, enabling far better fits to the observations. However, this means that our spectra are perhaps better considered as 12-hour summary spectra, representing the average spectral conditions of a pixel location over a 12-hour window. This also means that our methodology would ``miss'' any short-duration events that occur in only one or two of our two-hour time series. However, detection of such sporadic features is not the purpose of this method (and indeed is challenging for any technique). 

Extensive validation tests were performed with different averaging windows, finding that the averaging had the greatest impact on the low end of the frequency spectrum and then consequently has the greatest influence on the power-law index value. Lorentzian location values were essentially identical (to within a few seconds, at most) regardless of averaging, and thus studies focused solely on periodicities could likely employ less averaging. A slightly better power-law fit may be obtained by using longer time series (because additional low-frequency points are used in the fit), but with the trade-off of increased noise and a lower success rate in model fits. In general, we observed that variations in our averaging procedures resulted only in variations in the fitted power-law slopes, with more averaging tending to decrease the slope value, and thus implying that alternate methods may return slope values higher than those stated here. However, all variations remained self-consistent across all features in the region of interest \textit{i.e.} a feature with high power-law index (\textit{e.g.} loop) would always retain a high power-law index relative to all other features. Therefore, we do not claim to be able to determine the ``true'' value of any power-law slopes, but it is unlikely that such a ``true'' value exists in the context of a highly dynamic corona. 

The high-frequency end of the spectrum, and thus the Lorentzian (or Gaussian) parameters, are largely insensitive to the length of the time series used, being more reliant on a high temporal resolution to fully capture that end of the spectrum. The choice of a Lorentzian was driven by the desire to select a model that has some physical meaning in the context of (damped) oscillations. However, as noted, we obtain largely identical results with a Gaussian model \citep[per][]{Ireland15}, with a notable exception of sunspot cores, which are significantly better fit with a Lorentzian. A third alternate proposed model is the Kappa function used by \cite{Auchere16b}, who demonstrate the ``spectral humps'' can be the result of ``periodic trains of pulses of random amplitudes'' rather than oscillations centered at the peak of the apparent hump in the spectrum. We have tested this function in a more limited case and note results largely identical with both the Lorentzian and Gaussian models. Given that we are fitting similarly shaped functions to noisy observations, this result is perhaps unsurprising. We suspect -- but can not currently verify -- that many of our observed ``oscillations'' (\textit{e.g.} the widespread 11-minute oscillations in 171\,{\AA}) may indeed be a consequence of these pulse trains, as this certainly provides a plausible mechanism. However, in cases such as sunspots and the chromosphere, literature strongly supports our identification of the oscillatory properties we report here, and in the sunspot case, oscillations are clearly visible by-eye in animated sequences of observations. Therefore, in such cases, we would propose that the damped oscillation (Lorentzian) model provide a superior choice. Ultimately, a combination of Kappa and Lorentzian models may prove the optimum choice, although our results imply that making a decision between these models may be challenging in the presence of noise in the observations. Nonetheless, overcoming this challenge may significantly improve our understanding of the underlying physics occurring in the regions that exhibit such spectral features.

In this study, we selected a time interval without large-scale dynamic events, although note a minor B9.2 flare in the region around 04\,{UT}. As part of our validation, we noted that omission of the two-hour segment containing the flare results in essentially no change in the power spectra or fitted model parameters. Such short duration events are thus apparently not impactful so long as they do not substantially change the physical configuration of the corona in the flaring region. It is not possible to avoid all dynamic activity in the solar corona, as most features on all scales are constantly in motion. However, by avoiding major impulsive (\textit{e.g.} flares) or eruptive events (\textit{e.g.} filament eruptions), we are confident that our spectra do not include a bias from major events. However, we reiterate that our method is not appropriate for studies that wish to locate sporadic, short-duration events. Thus, the use of this technique requires consideration of the time-series length under investigation to ensure that dynamic events are included or excluded as desired. 

On smaller scales, particularly in the hotter corona, we observe coronal loops in apparent motion. In some cases, these may be loops that are physically moving, but they may also be representative of situations in which, for example, one loop is cooling while a nearby loop is heating, giving the appearance of physical motion of a single loop. There is no simple solution here -- the corona is inherently highly dynamic and loop motion, whether apparent or real, is not trivial to correct without complex feature tracking and modeling. Furthermore, the spectral properties of the AIA filters are such that some channels include many different emission lines that provide an additional layer of complexity for analysis. We have intentionally omitted this consideration from the presentation of these early results, but we note that future studies using this technique will certainly need to incorporate a thorough understanding of the different spectral components in each channel. Furthermore, care would need to be taken if this technique were used to follow the behavior of a single loop structure, for example, but this is a challenge faced by all such studies and is perhaps more of a data limitation than a methodology limitation.

An additional complication for our technique arises due to the possible superposition of features within the corona. To some extent, our approach assumes the corona to be a flat plane, whereas, of course, it is a dynamic, optically thin 3D structure. Thus, for example, large coronal-loop structures can be seen arcing high above the lower corona. In such a situation, even assuming a static loop, any given pixel will contain dynamic intensity (and spectral) contributions from both the loop and the underlying corona. Compounding this, the derotation algorithms applied are likely incapable of properly correcting an ``optically deep'' field of view. Also, as noted in Section~\ref{s:segment_discuss}, the fact that these loops are often actually \textit{not} stable can lead to high variations in derived power-law indices local to these features, depending on how the data are sampled (averaged). However, as our results show (particularly in the hot corona), loop structures are clearly defined in all aspects of our parameterizations, and our approach is again validated by the obvious visual correspondence between the parameterizations and the underlying observations.

\section{Model Components}
\label{s:Categories}

Components of coronal power spectra can be broadly described as ``turbulent'', ``periodic'', or ``quiescent'', with corresponding spectral characteristics of a power-law, a Lorentzian peak, and a flat power-law tail (considered to be primarily white noise). Per the plot captions in Figures~\ref{f:All_TS} and \ref{f:All_Fits}, we broadly categorize power spectra as i) Tail (white noise) dominated without a Lorentzian (panels a and b in Figures~\ref{f:All_TS} and \ref{f:All_Fits}); ii) Power-law dominated without Lorentzian peak (panel c in Figures~\ref{f:All_TS} and \ref{f:All_Fits}; iii) Power-law dominated with Lorentzian (panel d in Figures~\ref{f:All_TS} and \ref{f:All_Fits}). It is important to note that while the power-law and Lorentzian features are a consequence of physical processes, the power-law tail is primarily a consequence of reaching a noise floor (photon noise) in the observations. In Table~\ref{t:summary} we summarize the locations and properties of the Lorentzian (columns 2 and 3) and power-law (columns \,4--\,6) components of our parameterizations for each wavelength studied. 

\begin{table}[htbp]
\caption{Summary of properties and locations of Lorentzian and Power-law Features. For each channel (\textit{column 1}), we indicate the typical location of Lorentzian features (\textit{column 2}), the percentage of the region of interest that contains statistically significant Lorentzians (\textit{column 3}), the typical locations of high (\textit{column 4}) and low (\textit{column 5}) power-law indices, and the general range of power-law indices observed in that channel (\textit{column 6}). For brevity, we use ``CH'' to denote coronal hole. \newline}
\label{t:summary}

\resizebox{\columnwidth}{!}{\begin{tabular}{l||l|l||l|l|l}& \multicolumn{2}{c||}{Lorentzian Features} & \multicolumn{3}{c}{Power-law Features}\\
Channel &  Locations & Cover & High Index & Low Index & Range\\
\hline
193 & \begin{tabular}{@{}l@{}}Sporadic, \\ Sunspot \end{tabular} & 19.3\,\% & \begin{tabular}{@{}l@{}} Loops, center of \\ coronal cells \end{tabular} & \begin{tabular}{@{}l@{}}
  CH, cell boundaries,\\  filament channel \end{tabular} & \,1.1--\,2.5\\
\hline

171 & \begin{tabular}{@{}l@{}}Sporadic, \\ Sunspot \end{tabular}  & 37.9\,\% & Loops, footpoints & \begin{tabular}{@{}l@{}} CH, low signal \\ areas \end{tabular}  & \,1.2--\,2.2\\
\hline

304 & \begin{tabular}{@{}l@{}}Widespread, \\ Sunspot \end{tabular}  & 62.0\,\% & \begin{tabular}{@{}l@{}} Active region, \\ bright points \end{tabular} & \begin{tabular}{@{}l@{}} CH, widespread \\ sporadic locations \end{tabular}  & \,1.1--\,1.7\\
\hline

1700 & Global & 99.8\,\% & \begin{tabular}{@{}l@{}} Sunspot, \\ magnetic network \end{tabular}  & \begin{tabular}{@{}l@{}} Internal to magnetic \\ network \end{tabular} & \,1.1--\,1.5\\
\hline
\end{tabular}}
\end{table}

\subsection{Lorentzian}

Periodic features, represented by our Lorentzian model component, are observed throughout the solar atmosphere at numerous timescales, although our study is limited to those of 11 minutes or less. Generally, the Lorentzian component locations are found at the high-frequency end of the spectrum, such as shown in Figure~\ref{f:All_Fits}d, which corresponds to Point D in Figure~\ref{f:All_viz} and is situated near the sunspot in AR-1777. 

The first observation that we note is a general one -- namely, that at characteristically cooler/lower heights in the corona, significant Lorentzian components become more prevalent, reaching almost complete coverage in 1700\,{\AA}. This is evidenced by the increase in coverage (column 3 of Table~\ref{t:summary}) of a given channel by statistically significant Lorentzians. In 1700\,{\AA} these features are entirely global, but they become more sporadically located in characteristically hotter EUV channels.

The exception is the sunspot, whose periodicities permeate all of the channels investigated and are particularly well-represented by the Lorentzian (damped oscillation) model. This is a result that closely mirrors the findings of \cite{Reznikova12a}, who noted that ``... the strong magnetic field of a sunspot works as a waveguide for the acoustic waves propagating from the photosphere level and eventually reaching a 1 MK corona.'' At the photosphere and in 304\,{\AA}, we see almost a perfect series of concentric periodicities (as noted by \cite{Yuan14}, for example) around the sunspot; a feature we labeled a \textit{coronal bullseye}, with the three-minute central peak pervasive through to the 193\,{\AA} observations.  

Our presentation of this structure may be related to observations reported in \cite{Reznikova12a} and \cite{Reznikova12b}, although our approach differs in that instead of isolating specific frequencies, we instead include and visualize all frequencies. This enables the production of a full 2D map of the periodicities surrounding sunspots  (and around certain loop footpoint, as we have observed in datasets not presented here) and reveals structural patterns that may otherwise be missed. In 171\,{\AA} and 304\,{\AA}, the coronal bullseyes are well-structured and defined, with radially decaying periodicities from a central peak of three-minutes. In contrast, no such decay is observed in the 1700\,{\AA} observations; instead, the central three-minute region is entirely isolated from an outer five-minute region by a very thin, disorganized region of chaotic oscillations. A similar finding was reported by \cite{Tziotziou07}, who noted a rapid jump in the oscillation period at the umbral--penumbral boundaries in Doppler observations. Our results further this by noting that this ``jump'' region sometimes contains a steep gradient of periodicities, but it often shows a complete lack of any statistically significant periodicities.

All Lorentzian features presented here have a corresponding derived FWHM, reported in the temporal domain (in units of minutes). As previously noted, the relationship between Lorentzian location and FWHM is complex, and it is both feature and wavelength-dependent. In the hot corona, magnetic structures (sunspot, loops) have the smallest FWHM, with values similar to their periodicity. In the chromosphere (1700\,{\AA}), however, the magnetic networks have larger FWHM with values for any given location on the order of two to three times that of their periodicity. The complexity of these relationships is such that a dedicated study is warranted here in which features are perhaps organized by type, physical mechanism, and wavelength, and the relationship between their periodicity and width studied in detail to gain an understanding of the damping mechanism(s) involved, if any. While we have presented the FWHM parameter as a potential proxy for damping of oscillatory processes, it is important to reiterate that not all Lorentzian-like spectral features observed in the corona are necessarily a result of damped processes, but we feel that the FWHM is a potentially useful metric throughout the corona, regardless of the driving mechanisms. 

Finally, we reiterate that detected ``oscillatory'' features do not necessarily imply an underlying oscillatory process, with many such ``spectral hump'' features being equally well accounted for by the mechanisms described by \cite{Auchere16b}. A natural next step for this work is to establish which of the proposed models (Lorentzian, Gaussian, Kappa) account best for which AIA channels and solar features, as this should provide valuable insight into the physical drivers of these spectral features.

\subsection{Power Law}

Broadly speaking, most coronal regions can be described to some approximation by a simple power law, with power-law indices confined within the approximate range of 1.1 to 2.5 in the presented observations (but exceeding 3.0 in other data sets that we have surveyed). A typical power-law dominant spectrum is shown in Figure~\ref{f:All_Fits}c, corresponding to Point C in panel a of Figure~\ref{f:All_viz}, which is a small and bright loop region. As summarized in column 4 of Table~\ref{t:summary}, these highest power-law indices are primarily observed in coronal loops at high temperatures (on the order of $10^6$~K). This finding is consistent with the spectral index found in simulations of coronal loops \citep[e.g.][]{Rappazzo07,Rappazzo08,Muller05,Matsumoto16}. In 193\,{\AA} we also observe high power-law indices in the center of coronal cells, most likely a consequence of the bundles of magnetic-field lines located in such features \citep{Sheeley12}. In 304\,{\AA}, the highest indices are observed in the active region surrounding the sunspot and scattered across the entire field of view in apparent correspondence with bright points in the visual observation (which appear by-eye to be related to the underlying magnetic network). In 1700\,{\AA} we see that the highest indices correspond clearly and directly to the magnetic network (and again the sunspot), with essentially no exceptions.

Our interpretation is thus that strong concentrations of magnetic fields facilitate a rapid cascade of energy through the spectrum and result in relatively high power-law indices, mirroring those found in simulation-based studies (\textit{e.g.} \cite{Kitiashvili15}). Regions with relatively shallower slopes would presumably be experiencing processes that inhibit the cascade of energy. More generally, all solar power spectra appear to contain at least some component indicative of turbulence. Intuitively, this is to be expected as was noted by \cite{Ireland15}: ``[the power-law fit is] consistent with the idea that the solar atmosphere is heated everywhere by small energy deposition events.''   

The lowest power-law indices are found in the coronal hole (193, 171, 304\,{\AA}), the coronal cell boundaries and filament channel (193\,{\AA}), and many low-signal areas (193, 171, 304\,{\AA}). In 1700\,{\AA} the lowest indices are observed in the center of regions internal to the magnetic network. A common theme among all of these is relatively lower signal (and higher noise levels and thus longer tail, \textit{e.g.} Figure~\ref{f:All_Fits}c). The power-law fits for such spectra are often fit to relatively few points and therefore will have a higher level of uncertainty. However, and as discussed in the following section, specific and differing kinds of features (\textit{e.g.} filament \textit{versus} coronal hole) return self-consistent values across all data sets we have investigated (including many outside of that presented here), leading us to be confident that the power-law values that we report are meaningful, although subject to uncertainties we have already discussed. Further to this, as discussed in Section~\ref{s:segment_discuss}, the choice of averaging scheme for the observations can change the value of power-law indices [n] on average $\approx 10$\,\% but dependent on wavelength and feature. For example, more averaging tends to marginally decrease (on average) \textit{n}-values in 193\,{\AA}, but marginally increase them (on average) in the other channels, with the most dynamic regions of the corona (\textit{e.g.} those directly adjacent to moving loop structures) showing the largest changes. However, the majority of the region of interest remains stable in terms of \textit{n} and, for example, high-\textit{n} features always remain high relative to low-\textit{n} features regardless of the data preparation technique. Despite this feature-to-feature consistency, we do note that caution should be exercised in the literal interpretation of derived power-law indices as data preparation can globally bias all values.

\subsection{White Noise}

Quiet and low-signal regions of the corona exhibit an extended white noise (flat) spectrum above a certain frequency, referred to as the rollover frequency (Equation~\ref{eq:rollover_freq}). The interpretation here is that large rollover periods imply a smaller power-law frequency range and a larger white-noise (photon noise) frequency range in the spectra.  

Panels a and b of Figure~\ref{f:All_Fits} show the model fits extracted from regions whose spectra show a significant white-noise component. Specifically here, these time series and their corresponding spectra were obtained from points identified in Figure~\ref{f:All_viz}a that were within i) a filament and ii) a coronal hole. Other filaments that we have studied have similar traits in that their power-law slope is larger than those observed in coronal holes, particularly at cooler temperatures, but they both have wide frequency ranges with flat spectra. This observation is likely driven by improved signal-to-noise ratios in filaments \textit{versus} coronal holes, though may also relate to the closed-magnetic-field nature of the filament structure. Nonetheless, studies of the spectral turbulence (or lack thereof) in coronal holes may have application to studies of the turbulence in the fast solar-wind that originates from these otherwise spectrally quiescent regions. These results may also have relevance to the understanding of the interplay between magnetic and kinetic energy in the corona; simulations such as \cite{Rappazzo08} have found large power-law indices in magnetically-dominated regions and flat spectra in regions dominated by kinetic energy. 

\subsection{Periodic Penumbral Voids and Umbral-Penumbral Transitions}

In Figure~\ref{f:results_1700}c and d we observed the feature that we label a penumbral periodic void, or PPV -- an annular region surrounding a sunspot in which no statistically significant periodicities are observed. As noted elsewhere, we have observed PPVs surrounding every sunspot in 1600/1700\,{\AA} data that we have investigated (currently more than two dozen). These features are obviously a consequence of some process or processes that impede or prevent the establishment of stable oscillatory behavior, likely related to the local magnetic-field structure surrounding the sunspot, or perhaps indicative of an impairment of wave coherence around the sunspot \citep{Zhao06}.

\begin{figure*}[htbp]
\centering
{\includegraphics[scale=0.445]{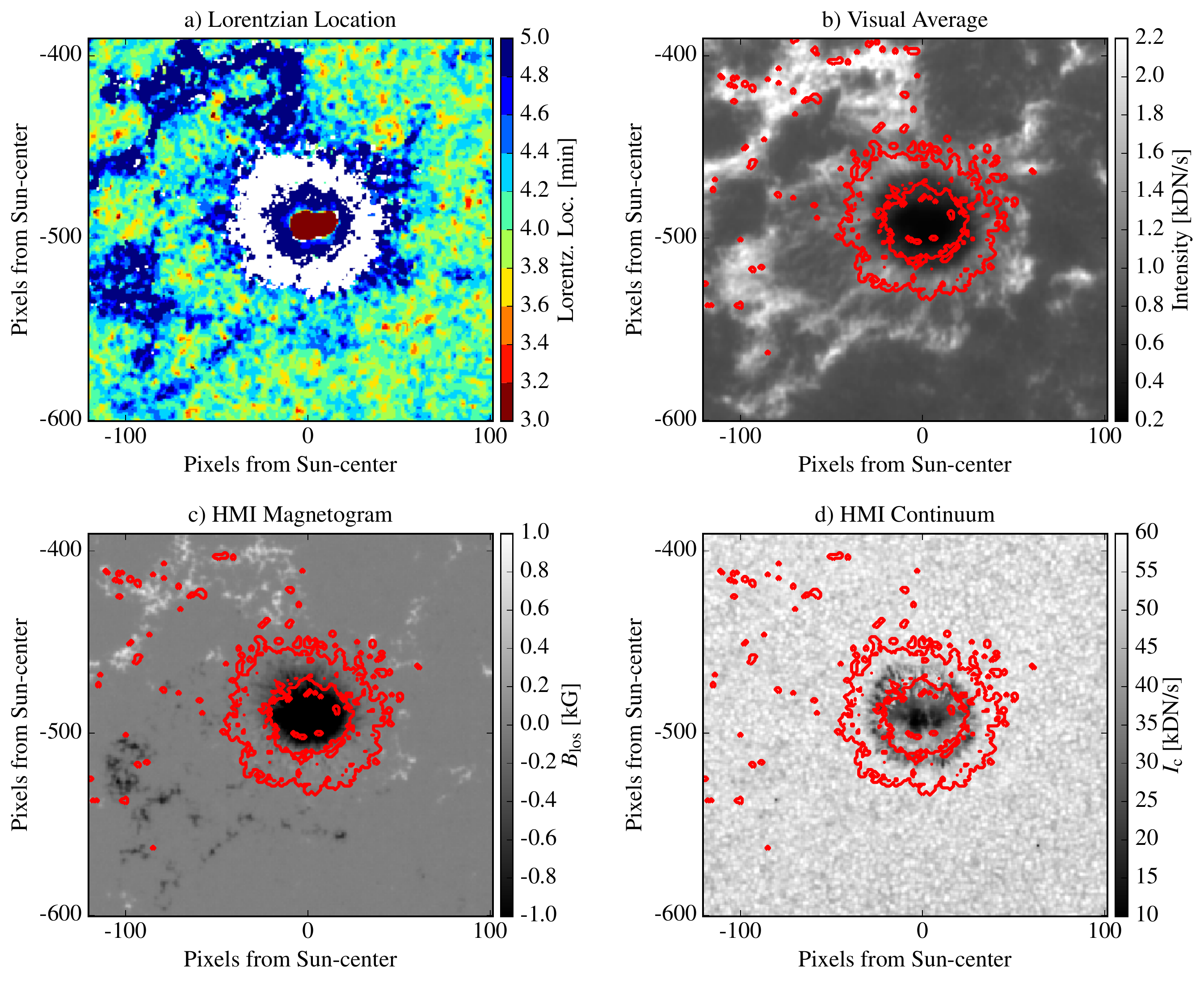}} 
\caption{(a) An enlarged view of the active region and PPV shown in Figure~\ref{f:results_1700}c and d, with spatial axes in units of pixels relative to solar disk center. (b) The visual average intensity image for this 1700\,{\AA} observation and panels c and d show the corresponding solar region as seen in the Helioseismic Magnetic Imager (HMI) Magnetogram and HMI Continuum observations, respectively. Overlaid on panels (b,c,d) is a red outline that corresponds to the borders of the white areas (the PPV) seen in panel a and the borders of the inner umbra--penumbra transition ring.}
\label{f:HMI}
\end{figure*}

In Figure~\ref{f:HMI} we provide a brief examination of both the PPV and an additional feature -- an inner umbra--penumbra transition ring. These features are highlighted by a red outline (actually the significance mask that we use to omit statistically insignificant results) overlaid upon the 1700\,{\AA} average-intensity image (Figure~\ref{f:HMI}b), and the corresponding HMI magnetogram and continuum  observations (Figures \ref{f:HMI}c and \ref{f:HMI}d). 

The PPV (white ring in Figure~\ref{f:HMI}a) is a broad feature both here and in the same 1600\,{\AA} observations (not shown), and it bounds the entire sunspot. In comparison with animations of the corresponding magnetogram observations, and as is somewhat evident from Figure~\ref{f:HMI}c, we note the PPV appears to correspond extremely well with the location of moving magnetic features \citep{Vrabec71, Harvey73, Wilson73}, the presence of which may be expected to impede coherent periodicities, either directly or as a consequence of the known chaotic or absent magnetic fields in regions close to sunspots \citep{Ryutova98}. A relationship to acoustic moats (or halos) \citep[e.g.][]{Lindsey98,Hanson15} is also a very plausible possibility. We note also that the PPV observed in 1700\,{\AA} is not observed in the overlying corona (304\,{\AA} or above), with instead that region of the corona showing strong periodic features (coronal bullseye). Thus it seems the oscillations that we observe in, say, 304\,{\AA}, are perhaps not driven from immediately below, but instead propagate through magnetic fields that come up from the sunspot and are draped over the PPV. Related to this is the observation that the PPV is essentially identical in both the 1600\,{\AA} and 1700\,{\AA} observations (the former not presented here), and thus given that both these channels are strongly continuum-dominated, the source of the PPV is itself likely rooted in the continuum, as opposed to having an origin in the weaker spectral lines within the relevant filters. 

The second feature we draw attention to in Figure~\ref{f:HMI}a is a so-called \textit{inner umbra--penumbra transition ring}: a very narrow ring of chaotic periodicities that act as a border between the three-minute umbral and five-minute penumbral regions (all entirely inside the PPV). The transition ring shown here fully encircles the sunspot umbra, with a mixture of very steep gradients of periodicities, or regions absent of any periodicity, observed along much of its circumference. These features are common to both the 1700\,{\AA} and 1600\,{\AA} observations, with minimal differences observed between these two channels for any given date/time range. 

It cannot be conclusively determined from our observations whether the absence of statistically significant periodicities in the umbra--penumbra transition ring arises from a lack of such signal (as is the case with PPVs), or results from interfering signals or colliding waves within our 12-hour window. As shown in Figure~\ref{f:HMI}d, the inner umbral--penumbral transition ring encompasses the sunspot core as observed in HMI continuum observations. This feature may, therefore, relate to the region at which inward and outward flows diverge \citep{Sheeley17}; again, such a region would not intuitively be expected to maintain a coherent periodicity. However, as noted above, multiple observed periodicities in such a spatially small region may be complicating the signal we observe here. Nonetheless, the chaotic spectra seen in this narrow region are fundamentally spectrally different to any other region observed throughout all investigated AIA channels thus far.

Oscillations and waves around sunspots have long been a key focus area, with a good recent review of the field appearing in Chapter 5 of \cite{Khomenko15}, with the overall picture being one of a complexity of enhanced and diminished powers in different oscillation modes. Literature searches did not yield a clear identification and/or description of the features that we label PPVs, with instead a diversity of studies (as cited) hinting at various aspects of this observation. Further analysis of these chromospheric features will require detailed study and models of the underlying magnetic fields, and perhaps considerations of moving magnetic features, running penumbral waves, and Doppler observations. Thus it is outside the scope of this initial survey of results. 

\section{Summary}

We have presented a methodology that enables the pixel-level spectral parameterization of solar EUV observations. This technique enables us to reduce EUV intensity time series to spectral model components that can be used to both separate out and explore the different underlying physical processes (\textit{e.g.} power-law \textit{versus} oscillatory) occurring in any given pixel-level location, resulting in a significant reduction in data volume while providing unique insights into coronal processes. We presented only results from two of the six model parameters -- the power-law index and the Lorentzian peak location -- with a third derived parameter representing the full-width at half-maximum (FWHM) of the Lorentzian derived from the model. We showed the spatial distribution of these parameter values map directly to visible features in the EUV images and outline, for example, coronal loops, bright points, coronal holes, sunspots, and the photospheric magnetic network. We also demonstrated the complex feature-dependent relationships between the location and width of observed Lorentzian features. The presented method could be extended to facilitate the understanding of the transfer of energy throughout the corona via a number of different studies. For example, with comprehensive calibration, the evolution of the power spectra for an individual identified structure (\textit{e.g.} a coronal-loop) could be investigated as a function of AIA wavelength, and compared to the spectra produced from the moderate breadth of numerical simulations of coronal loops.  

Using our methodology, we identify features in the locations of observed periodicity that we refer to as ``coronal bullseyes'' (circular periodic structures found over sunspots but also occasionally over coronal bright points), ``penumbral periodic voids'' (PPVs: circular rings surrounding sunspots, devoid of any inherent periodicity), and ``inner umbra--penumbra transition rings'' (very thin rings of chaotic periodicity acting as the boundary between three- and five-minute oscillations in sunspot cores). Sunspots are observed here to have strong and radially decaying periodicities with a bullseye-like structure with a three-minute central periodicity observed in all wavelengths. Well-defined PPV rings were observed in photospheric observations, and they appear to correspond to the sunspot penumbra and may be a consequence of moving magnetic features in magnetogram observations \citep{Harvey73}. We also identify a spectrally unique region of chaotic periodicities at the umbra--penumbra boundary, appearing as a narrow ring, which may be related to diverging flows around the sunspot core \citep{Sheeley17}. 

Our methodology identified a global 4.05/4.19 minute oscillation in the AIA 1600/1700\,{\AA} observations, in agreement with the wavelet analysis used by \cite{McIntosh04} that returned results to within 0.07 minutes of that detected by our analysis. More broadly, the 4-minute oscillation is a well-known observation, although a literature search did not yield any specific studies dedicated to exploring and characterizing the slightly different oscillations observed between 1600\,{\AA} and 1700\,{\AA}. These periodicities are explored further in an article currently in preparation by this article's authors \citep{Battams18a}. 

Regions of strong closed magnetic fields (\textit{e.g.} coronal loops) were shown to have spectra with large power-law indices, while coronal holes and filaments have a wide range of frequencies with flat spectra, but with steep initial power-law indices found in filaments, likely a result of their magnetic structure. Sporadic five-minute oscillations are seen in 304\,{\AA} that readily pervade to 171\,{\AA} and, to a lesser extent, 193\,{\AA}, although the nature of these and many other observed oscillations remains an open question \citep{Auchere16a,Auchere16b}. We have demonstrated (Section~\ref{s:segment_discuss}) that the determined periodicity for these apparent oscillations is largely unaffected by data preparation (segmenting and 3$\times$3 averaging). We have also quantified the uncertainties of power-law slopes arising from our averaging, finding that increased segmenting (averaging) does systematically bias power-law slopes, but it does so uniformly and to within reasonable tolerances. However, we reiterate that caution is warranted with a literal interpretation of power-law slopes.

Finally, we have shown that the FWHM of fitted Lorentzians -- an analog for the damping coefficients -- has a complex relation related to both the nature of the structures in which we observe them (\textit{e.g.} coronal loops, sunspot, magnetic network), as well as the wavelength channel in which we observe them, with all showing ratios between one and five (\textit{i.e.} Lorentzian widths,  are, at most, five times that of their period of oscillation). Such results warrant detailed investigation and could reveal valuable insight into the different physical processes driving, modulating and damping oscillations throughout the corona.

Our approach could also be used for detailed studies of the propagation of waves throughout the solar corona, with the ability to track signals at the pixel level over spatially large regions. There is also an opportunity to study wave propagation around sunspots, where in 1700\,{\AA} for example, we observe the ``penumbral periodic void'' features in which the spectrum becomes essentially a pure power law throughout a ring surrounding the sunspot, absent of any oscillatory behavior. Circular regions of both enhanced \citep[e.g.][]{Howe12} and diminished \citep[e.g.][]{Muglach03} spectral power have been noted in published literature, among a wealth of similar studies and results \citep{Khomenko15}. However, those observations are perhaps limited by the methodology used to uncover those phenomena, in which a narrow frequency range of the spectrum is considered. We expect that the presented method may aid in resolving discrepancies arising from previous studies by providing a spatial context and a complete spectral characterization over an entire region. Studies such as those by \cite{Lindsey98}, \cite{Tziotziou07}, \cite{Reznikova12a}, and \cite{Reznikova12b}, appear to have seen facets of PPVs and coronal bullseyes, but again our methodology enables a full 2D spatial investigation of periodicities in the corona, thus tying together such existing studies. 

\acknowledgments
K. Battams was supported by the NRL Edison Memorial Program and the Office of Naval Research. The authors wish to thank Jack Ireland for his many inputs during discussion and assistance with model-fitting validation. We are also grateful for the insights of our anonymous reviewers, whose comments have led to significant improvements in this study. 

\section*{Disclosure of Potential Conflicts of Interest}
The authors declare that they have no conflicts of interest.

\appendix

\section{Spectral Fitting}
\label{Appendix:SpectralFitting}

In this appendix, we give details on the considerations and issues involved with fitting a model to spectra computed from time series generated by extracting intensity values from a single pixel over a 12-hour time interval. The nominal image cadence is either 12- or 24-seconds with few missing images (at most $\approx 3\,\%$). The time series were placed on a uniform 12- or 24-second time grid and the gaps removed using linear interpolation prior to computing the spectra.

To estimate the spectral model parameters, many methods were considered in order to address the following issues:

\begin{enumerate}
\item Non-stationarity -- Over a 12-hour time period, the spectra at a given location may change from, for example, power-law + tail to power-law dominated. To address this, we can use shorter time segments to compute the spectra with the drawback of a possibly less accurate power-law index because the spectra will have fewer points at low frequencies.  
\item Noise -- The spectra for a given 12-hour time series typically has large noise amplitude, and as a result, ``failed'' fits often resulted. A failed fit is one in which the curve-fitting routine produces a spectrum that visually does not match the spectra in a sensible manner or does not fit at all. These fits are due to inherent limitations in nonlinear fitting algorithms. We considered two approaches to reducing the noise: i) computing the average of spectra derived from segments of the full-time series and ii) computing the average spectra in a 3$\times$3 pixel box, which has the drawback that neighboring pixels may not have the same spectral type.
\item Computation time -- Spectra with a large amount of noise take much longer to fit. As an example, when fits for 1600$\times$1600 spectra are computed using no averaging, the time was projected to be greater than $\approx 100$ hours compared to $\approx 4$ hours for the averaging method that was used.
\end{enumerate}

Based on these issues and considerations, the method used for computing model parameters in this work is given as follows. We note that this methodology was based on extensive numerical testing and experimentation along with visual inspection of the fits at individual spatial locations (to verify that the fits were consistent with what was expected visually).

\begin{enumerate}
\item For each pixel, average the spectra from six two-hour non-overlapping time segments from the full 12-hour interval;
\item Average nine spectra in a 3$\times$3 box to compute the final spectra for a pixel at the center of the box;
\item Compute parameter estimates using the Dog-Box method from the SciPy $\sf{optimize.curve\_fit}$ version 0.18.1 package for Python 3.6.4 with parameter bounds given in Table~\ref{t:bounds} and uncertainties corresponding to the standard deviation of the nine spectra values used in the averaging described in (ii). For the 1700\,{\AA} channel, we used an uncertainty at frequency $f_i$ proportional to $\log_{10}(f_{i+1}/f_{i})$ with the uncertainty for the highest frequency equal to that of the next-highest. This \textit{ad-hoc} approach was taken for 1700\,{\AA} because it led to fewer failed fits and better fits from a visual perspective; and
\item Use the best-fit parameters from the Dog-Box optimization as initial guesses for the TRF optimization method from SciPy's $\sf{optimize.curve\_fit}$ optimization package with the same parameter and uncertainties used for the Dog-Box optimization step.
\end{enumerate}

The first two steps were needed to reduce the noise in the spectra and decrease the amount of time needed for each optimization. Because we wanted to keep our spatial features as sharp as possible, we used only a 3$\times$3 spatial averaging window and then obtained additional smoothing from segmentation of the 12-hour interval.  This combination seemed to involve the least amount of averaging required to obtain few failed fits and for the computation to complete in a reasonable amount of time. Time segments of two hours in length were used because we found their power-law indices were similar to those obtained from using the full 12-hour segment, and when segments of one hour were used, the power-law indices began to show substantial differences.

The parameter ranges in Table~\ref{t:bounds} used for optimization were based on those used by \cite{Ireland15} and were refined to those presented through a process of trial and error. The lower bound for $C$ was chosen so that its middle value was near the center of the histogram peak in resulting distributions. Our value of $n=0.3$ is lower than that of \cite{Ireland15} because our region of interest included a coronal hole, which we observe to frequently have such low power-law indices.

The fourth step was introduced because although the Dog-Box method produced in general the best fits of the optimization methods in $\sf{SciPy's optimize.curve\_fit}$ optimization package, in some cases we found unphysical spikes in the histogram of the 1598$\times$1598 fits for one or more of the six model parameters of the for a given wavelength at values that were at the centers of the parameter bounds given in Table~\ref{t:bounds}. This was found to be due to the fact that the Dog-Box optimization method uses the centers of the parameter bounds as the initial guesses, resulting in early termination of the minimization algorithm if a local minimum of the function happened to be found near there. This issue was corrected by the use of a two-stage/step curve fit routine that used the Dog-Box method to compute initial parameter estimates that were then used as the initial estimates for a TRF optimization.

\begin{table}
\caption{Parameter bounds used in model fitting. The parameter $\beta$ is provided in both frequency [mHz] and temporal [minutes] units.}
\label{t:bounds}
\begin{tabular}{c||cc}     
  \hline                   
\multicolumn{3}{c}{Parameter Constraints} \\
\hline 
Parameter & Lower Bound & Upper Bound  \\
  \hline
$A$ & 0. & $2\times 10^{-3}$  \\
$n$ & 0.3 & 6.0  \\
$C$ & -0.01 & 0.01  \\
$\alpha$ & $10^{-5}$ & 0.2  \\
$\beta$ frequency [mHz] & 1.5 & 10 \\
$\beta$ period [min] & 11.1 & 1.66\\
$FWHM$ [min] & 0.9 & 110.8\\
  \hline
\end{tabular}
\end{table}

\section{Significance Calculation} 
\label{Appendix:Masking}

Model M1 has a total of $p_1=3$ adjustable parameters and Model M2 has $p_2=6$.  As a result, M2 is expected to provide a better fit to the spectra on average.  The $F$ test is used to determine when this is meaningful.  The $F$-statistic associated with this test, which applies when M1 is nested in M2, is

\begin{equation}
F = {\frac{\left(\frac{RSS_1-RSS_2}{p_{2}-p_{1}}\right)}{\left(\frac{RSS_2}{n-p_{2}}\right)}}
\label{eq:f_statistic}
\end{equation}
where $n$ is the number of data points used to estimate the model parameters and $RSS$ is the weighted sum of squared residuals.  The null hypothesis is that M1 fits the data as well as M2 (stated formally, that the $\alpha$ parameter in M2 is zero).  This hypothesis is rejected when the value of the $F$-statistic is above the threshold in the $F(p_2-p_1,n-p_2)$ distribution associated with a false rejection probability of $p<0.005$.  Stated informally, we claim a spectra has a periodic component when $p<0.005$ with the expectation that this claim is false at most $0.5$\,\% of the time.  The threshold of $p=0.005$ was chosen after testing values over two orders of magnitude.  We found that $p<0.005$ gave null hypothesis rejections that were most consistent with visual inspection of the two model fits.  

For consistency, the $p<0.005$ threshold was applied to all wavelengths. As noted in the main text, relaxation of the $p<0.005$ constraint to $p<0.05$ revealed additional structure in the coronal bullseye in 193\,{\AA} that is likely to be real because of the spatial coherence of the structure. However, the use of $p<0.05$ for all wavelengths would have added significantly more noise to the observations in the wavelengths that had more locations with spectra with a significant periodic component.

\begin{figure*}[htbp]
\centering
{\includegraphics[scale=0.43]{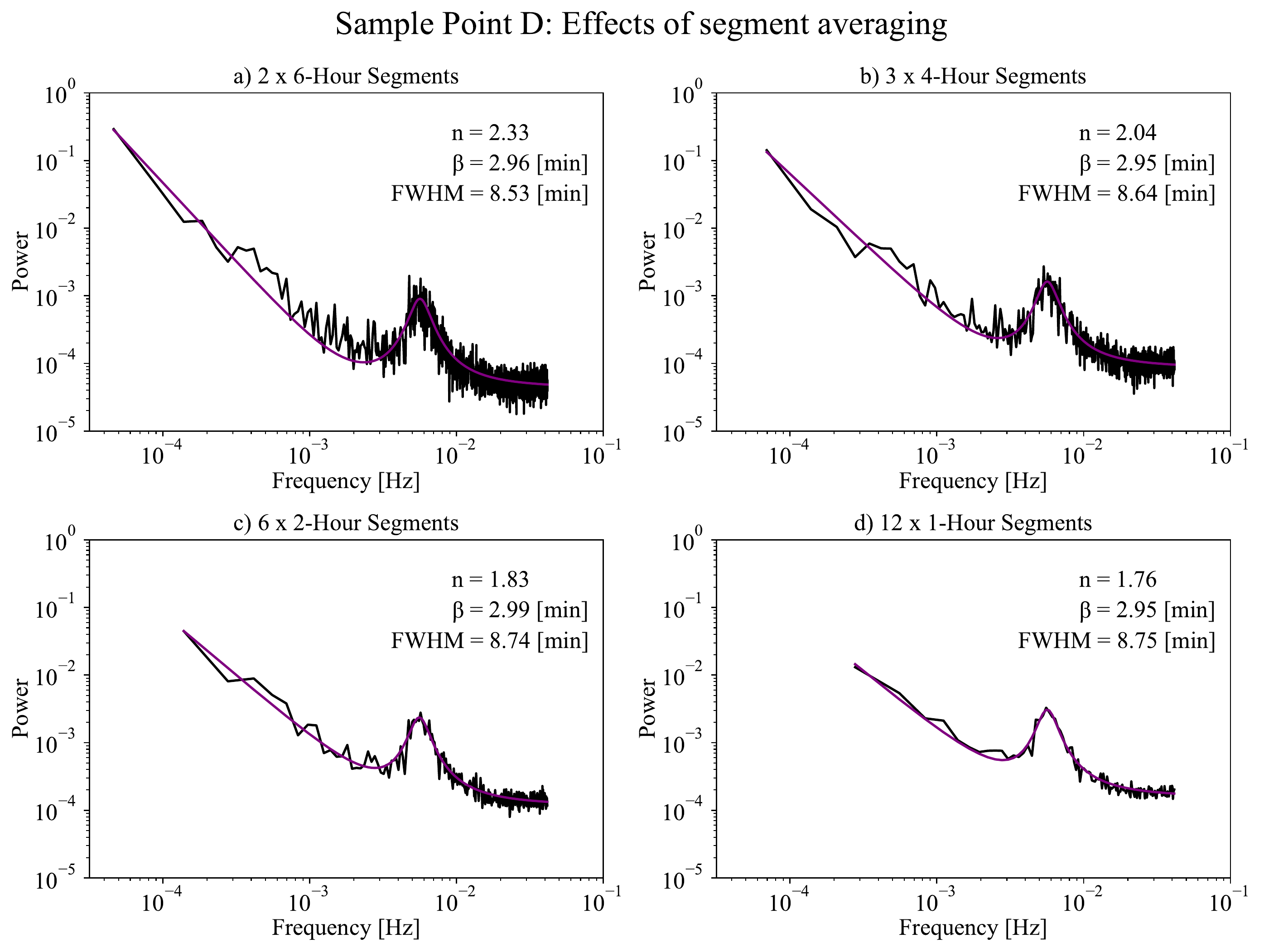}} 
{\includegraphics[scale=0.41]{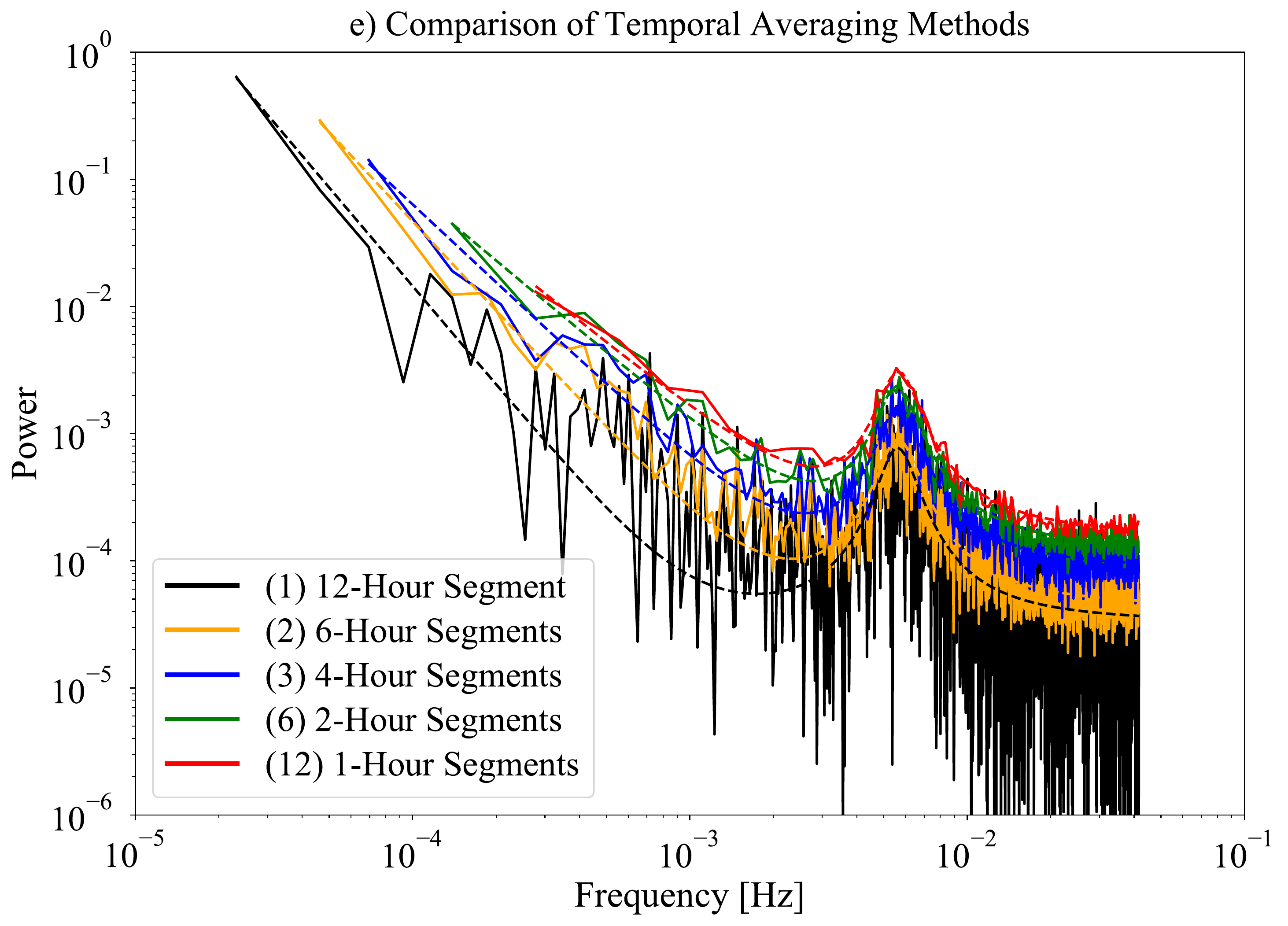}} 
\caption{A demonstration of the effects of the segmenting procedure for average of spectra to reduce noise. Each of the four upper panels represents a different level of segmenting a full 12-hour sequence using two six-hour sequences (\textit{upper-left}), three four-hour sequences (\textit{upper-right}), six two-hour sequences (\textit{lower-left, as used in this study}) and twelve one-hour sequences (\textit{lower-right}). These averaging choices correspond to those discussed in Section~\ref{s:segment_discuss}. The original data time series for these plots was obtained from Sample Point D (Figure~\ref{f:All_viz}a) in 171\,{\AA}, and was averaged in a 3$\times$3-pixel box per the methodology described in Section~\ref{s:fitting}. The large panel in the lower half of the image overlays all four averaged spectra on top of the raw (unaveraged) spectrum of the original time series (black).}
\label{f:appendix_seg_avg}
\end{figure*}

\section{Segment Averaging}
\label{Appendix:seg_avg}

To illustrate the impact of the segment averaging procedure on power spectra, Figure~\ref{f:appendix_seg_avg} has four power spectra obtained from location D in Figure~\ref{f:All_viz}a (171\,{\AA} with the different degrees of segment averaging discussed in Section~\ref{s:segment_discuss}. Specifically we present averaging of: two six-hour sequences (Figure~\ref{f:appendix_seg_avg}a), three four-hour sequences (Figure~\ref{f:appendix_seg_avg}b), six two-hour sequences (Figure~\ref{f:appendix_seg_avg}c) and 12 one-hour sequences (Figure~\ref{f:appendix_seg_avg}d). These spectra are additionally averaged using the 3$\times$3-pixel average procedure described in Section~\ref{s:fitting}. Figure~\ref{f:appendix_seg_avg}e presents these averaged spectra overlaid on a raw (unaveraged) spectrum obtained from the original 12-hour time series. 

Figure~\ref{f:appendix_seg_avg} illustrates our finding that the spectral averaging has the most impact on power-law indices, with \textit{n} varying from 2.33 to 1.76 for this particular location, but with the Lorentzian contribution essentially unchanged regardless of segmenting. In both cases, we note that noise is far lower in spectra in which the segmented averaging is applied, resulting in almost no failed model fits for large regions of interest. As discussed in Section~\ref{s:segment_discuss}, the decrease of 0.57 in power-law index seen in this example is atypically high for the 171\,{\AA} channel, where across the entire region of interest we observe that the median variation due to segmenting is 0.04 with a standard deviation of 0.30 (per Table~\ref{t:segment_stats}).

\section{Example Spectra: Poor and Good Fits in 171 and 1700\,{\AA}}
\label{Appendix:example_spectra}

\begin{figure}
  \centering
  \setlength{\tabcolsep}{0.5mm}
  \begin{tabular}{cc}
  \includegraphics[scale=0.215]{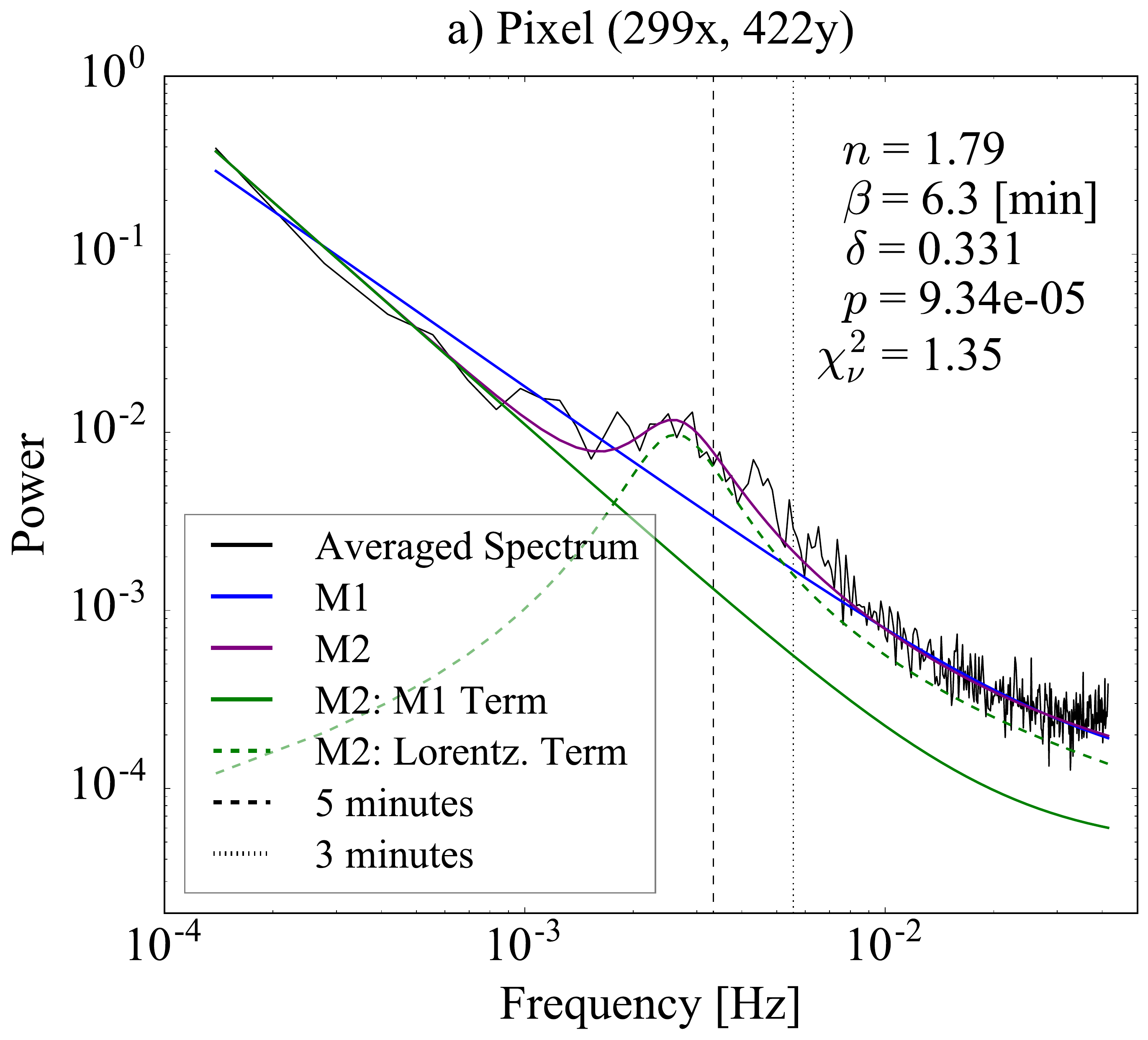}&\includegraphics[scale=0.215]{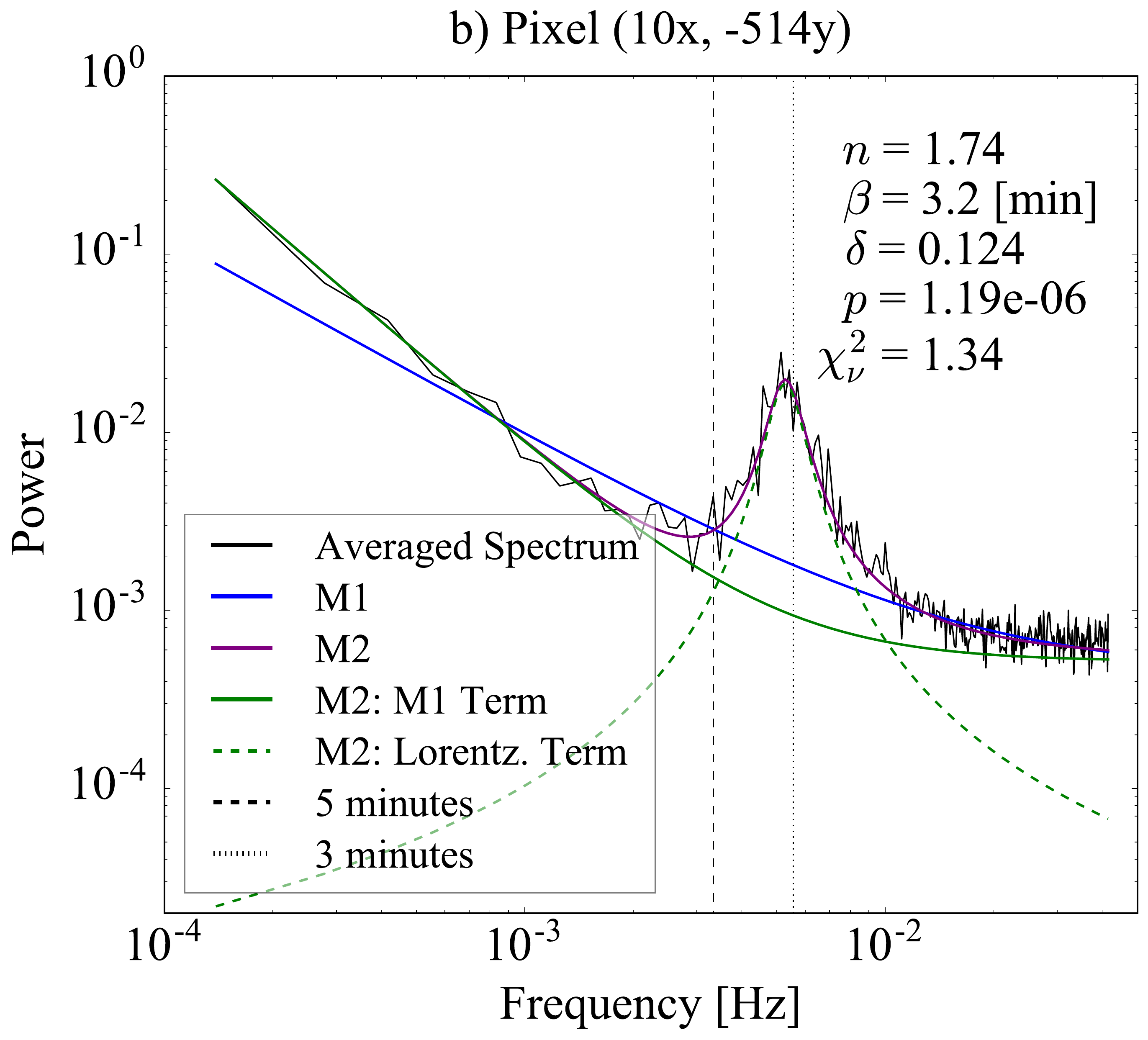}\\
  \includegraphics[scale=0.215]{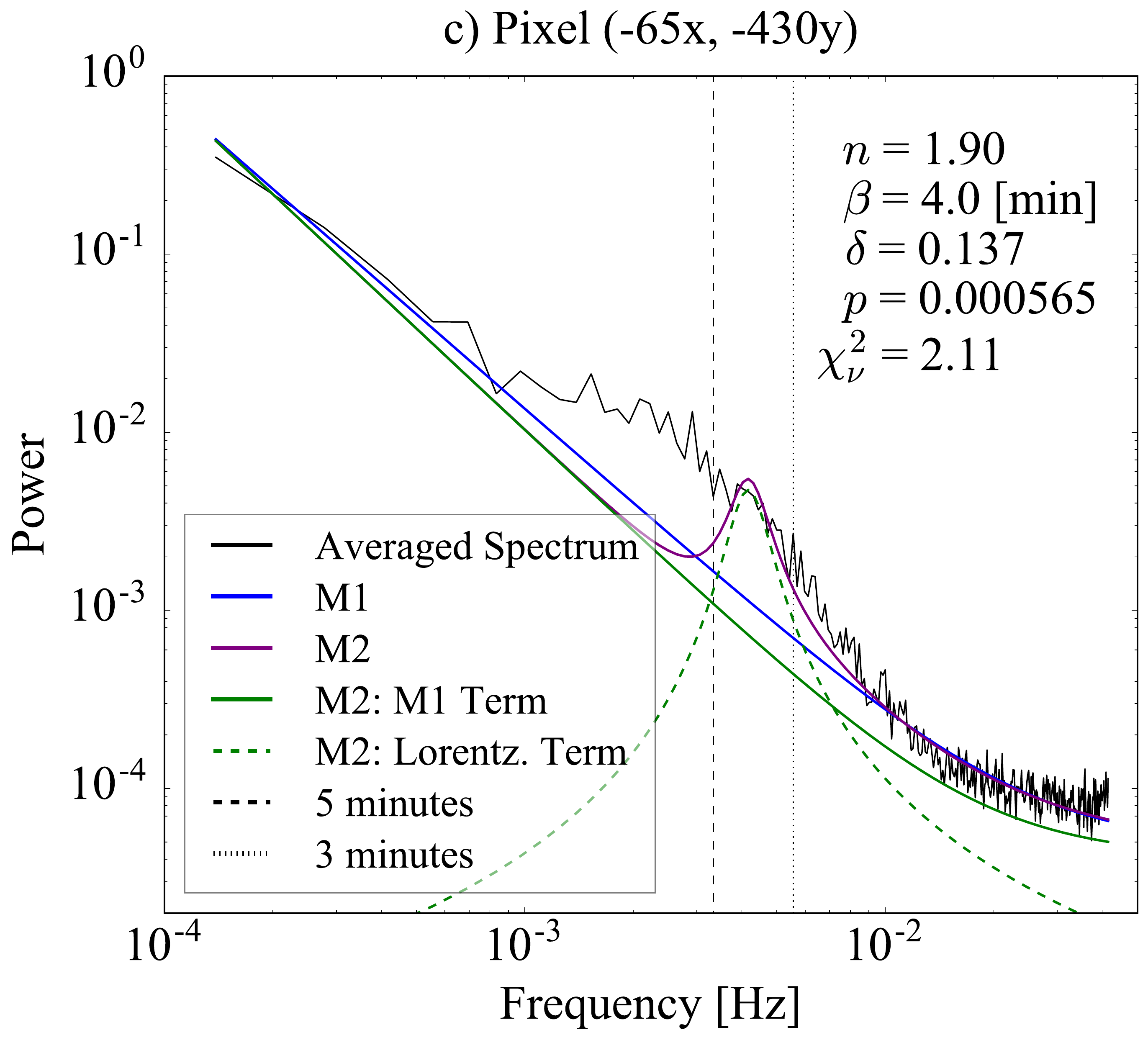}&\includegraphics[scale=0.215]{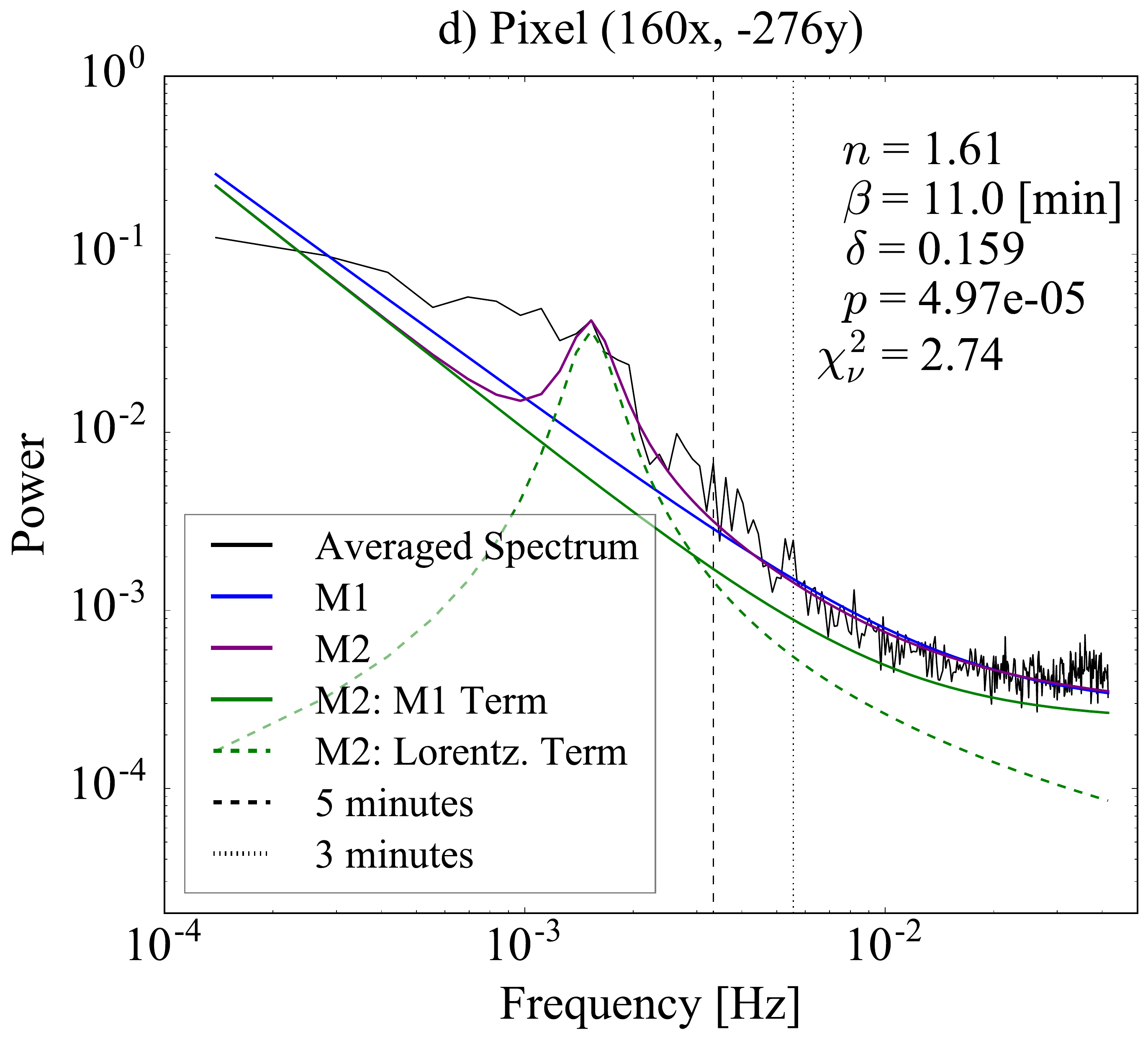} 
  \end{tabular}
  \caption{Examples of good (\textit{top row}) and poor (\textit{bottom row}) model fits in the 171\,{\AA} observations. Panel a corresponds to a bright footpoint just above the coronal hole; Panel b corresponds to a point inside the sunspot umbra; Panel c corresponds to a point within the magnetic network structure as identified in the magnetogram observations (white region of Figure~\ref{f:All_viz}b, just to the northwest of the active region). Panel d corresponds to the boundary of the coronal hole. All point coordinates use the same axes as those shown in Figure~\ref{f:All_viz}d.}
\label{f:171_examples}
\end{figure}

\begin{figure}
  \centering
  \setlength{\tabcolsep}{0.5mm}
  \begin{tabular}{cc}
  \includegraphics[scale=0.215]{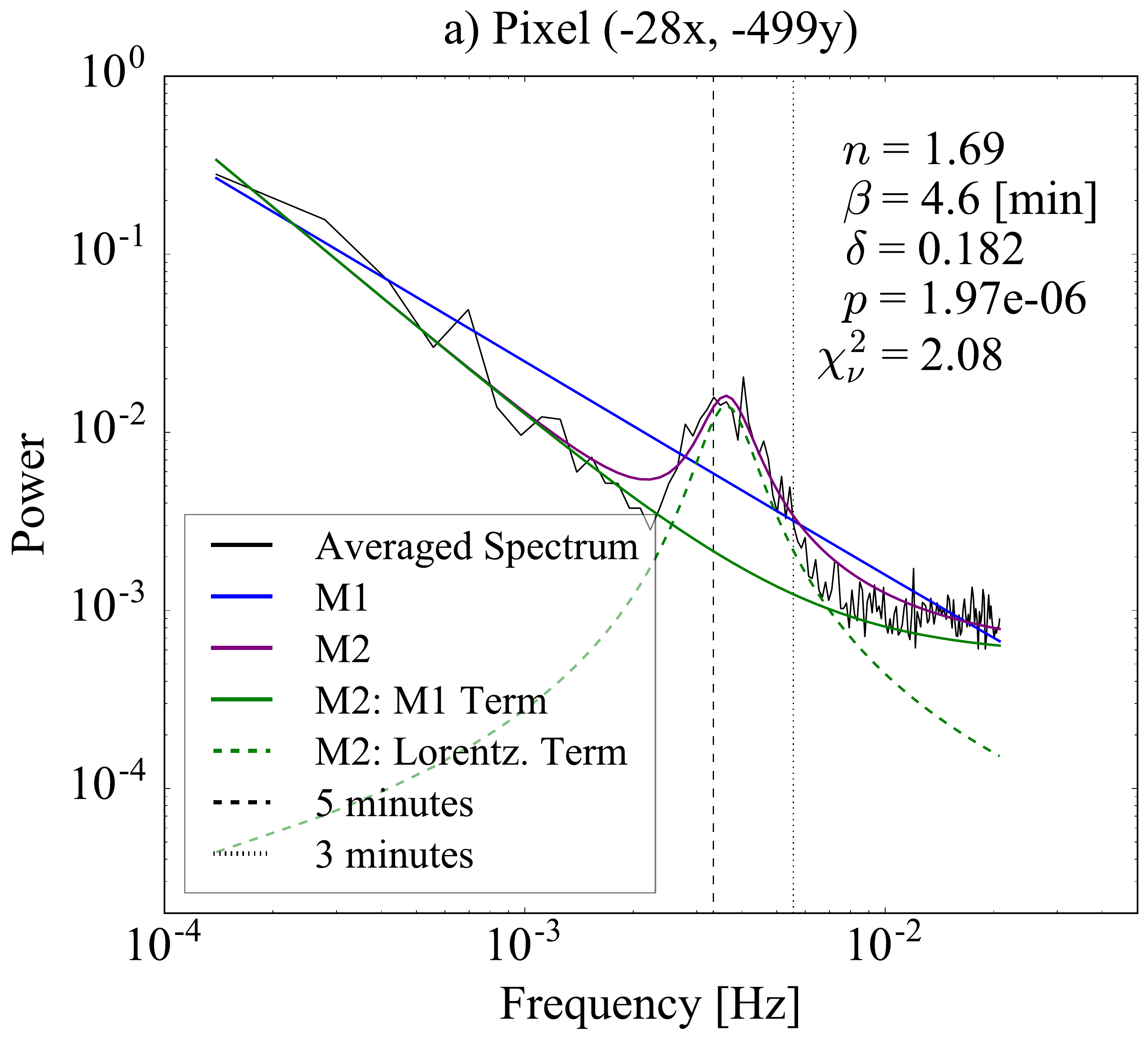}&\includegraphics[scale=0.215]{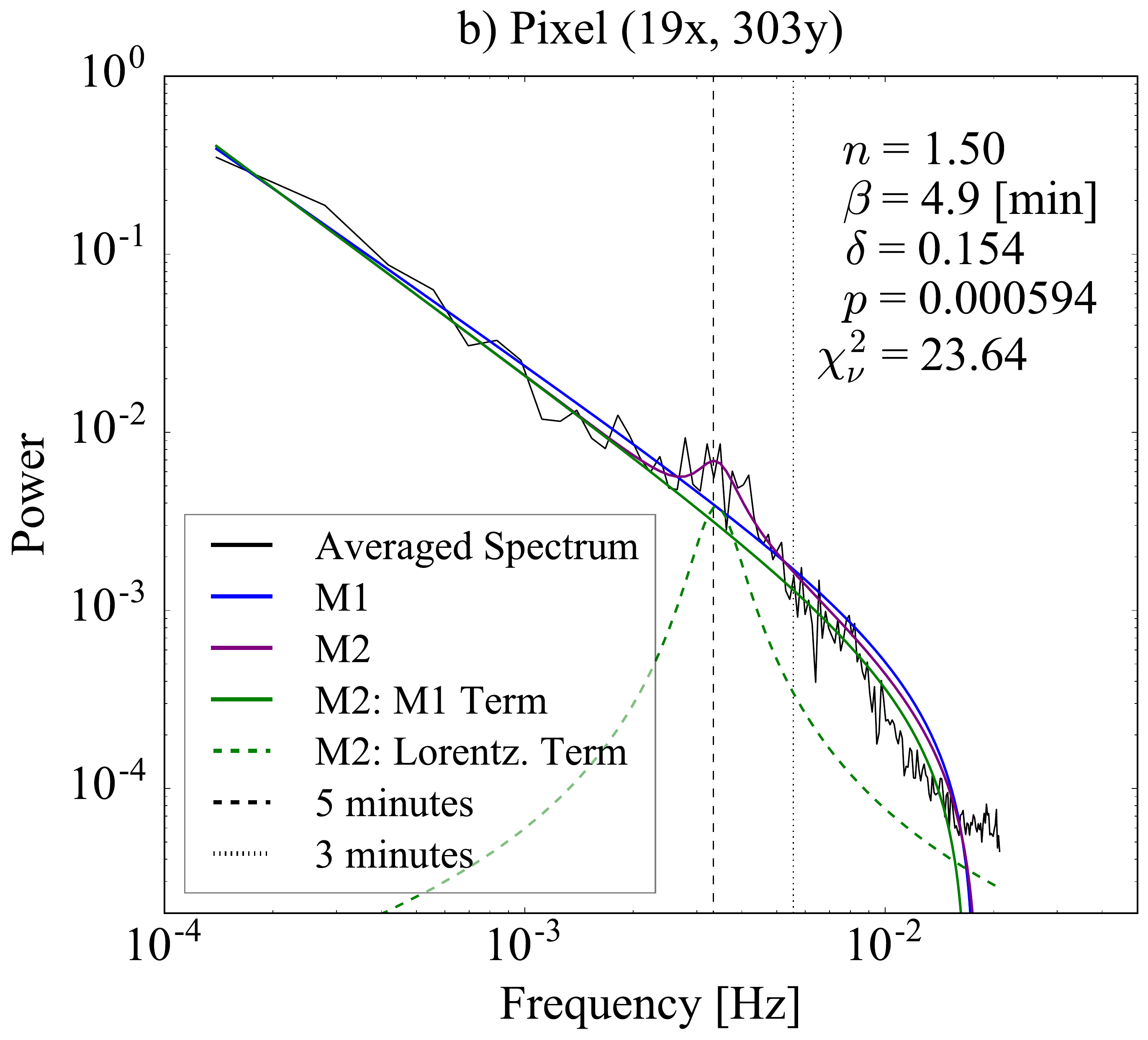}\\
  \includegraphics[scale=0.215]{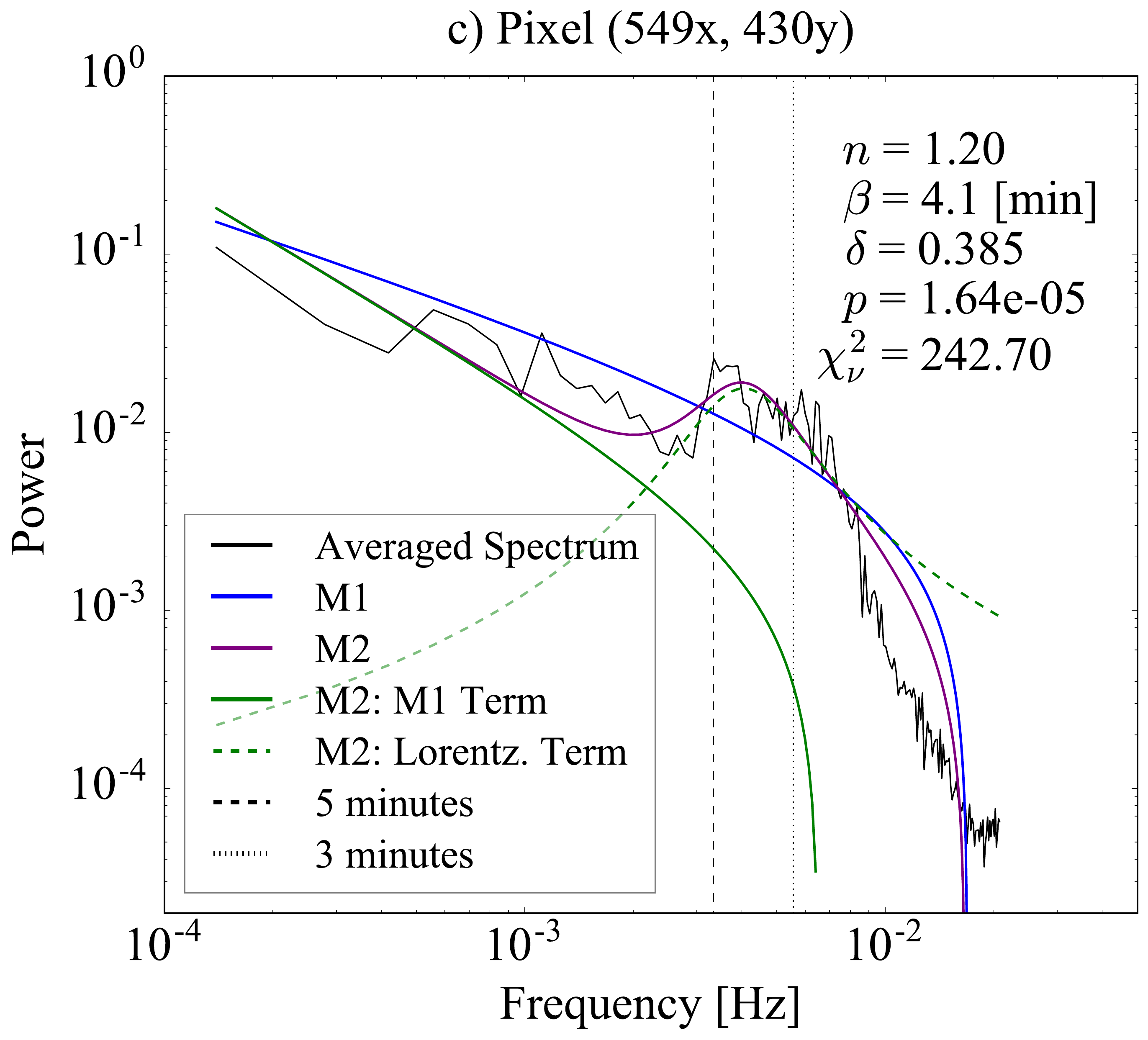}&\includegraphics[scale=0.215]{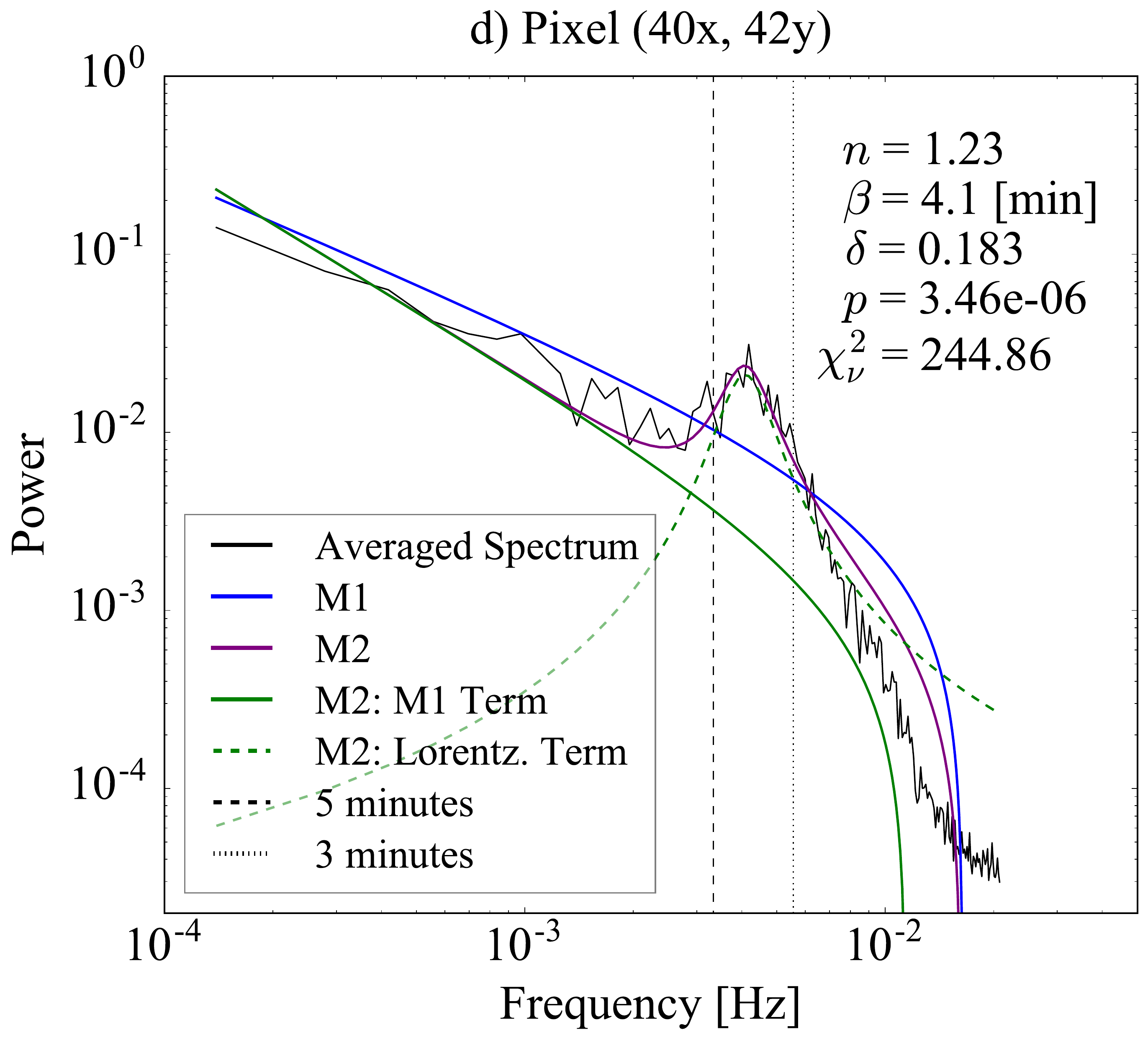} 
  \end{tabular}
  \caption{Examples of good (\textit{top row}) and poor (\textit{bottom row}) model fits in the 1700\,{\AA} observations. Panel a corresponds to the sunspot penumbra; Panel b corresponds to a point on the boundary of the magnetic network; Panel c is also a point on the boundary of the magnetic network; and Panel d corresponds to a location distant from both the magnetic network and the sunspot. All point coordinates use the same axes as those shown in Figure~\ref{f:All_viz}f.}
\label{f:1700_examples}
\end{figure}

Figures \ref{f:171_examples} and \ref{f:1700_examples} provide a limited set of examples of good (panels a and b) and poor (panels c and d) model fits in the 171\,{\AA} and 1700\,{\AA} observations respectively. The pixel locations use the same coordinate axes as those shown in Figure~\ref{f:All_viz}d. The presented fits do not necessarily represent the very worst or very best cases of each; our methodology provided us with approximately 6.4 million such fits, of which these are only representative sample in two of the four channels studied. 

In Figure~\ref{f:171_examples}a and b we show two examples of spectra that fit well to our M2 model, \textit{i.e.} include significant Lorentzian components. The spectrum in Figure~\ref{f:171_examples}a) corresponds to a bright footpoint just above the coronal hole, and Figure~\ref{f:171_examples}b to a point inside the sunspot umbra.

Figure~\ref{f:171_examples}c and d shows two examples of spectra in which we observe a very broad spectral hump that our model is unable to capture. Figure~\ref{f:171_examples}c corresponds to a point within the magnetic network structure as identified in the magnetogram observations (white region of Figure~\ref{f:All_viz}b, just to the northwest of the active region, and Figure~\ref{f:171_examples}d corresponds to the base of a plage. Both models capture the high-frequency part of the spectrum well, but they fail on the low-frequency observations. Such fits are not typical of our results in this channel, but they highlight the kinds of power spectra that do not fit our model well (both of which would likely be better represented by the Kappa model). It is worth noting that both of these particular spectra are considered statistically significant Lorentzians in our methodology, despite a seemingly poor fit. In limited tests we see that spectra like these are well-fit by the Kappa model employed by \cite{Auchere16a} with one interesting result that the $\rho$ term in the Kappa function, stated as equal to $T/120$, where T is the even spacing of the pulses (in hours), returns values very close to those as located by our Lorentzian component. However, we reiterate that this result is preliminary, and it may be misinterpreted given our limited investigations of the Kappa model.

In the upper row of Figure~\ref{f:1700_examples} we show two examples of reasonably well-fit spectra in 1700\,{\AA}.  The pixel locations use the same coordinate axes as those shown in Figure~\ref{f:All_viz}f. Figure~\ref{f:1700_examples}a corresponds to the sunspot penumbra and is typical of sunspot umbra and penumbra spectra, the majority of which are extremely well-fit, while Figure~\ref{f:1700_examples}b corresponds to a point just inside the magnetic network, very close to its boundary, in which the power spectrum is almost a simple power law with a small but clear ``bump'' at $\approx$ five-minutes. The reduced $\chi^2$ here is lowered by the falloff of the model at very high frequencies (a common issue with using our model to fit chromospheric power spectra). 

The lower row of Figure~\ref{f:1700_examples} shows two examples of poor model fits. Figure~\ref{f:1700_examples}c is a point very close to the boundary of the magnetic network, but just outside the network, presenting a power spectrum that would likely be better fit by a broken power law, and Figure~\ref{f:1700_examples}d is located distant from both the magnetic network and the sunspot, and highlights a spectrum in which the model fit is visually quite good yet misses most of the very high-frequency points ($\lesssim$ 2.5-minutes, which generally we do not care about) and several points above the $\approx$ five-minute line, resulting in a very poor $\chi^2$ value of 244.86. Despite this, we can see the Lorentzian component clearly fits well the apparent oscillation at $\approx$ 4.1-minutes, and this supports our confidence in identifying global oscillations in both 1700 and 1600\,{\AA} concurrent with those identified in other studies. As a general rule, the $\chi^2$ values for 1700\,{\AA} are impacted most by the model missing the very highest frequencies. The impact of this, however, may be arguably low given that we do not focus on any features with periodicities shorter than three-minutes, and the 24-second cadence of the 1700\,{\AA} observations means that aliasing artifacts could reasonably be expected in part of this spectrum. Nonetheless this model is clearly not optimal for describing the majority of chromospheric power spectra.

\clearpage

%
%

%
%
\bibliographystyle{spr-mp-sola}
\bibliography{SolarSpectra}  
%
%
%
%

\end{article} 
\end{document}

%% file: defs.tex
\newcommand{\BibTeX}{\textsc{Bib}\TeX}
\newcommand{\etal}{{\it et al.}}
\newcommand{\mdash}{{-}}

\newcommand{\deriv}[2]{\frac{{\mathrm d} #1}{{\mathrm d} #2}}
\newcommand{\rmd}{ {\ \mathrm d} }
\renewcommand{\vec}[1]{ {\mathbf #1} }
\newcommand{\uvec}[1]{ \hat{\mathbf #1} }
\newcommand{\pder}[2]{ \f{\partial #1}{\partial #2} }
\newcommand{\grad}{ {\bf \nabla } }
\newcommand{\curl}{ {\bf \nabla} \times}
\newcommand{\vol}{ {\mathcal V} }
\newcommand{\bndry}{ {\mathcal S} }
\newcommand{\dv}{~{\mathrm d}^3 x}
\newcommand{\da}{~{\mathrm d}^2 x}
\newcommand{\dl}{~{\mathrm d} l}
\newcommand{\dt}{~{\mathrm d}t}
\newcommand{\intv}{\int_{\vol}^{}}
\newcommand{\inta}{\int_{\bndry}^{}}
\newcommand{\avec}{ \vec A}
\newcommand{\ap}{ \vec A_p}
\newcommand{\bb}{ \vec B}
\newcommand{\jj}{ \vec j}
\newcommand{\rr}{ \vec r}
\newcommand{\xx}{ \vec x}

\newcommand{\adv}{    {\it Adv. Space Res.}} 
\newcommand{\annG}{   {\it Ann. Geophys.}} 
\newcommand{\aap}{    {\it Astron. Astrophys.}}
\newcommand{\aaps}{   {\it Astron. Astrophys. Suppl.}}
\newcommand{\aapr}{   {\it Astron. Astrophys. Rev.}}
\newcommand{\ag}{     {\it Ann. Geophys.}}
\newcommand{\aj}{     {\it Astron. J.}} 
\newcommand{\apj}{    {\it Astrophys. J.}}
\newcommand{\apjl}{   {\it Astrophys. J. Lett.}}
\newcommand{\apss}{   {\it Astrophys. Space Sci.}} 
\newcommand{\cjaa}{   {\it Chin. J. Astron. Astrophys.}} 
\newcommand{\gafd}{   {\it Geophys. Astrophys. Fluid Dyn.}}
\newcommand{\grl}{    {\it Geophys. Res. Lett.}}
\newcommand{\ijga}{   {\it Int. J. Geomagn. Aeron.}}
\newcommand{\jastp}{  {\it J. Atmos. Solar-Terr. Phys.}} 
\newcommand{\jgr}{    {\it J. Geophys. Res.}}
\newcommand{\mnras}{  {\it Mon. Not. Roy. Astron. Soc.}}
\newcommand{\nat}{    {\it Nature}}
\newcommand{\pasp}{   {\it Pub. Astron. Soc. Pac.}}
\newcommand{\pasj}{   {\it Pub. Astron. Soc. Japan}}
\newcommand{\pre}{    {\it Phys. Rev. E}}
\newcommand{\solphys}{{\it Solar Phys.}}
\newcommand{\sovast}{ {\it Soviet  Astron.}} 
\newcommand{\ssr}{    {\it Space Sci. Rev.}} 


 

